\documentclass[12pt]{article}

\usepackage{amsfonts, amsmath, amssymb}
\usepackage{cancel,slashed}
\usepackage{bbm}
\usepackage{amssymb}
\usepackage{graphicx}
\usepackage{cite}

\newcommand{\bmat}{\left(\begin{array}}
\newcommand{\emat}{\end{array}\right)}
\newcommand{\uno}{\mathbbm{1}}

\def\p{\partial}
\def\a{\alpha}

\def\b{\beta}
\def\g{\gamma}
\def\d{\delta}
\def\th{\theta}
\def\om{\omega}

\def\vphi{\varphi}
\def\-{\hphantom{-}}

\def\s2{\frac{1}{\sqrt2}}

\def\oh{\frac{1}{2}}
\def\beq{\begin{equation}}
\def\eeq{\end{equation}}
\def\beqa{\begin{eqnarray}}
\def\eeqa{\end{eqnarray}}

\def\im{{\rm Im \,}}

\def\tr{{\rm tr \,}}
\def\Tr{{\rm Tr \,}}

\def\ca{{\mathcal A}}

\def\cd{{\mathcal D}}

\def\cn{{\mathcal N}}

\def\car{{\mathcal R}}
\def\co{{\mathcal O}}

\def\Dsl{\,\raise.15ex\hbox{/}\mkern-13.5mu D} 
\def\r#1{\mbox{{\bf #1}}}

\def\r#1{{\bf #1}}
\def\br#1{{\bf \overline{#1}}}

\def\e{\epsilon}
\def\mff{\mathfrak{F}}
\def\mfa{\mathfrak{A}}

\def\CL {{\cal L}}
\def\CA {{\cal A}}

\def\CO {{\cal O}}

\def\im{\mbox{Im }}
\def\tr{\mbox{Tr}}
\def\str{\mbox{STr}}

\def\be{\begin{equation}}
\def\ee{\end{equation}}
\def\bea{\begin{eqnarray}}
\def\eea{\end{eqnarray}}
\def\raw{\rightarrow}

\def\IN{\mathbb{N}}

\def\oh{\frac{1}{2}}
\def\a{{\alpha}}
\def\b{{\beta}}
\def\d{{\delta}}
\def\eps{{\epsilon}}
\def\th{{\theta}}

\def\lam{{\lambda}}

\def\om{{\omega}}
\def\sig{{\sigma}}

\def\g{{\gamma}}

\def\vphi{{\varphi}}
\def\p{{\partial}}

\newcommand{\mfn}[1]{\mbox{\footnotesize$#1$}}
\def\vec#1{{\overrightarrow{#1}}}
\def\str{\mbox{STr}}

\def\w{{\wedge}}
\def\ph{{\partial_{\langle A\rangle}}}
\def\pa{{\bar \partial_{\langle A\rangle}}}

\def\lp{{\langle\Phi\rangle}}
\def\lpc{{\langle\Phi_{xy}\rangle}}

\def\sm2{{\mbox{\small 2}}}

\begin{document}
\pagestyle{plain}

\makeatletter
\@addtoreset{equation}{section}
\makeatother
\renewcommand{\theequation}{\thesection.\arabic{equation}}
\pagestyle{empty}
\rightline{ IFT-UAM/CSIC-12-110}
\vspace{0.5cm}
\begin{center}
\LARGE{ Non-perturbative effects and  Yukawa hierarchies  \\
 in   F-theory  $SU(5)$ Unification 
\\[10mm]}
\large{A. Font$^1$, L. E. Ib\'a\~nez$^{2,3}$, F. Marchesano$^2$ and D. Regalado$^{2,3}$\\[6mm]}
\small{
${}^1$  Departamento de F\'{\i}sica, Centro de F\'{\i}sica Te\'orica y Computacional \\[-0.3em]
 Facultad de Ciencias, Universidad Central de Venezuela\\[-0.3em]
 A.P. 20513, Caracas 1020-A, Venezuela  \\[2mm] 
${}^2$ Instituto de F\'{\i}sica Te\'orica UAM-CSIC, Cantoblanco, 28049 Madrid, Spain \\[2mm] 
${}^3$ Departamento de F\'{\i}sica Te\'orica, 
Universidad Aut\'onoma de Madrid, 
28049 Madrid, Spain
\\[8mm]} 
\small{\bf Abstract} \\[5mm]
\end{center}
\begin{center}
\begin{minipage}[h]{15.0cm} 
Local $SU(5)$ F-theory models lead naturally to Yukawa couplings for the third generation
of quarks and leptons, but inducing Yukawas for the lighter generations
has proven elusive. Corrections coming from gauge fluxes fail to generate the required couplings,
and naively the same applies to instanton effects. We nevertheless revisit  the effect of instantons in F-theory
GUT constructions and find that contributions previously ignored in the literature induce the
leading non-perturbative corrections to the Yukawa couplings. We apply our results to the case of
$10\times {\bar 5}\times {\bar 5}$ couplings in local $SU(5)$ F-theory GUTs, showing
that non-perturbative effects naturally lead to hierarchical Yukawas.
The hypercharge flux required to break $SU(5)$ down to
the SM does not affect the holomorphic Yukawas but
does modify the profile of the wavefunctions, explaining the
difference between the D-quark and lepton couplings at the unification scale.
The combination of non-perturbative corrections and
magnetic fluxes allows to describe the  measured lepton and
D-quark masses  of second and third generations  in the SM.

\end{minipage}
\end{center}
\newpage
\setcounter{page}{1}
\pagestyle{plain}
\renewcommand{\thefootnote}{\arabic{footnote}}
\setcounter{footnote}{0}


\tableofcontents


\section{Introduction}

One of the most difficult puzzles of the  Standard Model (SM) is the structure of fermion masses and mixings. If the SM arises as a low-energy limit
of an underlying string theory \cite{thebook} , it should be possible to understand  this structure in terms of more fundamental parameters characterizing 
the string vacuum.  In particular the values of Yukawa couplings in string compactifications are determined by the geometric properties of
extra compactified dimensions. An explicit computation of Yukawa coupling constants seems then quite difficult since we would need
detailed information about the geometry of the corresponding compact space.

It has been however realized  \cite{aiqu} that, due to the localization properties of branes, some quantities of physical interest do not depend on the
full geometry of the compactification space but rather on local information around the region in which the SM fields are localized. 
This is particularly the case of Type IIB string compactifications with the SM fields localized on D$p$-branes with $p\leq 7$ and also
on F-theory constructions. In these Type IIB compactifications the Yukawa couplings are obtained as overlap integrals involving the 
three wavefunctions of the quark/lepton  and Higgs  states involved.  In particular, in the case of F-theory $SU(5)$ GUT's the 
quark/lepton fields are localized on  complex {\it matter curves} with ${\bf {\bar 5}},{\bf 10}$ quantum numbers. Yukawa couplings 
appear at the points of intersection of three matter curves corresponding to a right-handed fermion, a left-handed fermion
and a Higgs multiplet transforming also as a 5-plet.  With the wavefunctions of the three fields localized on the matter curves, the
overlap integral is then dominated by the local properties of the wavefunction around the intersection point and hence no 
global information is required to compute the holomorphic Yukawa couplings. This opens the door to the explicit computation of Yukawa 
couplings in  string compactifications with non-trivial curved compact spaces.

Particularly interesting from this point of view are the mentioned  local F-theory $SU(5)$ GUT models
\cite{dw1,bhv1,bhv2,dw2},  which have been recently the subject of intense study (for reviews see e.g. \cite{ftheoryreviews}).
These models are able to combine advantages of heterotic compactifications (gauge coupling unification) and Type IIB orientifolds
(localization of the SM fields and moduli fixing through closed string fluxes).  In these F-theory models the local dynamics on the matter curves
is governed by the  8d effective action of 7-branes, and one can obtain explicit local expressions for the wavefunctions of the matter fields.
The Yukawa couplings arise at the triple intersection of matter curves \cite{dw1,bhv1,bhv2,dw2,hktw,hktw2}
, and can be computed from a superpotential of the form 
$W\simeq  \int \tr \left( F \wedge \Phi \right)$, where $F=dA -i A\wedge A$ is the field strength of the 8-dimensional gauge fields and $\Phi$ is a field parametrizing 
fluctuations in the transverse dimensions to the branes, and the integral extends over the 7-brane worldvolume where the $SU(5)$ degrees of
freedom are localized.  Within this simple scheme one finds that  only one generation of quark/lepton fields gets a Yukawa coupling and may eventually
become massive \cite{hv08,fi1}. This is analogous to the result  obtained in Type II toroidal orientifolds
in which the Yukawas may be computed  explicitly \cite{yukint,magnus,ms04}.
This is an interesting starting point since indeed in the SM the third generation is much heavier than the rest and one may think that some 
additional  corrections could give masses to the first two generations.\footnote{For different approaches to the generation of hierarchies of fermion masses in F-theory unification see e.g.  \cite{DudasPalti,Ross,Krippendorf}.} It was first thought \cite{hv08}  that the presence of the world-volume fluxes required 
both to get chirality and break the $SU(5)$ symmetry down to the SM could be the source of these corrections. 
However it was soon realized   \cite{cchv09,cp09,fi09}
that open string fluxes do not modify at all the holomorphic Yukawa couplings and hence cannot 
give rise to Yukawa couplings for the lighter generations.

In \cite{mm09,afim} it was pointed out that non-perturbative effects from distant D3-instantons could be the source of the required corrections.
The most obvious such corrections where found to be proportional to $\epsilon \int  \tr \left(\Phi F\wedge F\right)$ and  turn out to have 
an alternative useful description in terms of  non-commutative geometry \cite{cchv09}. 
Indeed such corrections where shown to lead to the required
corrections in  a simple toy model with gauge group $U(3)$ \cite{afim}. It was however already pointed out in this reference  that in realistic cases, namely 
$SU(5)$ GUT's  with an enhanced symmetry group $SO(12)$ or $E_{6,7,8}$ at the Yukawa triple intersection point, such corrections 
identically vanish, since the cubic trace is zero in orthogonal and exceptional groups. This seemed again to make problematic the generation 
of fermion hierarchies in F-theory GUT constructions.

In this paper we reexamine all these issues and point out that non-perturbative D3-instanton effects give rise to additional
corrections, some of them previously overlooked.  Such corrections to the superpotential have  the form
\be
W_{\rm np} =\, m_*^4\left[  \frac{\eps}{2} \sum_{n \in \IN} \int  \theta_n \,  \str \left( \Phi^n F \wedge F\right)\right]
\label{suponpintro}
\ee
where $\epsilon$ is a small parameter and $\theta_n$ are holomorphic functions of the local coordinates.  
As mentioned above the contribution with $n=1$ vanishes for the realistic cases where the Yukawa enhancement groups are $SO(12)$ or $E_6$.
 We hence study in detail the remaining leading corrections to the Yukawa couplings induced by the $n=0$ and $n=2$ terms, applied to the 
 $SO(12)$ case which is relevant for the Yukawa couplings of charged leptons and D-quarks.

Describing non-perturbative corrections as in  (\ref{suponpintro}) simplifies the procedure to compute corrected Yukawas. More precisely, one may apply dimensional reduction techniques to express them in terms of a triple overlap of zero mode internal wavefunctions. In this sense, the presence of $W_{\rm np}$ has a two-fold effect. On the one hand it modifies the zero mode internal wavefunction profile and on the other hand it induces new 8d couplings that upon dimensional reduction become new sources of Yukawa couplings. As in \cite{afim}, taking both effects into account gives an interesting Yukawa pattern, in which the holomorphic Yukawas depend on $\th_n$ but are independent of worldvolume fluxes. While in \cite{afim} this result can be guessed based on a dual non-commutative description of the 7-brane superpotential \cite{cchv09,mm09}, for the general case (\ref{suponpintro}) such description is not available. Nevertheless, one can still generalize the results of \cite{cchv09} to obtain a residue formula that computes holomorphic Yukawas, and where the flux-independence of the latter is manifest.

Interestingly enough we find that a hierarchy of mass eigenvalues of the form  $(1,\epsilon,\epsilon^2)$ is automatically present,
explaining the observed hierarchical structure. Here $\epsilon\simeq 10^{-2}$ is a small non-perturbative parameter measuring the 
size of the effects induced by the distant instantons. 
 The holomorphic Yukawa couplings obtained are identical for both D-quarks and leptons,
since they live in the same  $SU(5)$ representations and the holomorphic Yukawas are flux independent. This looks problematic since running up in energies the observed D-quark and lepton masses, unification  of Yukawa couplings $Y_{b,s,d}=Y_{\tau,\mu,e}$ does not hold experimentally, rather leptons of the second and third generations tend to have larger Yukawa couplings than the respective D-quarks at the unification scale. We find however that the hypercharge flux required for the $SU(5)\rightarrow SM$ symmetry breaking may explain this difference. Roughly speaking, the difference may be understood as arising from the fact that the wavefunctions for leptons are more localized than those of D-quarks, due to the fact that they have larger hypercharge quantum numbers.

The structure of this paper is as follows. In Section \ref{s:reviewF} we review the construction of local F-theory GUTs. In section \ref{s:so12} we construct a local $SU(5)$ GUT model with enhanced $SO(12)$ symmetry, which describes the Yukawa couplings of charged leptons and D-quarks. The spectrum of zero modes reproduces the matter content of the MSSM, but the Yukawa couplings exhibit the rank-one structure mentioned above. In section \ref{s:nplocal} we introduce the non-perturbative effects that will give rise to the superpotential (\ref{suponpintro}), and compute the corrected zero mode equations. Such equations are solved in section \ref{s:npwave} for the model of $SO(12)$ enhancement constructed before, while the corresponding Yukawas are computed in section \ref{s:yc}. The discussion of these last two sections is slightly technical, and the reader not interested in such details may safely skip to section \ref{s:pheno}, where a phenomenological analysis of the final Yukawa couplings is performed. In particular, we confront our results with the measured masses of D-quarks and charged leptons, showing that a natural hierarchy of masses arises and that the effect of the hypercharge flux allows us to understand the ratios between them.

Several technical details have been relegated to the appendices. Appendix \ref{ap:wave} solves the zero mode wavefunctions for 
the $SO(12)$ model in absence of non-perturbative effects, and compute the wavefunction normalization factors which encode the hypercharge flux dependence of the Yukawa couplings. Appendix \ref{ap:flux} discusses in some detail the choice of worldvolume fluxes made for this $SO(12)$ model, motivating them via the notion of local chirality in F-theory. Appendix \ref{ap:supo} derives the non-perturbative superpotential (\ref{suponpintro}), and shows that the D-term is not corrected. Finally, in appendix \ref{ap:res} we derive a residue formula for the non-perturbative Yukawa couplings, that allows to cross-check and extend the results obtained in the main text.

\section{Review of local F-theory models}
\label{s:reviewF}

Following the general scheme of \cite{bhv1,bhv2,dw1,dw2} (see also \cite{collinucci09,bgjw09,mss2,Cordova,mss,gkw09,grimm,dp10}), in order to construct a local F-theory GUT model one may consider a stack of 7-branes wrapping a compact divisor $S_{\rm GUT}$ of the threefold base $B$ of an elliptically-fibered Calabi-Yau fourfold. The gauge degrees of freedom that arise from $S_{\rm GUT}$ are specified by the particular set of $(p,q)$ 7-branes that are wrapped on $S_{\rm GUT}$ or, in geometrical terms, by the singularity type of the elliptic fiber on top of such 4-cycle. Hence, one may easily engineer local models where the GUT gauge group $G_{\rm GUT}$ is given by 
$SU(5)$, $SO(10)$ or even $E_6$. 

Besides the stack of 7-branes on $S_{\rm GUT}$, a semi-realistic F-theory model will contain further 7-branes that wrap another set of divisors $S_i$, which intersect $S_{\rm GUT}$ on certain curves $\Sigma_i$. On top of the latter set of curves of $S_{\rm GUT}$ the singularity type of the elliptic fiber is enhanced, in the sense that the Dynkin diagram that is associated to the singularity corresponds to a higher rank Lie group $G_{\Sigma_i}$  that contains $G_{\rm GUT}$. In practice, this implies that new degrees of freedom appear at the intersection of the 7-branes, more precisely chiral matter multiplets in a certain representation of $G_{\rm GUT}$, localized at the so-called matter curves $\Sigma_i$. 

Finally, two or more matter curves may meet at a point $p \in S_{\rm GUT}$ and at that point the singularity is promoted to an even higher one, such that the corresponding Lie group $G_p$ not only contains $G_{\rm GUT}$ but also each of the $G_{\Sigma_i}$ involved. This time there are no new degrees of freedom arising at the point $p$, but rather contact interactions involving the chiral multiplets from each curve $\Sigma_i$. Of particular interest are those cases where three matter curves meet at $p$, as they give rise to Yukawa couplings between chiral multiplets of the GUT matter fields. 

Of course, in the process of describing a local model one must not only specify the gauge group $G_{\rm GUT}$, but also the enhanced group $G_{\Sigma_i}$ at each of the matter curves. This information and the intersection loci of matter curves determines the groups $G_p$ at each point $p$ where Yukawa couplings develop. Typically, starting with a GUT gauge group such as $G_{\rm GUT} = SU(5)$ one may end up with enhanced groups at Yukawa points $p$ such as $SO(12)$, $E_6$, $E_7$ or $E_8$. In the next section we will analyze a local model that describes the case where $G_{\rm GUT} = SU(5)$ and $G_p = SO(12)$, which corresponds to the setup describing down-type Yukawas for a local $SU(5)$ F-theory model. 

While the above geometric picture is already quite illuminating, one of the most powerful results of \cite{bhv1,bhv2,dw1,dw2} is to provide a simple framework to compute the matter content arising at each curve $\Sigma_i$ and the Yukawa couplings at their triple intersections.  Such framework makes use of a 8d effective action related to a stack of 7-branes which, upon dimensional reduction on a 4-cycle $S$, provides all the dynamics of the 4d degrees of freedom \cite{bhv1}.\footnote{See \cite{cp09} for a derivation in terms of a 8d SYM Lagrangian.} In particular, the Yukawa couplings between 4d chiral fields arise from the superpotential
\be
W\, =\, m_*^4 \int_{S} \tr \left( F \wedge \Phi \right)
\label{supo7}
\ee
where $m_*$ is the F-theory characteristic scale, $F = dA - i A \wedge A$ is the field strength of the 8d gauge vector boson $A$ arising from 7-branes, and $\Phi$ is a (2,0)-form on the 4-cycle $S$ describing its transverse geometrical deformations. Near the Yukawa point $p$, we can take $A$ and $\Phi$ to transform in the adjoint of the enhanced non-Abelian group $G_p$, which in our case will be given by $SO(12)$. Further dynamics of this system is encoded in the D-term
\be
D\, =\, \int_S \omega \wedge F + \frac{1}{2}  [\Phi, \bar{\Phi}]
\label{FI7}
\ee
where $\omega$ stands for the fundamental form of $S$. Together with the superpotential, this D-term relates the spectrum of 4d zero and massive modes to a set of internal wavefunctions along $S$, and the couplings between these 4d modes to the overlapping integrals of such wavefunctions.

Notice that from this latter viewpoint we seem to have a single divisor $S$ with a higher gauge group $G_p$. One must however take into account that both $\Phi$ and $A$ have a non-trivial profile. On the one hand the nontrivial profile for $\Phi$ (more precisely the fact that the rank of $\langle \Phi \rangle$ jumps at the curves $\Sigma_i$) takes into account the fact that we do not have a single divisor $S$, but rather a set of intersecting divisors $S_{\rm GUT}$ and $S_i$. A non-vanishing $\langle \Phi \rangle$ then breaks the would be gauge group $G_p$ to the subgroup $G_{\rm GUT} \times \prod_i G_i$, with $G_i$ the gauge groups of the 7-branes wrapping the divisors $S_i$, typically chosen to be $U(1)$. 

On the other hand, the effect of $\langle A \rangle$ is to provide a 4d chiral spectrum and to further break the GUT gauge group $G_{\rm GUT}$ down to the subgroup that commutes with $\langle A \rangle$, as it is usual in  compactifications with magnetized D-branes \cite{magnus,japan,ConlonWF,DiVecchia,cm09}. As a result, one may obtain a 4d MSSM spectrum from the above construction by first engineering the appropriate GUT 4d chiral spectrum via $\langle \Phi \rangle$ and an $\langle A \rangle$ which commutes with $G_{\rm GUT}$, and by then turning on an extra component of $\langle A \rangle$ along the hypercharge generator $Q_Y$ in order to break $G_{\rm GUT} \raw G_{MSSM}$ \cite{bhv2}. Generically, the presence of a non-vanishing field strength $F_Y$ along the hypercharge generator is the only way to break the GUT gauge group down to the MSSM one. As a result, all the physics of the MSSM that differ from the parent GUT physics must depend on the data that describe $\langle F_Y \rangle$.

Finally, in addition to the above set of divisors hosting the MSSM gauge and matter content, there will be in general other divisors also wrapped by branes which may source non-perturbative effects. Typical examples are 7-branes with a gauge hidden sector of the theory that undergoes a gaugino condensate, or Euclidean 3-branes with the appropriate structure of zero modes to contribute the the superpotential of the 4d effective theory. Such ingredients are usually not considered in the construction of F-theory local models, and indeed they will not be present in the $SO(12)$ model described in the next section. However, as we will review in section \ref{s:nplocal}, they are crucial in endowing F-theory local models with more realistic Yukawa couplings. In fact, one of the main results of this work is to show this point for the class of $SO(12)$ local models  that we now proceed to describe.


\section{The {SO(12)} model}
\label{s:so12}

In this section we describe in detail the $SO(12)$ local model which we will analyze in the rest of the paper. Following the common practice in the F-theory literature, we will first specify the structure of 7-brane intersections and matter curves that breaks the $SO(12)$ symmetry down to $SU(5)\times U(1)^2$, and then add the worldvolume flux that induces 4d chirality and breaks the $SU(5)$ GUT spectrum down to the MSSM. 

While in this section we will use the language of F-theory local models, it is important to notice that the model at hand admits a more intuitive description in the framework of intersecting D7-branes in type IIB orientifolds. We will exploit such vantage point in the next section, in order to gain some insight on the non-perturbative corrections that can affect our local model. 

\subsection{Matter curves}

Following the general framework described in the previous section, let us consider a local model where the symmetry group at the intersection point of three matter curves is $G_p = SO(12)$. Away from this point, this group is broken to a subgroup because $\langle\Phi\rangle \neq 0$. One can then engineer a $\langle \Phi \rangle$ such that generically $SO(12)$ is broken to $SU(5)\times U(1)^2$, except for some complex curves where there is an enhancement to either $SO(10) \times U(1)$ or $SU(6) \times U(1)$. In this way, we can identify $G_S = SU(5)$ as the GUT gauge group and the enhancement curves as matter curves where chiral matter wavefunctions are localized. 

In order to make the above picture more precise let us consider the generators of $SO(12)$, in terms of which we can express the particle spectrum of our local GUT model. 
These generators can be decomposed as $\{H_i, E_\rho\}$, where the $H_i$, $i=1,\cdots, 6$, belong to the Cartan subalgebra of $SO(12)$ and the $E_\rho$ are step 
generators.\footnote{Throughout this work we use the standard form of the $SO(2N)$ generators in the fundamental representation \cite{georgi}.}
Recall that
\be
[H_i, E_\rho] = \rho_i E_\rho
\label{hec}
\ee
where $\rho_i$ is the $i$-th component of the root $\rho$. The 60 non-trivial roots are given by
\be
(\underline{\pm 1, \pm 1, 0,0,0,0}) 
\label{so12r}
\ee
where the underlying means all possible permutations of the vector entries.

Let us now choose the vev of the transverse position field $\Phi= \Phi_{xy} dx \wedge dy$ to be
\be
\langle \Phi_{xy} \rangle = m^2 \big( x Q_x + y Q_y \big)
\label{pvev}
\ee
where $m^2$ is related to the intersection slope between 7-branes as explained in section \ref{s:nplocal}, 
and the charge operators $Q_x$ and $Q_y$ are the following combinations of generators of elements of the $SO(12)$ Cartan subalgebra
\be
Q_x = -H_1 \quad ; \quad Q_y=\frac12\left(H_1 + H_2 + H_3 + H_4 + H_5 + H_6\right)
\label{qxy}
\ee
This choice of $\langle \Phi \rangle$ describes a $SO(12)$ local model that is similar to the $U(3)$ toy model analyzed in \cite{afim} in several aspects. This will allow us to apply several useful results of \cite{afim} to the more realistic case at hand.

Given (\ref{pvev}) one can understand the $SO(12)$ symmetry breaking pattern described above as follows. In general the step generators $E_\rho$ satisfy
\be
[\langle \Phi_{xy}\rangle , E_\rho] = m^2 q_\Phi(\rho) E_\rho
\label{pec}
\ee
with $q_\Phi$ a holomorphic function of the complex coordinates $x$, $y$ of the 4-cycle $S$. The subgroup of $SO(12)$ not broken by the presence of this vev corresponds to those generators that commute with $\langle \Phi \rangle$ at any point in $S$. This set is given by the Cartan subalgebra of $SO(12)$ and to those step generators $E_\rho$ such that $q_\Phi(\rho)=0$ for all $x,y$. It is easy to see that such unbroken roots are given by
\be
(0, \underline{1, -1, 0,0,0}) 
\label{ubr}
\ee  
together with the Cartan generators. Therefore, from the symmetry group $SO(12)$ only the subgroup $SU(5)\times U(1)^2$ remains as a gauge symmetry, and we can identify $G_S = SU(5)$ as our GUT gauge group. 

On the other hand, the broken generators of $SO(12)$, that have $q_\Phi\not=0$ for generic $x,y$, allow us to understand the pattern of matter curves and to classify the charged matter localized therein. Such broken roots and their charges $q_\Phi$ are displayed in table \ref{t1}.

\begin{table}[htb] 
\renewcommand{\arraystretch}{1.25}
\begin{center}
\begin{tabular}{|c|c|c|c|c|c|c|}
\hline
$\rho$ & root &  $q_\Phi $ & $s_{xx}$ & $s_{xy}$ & $s_{yy}$ & $SU(5)$ rep.  \\
\hline
$a^+$ & $(1, \underline{-1, 0, 0,0,0})$ & $-x$ & $\frac13$ & -$\frac14$ & $\frac14$ & $\br{5}$ \\
\hline
$a^-$ & $(-1, \underline{1, 0, 0,0,0})$ & $x$ & $\frac13$ & -$\frac14$ & $\frac14$ & $\r{5}$ \\
\hline
$b^+$ & $(0, \underline{1, 1, 0,0,0})$ & $y$ & \mfn{0} & \mfn{0} & $\frac1{12}$ & $\r{10}$ \\ 
\hline
$b^-$ & $(0, \underline{-1, -1, 0,0,0})$ & $-y$ & \mfn{0} & \mfn{0} & $\frac1{12}$ & $\br{10}$ \\ 
\hline
$c^+$ & $(-1, \underline{-1, 0, 0,0,0})$ & $x-y$ & $\frac13$ & -$\frac1{12}$ & $\frac1{12}$ & $\br{5}$ \\
\hline
$c^-$ & $(1, \underline{1, 0, 0,0,0})$ & $-(x-y)$ & $\frac13$ & -$\frac1{12}$ & $\frac1{12}$ & $\r{5}$ \\
\hline
\end{tabular}
\end{center}
\caption{\small Data of broken generators}
\label{t1}
\end{table}

From this table we see that there are three complex curves within $S$ where the bulk symmetry $SU(5)\times U(1)^2$ is enhanced, in the sense that there $q_\Phi =0$ for an additional set of roots. Concretely, for $x=0$ there are 10 additional roots that together with those in (\ref{ubr}) complete the $SU(6)$ root system. We have labeled such matter curve as $\Sigma_a$, so that in the language of the previous section we would have that $G_{\Sigma_a} = SU(6) \times U(1)$. These extra set of roots whose $q_\Phi$ vanishes at $\Sigma_ a$ can be split into subsets that have different $q_\Phi$ away from $\Sigma_a$. It is easy to convince oneself that each of these subsectors must fall into complete weight representations of $SU(5)$, which in turn correspond to the matter localized at the curve. In the case of $\Sigma_a$, there are two sectors $a^+$ and $a^-$ that correspond to the representations {\bf 5} and ${\bf \bar{5}}$ of $SU(5)$, respectively, as shown in table \ref{t1}. 

Similarly to $\Sigma_a$, at the curve $\Sigma_b = \{y=0 \}$ there are 20 extra unbroken roots and $SU(5) \times U(1)^2$ is enhanced to $SO(10) \times U(1)$, giving rise to the representations {\bf 10} and ${\bf \overline{10}}$. The third matter curve is given by $\Sigma_c = \{x=y\}$, where there is also an enhancement to $SU(6) \times U(1)$.     

Finally, let us consider a set of quantities that only depend on each root sector of the model. These are the symmetrized products\footnote{$S(A_1 \cdots A_N) = \frac1{N!} \left(A_1 \cdots A_N + {\rm all \ permutations}\right)$}
\be
S_{mn}(\rho) = S(E_\rho Q_m Q_n) \quad ; \quad m,n=x,y
\label{s3a}
\ee
where the generators are taken in the fundamental representation of $SO(12)$. As we will see, the equations of motion satisfied by the zero modes at the matter curves will depend on these quantities. For the broken roots we obtain
\be
S_{mn}(\rho) = s_{mn}(\rho) E_\rho
\label{s3b}
\ee
where the $s_{mn}=s_{nm}$ are constants also displayed in table \ref{t1}.

\subsection{Worldvolume flux}

To obtain a 4d chiral model the above pattern of matter curves is not enough, and it is necessary to add a non-trivial background worldvolume flux $\langle F \rangle$ to our local F-theory model. Just like the position field, such flux is usually chosen along the Cartan subalgebra of $SO(12)$, so that it commutes with $\langle \Phi_{xy} \rangle$ and the equations of motion of our system are simplified. Moreoever, considering a component of $\langle F \rangle$ along the hypercharge generator $Q_Y$ allows to break the GUT gauge group $SU(5)$ down to $SU(3) \times SU(2) \times U(1)_Y$, this being in fact the only way to achieve GUT symmetry breaking for the most generic class of F-theory GUT models. 

In order to construct a worldvolume flux with the desired properties we proceed in three steps. First we add a flux $\langle F_1 \rangle$ analog of the one introduced in the $U(3)$ toy model of \cite{afim}. Just like in there, this flux will create chirality on the curves $\Sigma_a$ and $\Sigma_b$, selecting the sectors $a^+$ and $b^+$ as the ones that contain the chiral matter of the model, as opposed to $a^-$ and $b^-$. Then we add an extra piece $\langle F_2 \rangle$ such that the matter curve $\Sigma_c$ also contains a chiral spectrum: a typical requirement to achieve an acceptable Higgs sector. None of these previous fluxes further break the gauge group $SU(5) \times U(1)^2$ so, finally,  we include a flux $\langle F_Y \rangle$ along the hypercharge generator $Q_Y$ that breaks $SU(5)$ down to $SU(3) \times SU(2) \times U(1)_Y$.

To proceed we then consider the flux
\be
\langle F_1 \rangle\, =\,i \left( M_x\, dx \wedge d\bar{x} + M_y\, dy \wedge d\bar{y}\right) \, Q_F
\label{f1vev}
\ee
where
\be
Q_F= \frac12 (H_1-H_2 - H_3 - H_4 - H_5 - H_6) = - Q_x - Q_y
\label{qdef}
\ee
which is the analog of the flux introduced in the $U(3)$ toy model of \cite{afim}. To analyze the effect of this flux it is convenient to define the $Q_F$-charge of the roots $E_\rho$ according to
\be
[Q_F, E_\rho] = q_F(\rho) E_\rho
\label{qrho}
\ee
The roots in (\ref{ubr}) are clearly neutral under this flux component $\langle F_1\rangle$, and so the gauge symmetry $SU(5)$ is not broken further by its presence. The roots in the sectors $a$ and $b$ are however not neutral. Hence, if the integral of (\ref{f1vev}) over each of these curves does not vanish, they will each host a chiral sector of the theory. In the following we will assume that this is the case and that $\langle F_1\rangle$ induces a net chiral spectrum of three ${\bf \bar{5}}$'s in the curve $\Sigma_a$ and three ${\bf 10}$'s in the curve $\Sigma_b$. If this chiral spectrum can be understood in terms of local zero modes in the sense of \cite{palti12}, then such chiral modes should arise in the sectors $a^+$ and $b^+$ of table \ref{t1}, respectively, and by the results of appendix \ref{ap:wave} one should choose $M_x < 0 < M_y$ to describe them locally. 

Notice that the roots belonging to the $c$ sector are neutral under (\ref{qdef}), and so the spectrum arising from the curve $\Sigma_c$ is unaffected by the presence of $\langle F_1 \rangle$. As the $SO(12)$ triple intersection point is where down-like Yukawa couplings arise from, we do need one ${\bf \bar{5}}$ in such curve, but however no ${\bf 5}$ so that no undesired ${\bf \bar{5}}{\bf 5}$ mass terms appear. This chiral spectrum on the sector $c$ can be achieved by adding the following extra piece of worldvolume flux
\be
\langle F_2 \rangle\, =\, i  \left( dx \wedge d\bar{y} +  dy \wedge d\bar{x}\right) \left(N_a Q_x + N_b Q_y  \right)
\label{fluxcso12}
\ee
It is easy to check that the particles localized at the matter curve $\Sigma_c$ are now non-trivially charged under the flux background, and that a local chiral spectrum can be achieved if we choose $N_a \neq N_b$. In particular, as shown in appendix \ref{ap:flux} for $N_a > N_b$ one obtains net local chirality in the sector $c^+$, yielding the desired ${\bf \bar{5}}$ which is the $SU(5)$ down Higgs. Notice that those particles at the curves $\Sigma_a$ and $\Sigma_b$ are also charged under (\ref{fluxcso12}). However, by construction the number of (local) families in such curves is independent of the flux $\langle F_2\rangle $, as also shown in appendix \ref{ap:flux}.

Let us finally add a third piece of worldvolume flux which, unlike (\ref{f1vev}) and (\ref{fluxcso12}), will break the $SU(5)$ gauge group down to the MSSM. As usual, such flux should be turned along the hypercharge generator $Q_Y$, and a rather general choice is given by
\be
\langle F_Y \rangle\, =\, i \left[ \left( dx \wedge d\bar{y} +  dy \wedge d\bar{x}\right) N_Y + \left(dy \wedge d\bar{y} - dx \wedge d\bar{x} \right)  \tilde{N}_Y \right]Q_Y
\label{fluxhyp}
\ee
where
\be
Q_Y\, =\, \frac{1}{3} \left(H_2 + H_3 + H_4\right) - \oh \left(H_5 + H_6\right) 
\ee
We have chosen the hypercharge flux to be a primitive (1,1)-form, so that it satisfies automatically  the equations of motion for the background. Note that (\ref{fluxhyp}) has two components that are easily comparable with the previous flux components (\ref{f1vev}) and (\ref{fluxcso12}). The first component, proportional to the flux density $N_Y$, is quite similar to $\langle F_2 \rangle$. Indeed, as happens for (\ref{fluxcso12}) its pullback vanishes over the matter curves $\Sigma_a$ and $\Sigma_b$, and so it does not contribute to the (local) index that computes the number of chiral families in the sectors $a$ and $b$. The second component, proportional to $\tilde{N}_Y$, may in principle affect the chiral index over the curves $\Sigma_a$ and $\Sigma_b$ but, following the common practice in the GUT F-theory literature, we will assume that this is not the case. Globally one requires that
\be
\int_{\Sigma_a} \langle F_Y \rangle \, =\, \int_{\Sigma_b} \langle F_Y \rangle \, =\, 0
\ee
so that three complete families of quarks and leptons remain at the curves $\Sigma_a$ and $\Sigma_b$ after introducing the hypercharge flux. Locally, we demand that the local zero modes still arise from the sectors $a^+$ and $b^+$, and this amounts to require flux densities such that
\be
M_x + q_Y \tilde{N}_Y < 0 < M_y + q_Y \tilde{N}_Y
\label{condify}
\ee
for every possible hypercharge value $q_Y$ in the sectors $a^+$ and $b^+$, see table \ref{t2} below. 

While innocuous for the matter spectrum at the curves $\Sigma_a$, $\Sigma_b$, the hypercharge flux is supposed to modify the chiral spectrum of curve $\Sigma_c$, in order to avoid the doublet-triplet splitting problem of $SU(5)$ GUT models \cite{bhv2}. Indeed, one typically assumes that $\int_c \langle F_Y \rangle \neq 0$, and since (\ref{fluxhyp}) couples differently to particles with different hypercharge, this implies a different chiral index for the doublet and for the triplet of ${\bf \bar{5}}$. Locally, we have that the total flux seen near the Yukawa point by the doublets on the sector $c^+$ is
\be
F_{{\rm tot}, {\bf 2}}\, =\, N_Y + 2 (N_a - N_b)
\ee
while the flux seen by the triplets is 
\be
F_{{\rm tot}, {\bf 3}}\, =\, -\frac{2}{3} N_Y + 2(N_a - N_b)
\ee
Hence, in order to have a vector-like sector of triplets in the local model we can set 
\be
N_Y\, =\, 3 (N_a - N_b)
\label{cond}
\ee
and then assume that such vector-like spectrum is massive. Notice that this condition still yields a chiral sector for the doublets and so forbids a $\mu$-term for them. Indeed, imposing (\ref{cond}) we have that 
\be
F_{{\rm tot}, {\bf 2}}\, =\, \frac{5}{3} N_Y
\label{relH2}
\ee
which in general will induce a net chiral spectrum of doublets in the curve $\Sigma_c$. Hence, imposing (\ref{cond}) the combined effect of $\langle F_2 \rangle$ and $\langle F_Y \rangle$ is such that  doublets of ${\bf \bar{5}}$ in the sector $c$ feel a net flux, while triplets do not. One may then choose the flux density $N_Y$ such that it yields a single pair of MSSM down Higgses at the curve $\Sigma_c$. 

To summarize, the total worldvolume flux on this local $SO(12)$ model is given by
\bea
\label{totalflux}
\langle F \rangle& = & i(dy\wedge d\bar y - dx\wedge d\bar x) Q_P + i(dx\wedge d\bar y+dy\wedge d\bar x)Q_S \\ \nonumber
 & & +\,  i(dy\wedge d\bar y + dx\wedge d\bar x) M_{xy} Q_F
\eea
 where we have defined
\bea
Q_P & = & M Q_F + \tilde{N}_Y Q_Y \\
\label{qpdef}
Q_S & = & N_aQ_x+N_bQ_y+N_YQ_Y
\label{qsdef}
\eea
and
\be
M\, \equiv \, \frac{1}{2} (M_y - M_x) \quad \quad \quad M_{xy}\, \equiv\, \frac{1}{2} (M_y + M_x)
\label{DterM}
\ee
Note that the combination of flux densities $M_{xy}$ corresponds to an FI-term, which will be set to vanish whenever supersymmetry is imposed.   

Just like in \cite{afim}, we can now express the vev of the corresponding vector potential $A$ in the holomorphic gauge defined in \cite{fi1}, namely as
\begin{equation}
\langle A \rangle^{\text{hol}} = i\left[(Q_P-M_{xy}Q_F) \bar x - \bar y Q_S\right] dx-i \left[(Q_P+M_{xy}Q_F) \bar y+ \bar x Q_S \right]dy
\label{aholo}
\end{equation}
this being the quantity that will enter into the equation of motion for the zero mode wavefunctions at the curves $\Sigma_a$, $\Sigma_b$ and $\Sigma_c$. The combined effect of the background $\langle \Phi \rangle$ and $\langle A \rangle$ breaks $SO(12)$ to $SU(3)\times SU(2) \times U(1)^3$, and as a result the sectors $a$, $b$ and $c$ split into further subsectors compared to table \ref{t1}. The content of charged particles under the surviving gauge group is shown in table \ref{t2}, where we have also displayed the charges of each sector under the operators $Q_F$, $Q_x$, $Q_y$ and $Q_Y$. We have also included the values of  $q_S$ and $q_P$, which are defined as
\be
[Q_S, E_\rho] = q_S(\rho) E_\rho \quad \quad \quad [Q_P, E_\rho] = q_P(\rho) E_\rho
\label{srho}
\ee
and which, unlike the other charges, depend on the flux densities of the model.  As discussed below and in appendix \ref{ap:wave}, each of these sectors obeys a different zero mode equation, and so it is described by a different wavefunction.

\begin{table}[htb] 
\footnotesize
\renewcommand{\arraystretch}{1.25}
\setlength{\tabcolsep}{5pt}
\begin{center}
    \begin{tabular}{ | c | c || c | c | c | c | c | c || c| c|}
    \hline
    Sector & Root & $SU(3)$ & $SU(2)$ & $q_Y$ & $q_F$ & $q_x$ & $q_y$  & $q_S$ & $q_P$ \\ \hline
    $a_1^+$ & $(1,\underline{-1,0,0},0,0)$ & $\bar{\mathbf 3}$ &$\mathbf 1$ & $-\frac{1}{3}$& $1$ & $-1$ & $0$  & $-N_a -\frac{1}{3} N_Y$  & $M -\frac{1}{3} \tilde{N}_Y$ \\ \hline
    $a_1^-$ & $(-1,\underline{1,0,0},0,0)$ & $\mathbf 3$ &$\mathbf 1$ & $\frac{1}{3}$& $-1$ & $1$ & $0$  & $N_a +\frac{1}{3} N_Y$ &  $-M + \frac{1}{3} \tilde{N}_Y$  \\ \hline
    $a_2^+$ & $(1,0,0,0,\underline{-1,0})$ & $\mathbf 1$ & $\mathbf 2$ & $\frac{1}{2}$ & $1$ & $-1$ & $0$ & $-N_a+\frac{1}{2}N_Y$ &  $M +\frac{1}{2} \tilde{N}_Y$ \\ \hline
    $a_2^-$ & $(-1,0,0,0,\underline{1,0})$ & $\mathbf 1$ & $\mathbf 2$  & $-\frac{1}{2}$ & $-1$ & $1$ & $0$ & $N_a-\frac{1}{2}N_Y$ &  $-M -\frac{1}{3} \tilde{N}_Y$ \\ \hline
    $b_1^+$ &$(0,\underline{1,1,0},0,0)$ & $\bar{\mathbf 3}$ & $\mathbf 1$  & $\frac{2}{3}$ & $-1$ & $0$ & $1$ & $N_b+\frac{2}{3}N_Y$ &  $- M +\frac{2}{3} \tilde{N}_Y$ \\ \hline
    $b_1^-$ &$(0,\underline{-1,-1,0},0,0)$ & $\mathbf 3$ & $\mathbf 1$ & $-\frac{2}{3}$ & $1$ & $0$ & $-1$  & $-N_b-\frac{2}{3}N_Y$ & $M - \frac{2}{3} \tilde{N}_Y$  \\ \hline
    $b_2^+$ & $(0,\underline{1,0,0},\underline{1,0})$ & $\mathbf 3$ & $\mathbf 2$ & $-\frac{1}{6}$ & $-1$ & $0$ & $1$  & $N_b-\frac{1}{6}N_Y$ & $- M - \frac{1}{6} \tilde{N}_Y$ \\ \hline
    $b_2^-$ & $(0,\underline{-1,0,0},\underline{-1,0})$ & $\bar{\mathbf 3}$ & $\mathbf 2$ & $\frac{1}{6}$ & $1$ & $0$ & $-1$ & $-N_b-\frac{1}{6}N_Y$ & $ M + \frac{1}{6} \tilde{N}_Y$  \\ \hline
    $b_3^+$ &$(0,0,0,0,1,1)$ & $\mathbf 1$ & $\mathbf 1$ & $-1$ & $-1$ & $0$ & $1$ & $N_b-N_Y$ & $ - M - \tilde{N}_Y$\\ \hline
    $b_3^-$ &$(0,0,0,0,-1,-1)$ & $\mathbf 1$ & $\mathbf 1$ & $1$ & $1$ & $0$ & $-1$ & $-N_b+N_Y$ &  $M + \tilde{N}_Y$ \\ \hline
    $c_1^+$ & $(-1,\underline{-1,0,0},0,0)$ &$\mathbf{ \bar 3}$ & $\mathbf 1$ & $-\frac{1}{3}$ & $0$ & $1$ & $-1$ & $N_a-N_b-\frac{1}{3}N_Y$ & $-\frac{1}{3} \tilde{N}_Y$ \\ \hline
    $c_1^-$ & $(1,\underline{1,0,0},0,0)$ &$\mathbf 3$ & $\mathbf 1$ & $\frac{1}{3}$ & $0$ & $-1$ & $1$ & $-N_a+N_b+\frac{1}{3}N_Y$ & $\frac{1}{3} \tilde{N}_Y$  \\ \hline
    $c_2^+$ & $(-1,0,0,0,\underline{-1,0})$ & $\mathbf 1$ & $\mathbf 2$ & $\frac{1}{2}$ & $0$ & $1$ & $-1$ & $N_a-N_b+\frac{1}{2}N_Y$ & $\frac{1}{2} \tilde{N}_Y$ \\ \hline
    $c_2^-$ & $(1,0,0,0,\underline{1,0})$ & $ \mathbf 1 $ & $\mathbf 2$  & $-\frac{1}{2}$  & $0$ & $-1$ & $1$& $-N_a+N_b-\frac{1}{2}N_Y$ & $-\frac{1}{2} \tilde{N}_Y$ \\ \hline
    X$^+$,Y$^+$ & $(0,\underline{1,0,0},\underline{-1,0})$ & $\mathbf 3$ & $\mathbf 2$ & $\frac{5}{6}$ & $0$ & $0$ & $0$ & $\frac{5}{6}N_Y$ & $\frac{5}{6}\tilde{N}_Y$ \\ \hline
    X$^-$,Y$^-$ & $(0,\underline{-1,0,0},\underline{1,0})$ & $\bar{\mathbf 3}$ & $\mathbf 2$ & $-\frac{5}{6}$  & $0$ & $0$ & $0$ & $-\frac{5}{6}N_Y$ & $-\frac{5}{6}\tilde{N}_Y$ \\
    \hline
    \end{tabular}
\end{center}
\caption{Different sectors and charges for the $SO(12)$ model.}
\label{t2}
\end{table}

\newpage

\subsection{Perturbative zero modes}
\label{ss:pzm}

Given the above background, and ignoring for the time being non-perturbative effects, one may solve for the zero mode wavefunctions on each of the sectors of table \ref{t2}. One obtains in this way the internal profile for each of the 4d $\cn=1$ chiral multiplets that arise from the matter curves $\Sigma_a$, $\Sigma_b$ and $\Sigma_c$, and in particular for the 4d chiral fermions of the MSSM. 

Following \cite{afim} one may consider the 7-brane action derived in \cite{bhv2}, more precisely the piece bilinear in fermions, and extract the equation of motion for the 7-brane fermionic zero modes. These equations can then be written in a Dirac-like form as 
\be
{\bf D_A} \Psi \, =\, 0
\label{Dirac9}
\ee
with
\be
{\bf D_A}\, =\, 
\left(
\begin{array}{cccc}
0 & D_x & D_y & D_z \\
-D_x & 0 & -D_{\bar{z}} & D_{\bar{y}} \\
-D_y & D_{\bar{z}} & 0 & -D_{\bar{x}} \\
-D_z & -D_{\bar{y}} & D_{\bar{x}} & 0
\end{array}
\right)
\quad \quad
\Psi\, =\, \Psi_\rho E_\rho \, =\, 
\left(
\begin{array}{c}
- \sqrt{2}\, \eta \\ \psi_{\bar{x}} \\ \psi_{\bar{y}} \\ \chi_{xy}
\end{array}
\right)\label{matrixDirac}
\ee
where the four components of $\Psi$ represent 7-brane fermionic degrees of freedom. As pointed out in \cite{bhv1}, these fermionic modes pair up naturally with the 7-brane bosonic modes that arise from background fluctuations 
\be
\Phi_{xy}\, =\, \langle \Phi_{xy} \rangle + \varphi_{xy} \quad \quad A_{\bar{m}}\, =\, \langle A_{\bar{m}} \rangle + a_{\bar{m}}
\label{fluctuation}
\ee
More precisely we have that  $(a_{\bar{m}}, \psi_{\bar{m}})$ and $(\varphi_{xy}, \chi_{xy})$ form 4d $\cn=1$ chiral multiplets. In addition, $(A_\mu, \eta)$ form a 4d $\cn=1$ vector multiplet that should include the gauge degrees of freedom of the model. One can see that these bosonic modes feel the same zero mode equations that their fermionic partners, and so solving (\ref{Dirac9}) gives us the wavefunction for the whole $\cn=1$ chiral multiplet. 

As we started from an $SO(12)$ gauge symmetry, $\Psi$ has gauge indices in the adjoint of $SO(12)$. Each covariant derivatives in ${\bf D_A}$ acts non-trivially on such indices, since they are defined as $D_m = \p_m - i [\langle A_m\rangle, \cdot]$ for those coordinates $m = x, y, \bar{x}, \bar{y}$ along the GUT 4-cycle $S$, and as ${D}_{\bar{z}} = -i [\langle \Phi_{xy} \rangle, \cdot]$ for the transverse coordinate $z$ \cite{afim}. It is then clear that each sector of table \ref{t2} will see a different Dirac equation. Hence, in order to solve for the zero mode wavefunctions of our model, we fix the roots $E_\rho$ to lie within a particular sector of table \ref{t2} and then solve sector by sector. 

Within each sector $\rho$, eq.(\ref{Dirac9}) is specified in terms of the following quantities: $q_\Phi(\rho)$, $q_S(\rho)$, $q_P(\rho)$ and $q_F(\rho)$.  The zero mode computation for each sector is done in detail in appendix \ref{ap:wave}, and one can see that each of these solutions is of the form 
\be
\Psi_\rho\, =\, 
\left(
\begin{array}{c}
0 \\ \vspace*{.1cm}
 -\displaystyle{\frac{i\lam_{\bar{x}}}{m^2}} \\ \displaystyle{\frac{i\lam_{\bar{y}}}{m^2}} \\ 1
\end{array}
\right) \chi_\rho^i E_\rho, \quad \quad \quad \chi_\rho^i \, =\, e^{- q_\Phi (\lam_{\bar{x}} \bar{x} - \lam_{\bar{y}} \bar{y})} f_i (\lam_{\bar{x}} y + \lam_{\bar{y}} x) 
\label{gensolpsi}
\ee
where we have solved for the zero modes in the holomorphic gauge of eq.(\ref{aholo}). Here $f_i$ are holomorphic functions of the variable $\lam_{\bar{x}} y + \lam_{\bar{y}} x$, with  $\lam_{\bar{x}}$,  $\lam_{\bar{y}}$ constants that depend on the flux densities $q_P$, $q_S$, $M_{xy}$ and the mass scale $m^2$, and which are different for each sector $\rho$. The index $i$ runs over the different holomorphic functions that are present on each sector, or in other words over the families of zero modes localized in the same curve. Finally, recall that $q_\Phi$ is a holomorphic function of the 4-cycle coordinates $x,y$. We have summarized the values of these quantities for each sector of our model in table \ref{t3},
\begin{table}[htb] 
\renewcommand{\arraystretch}{1.25}
\begin{center}
\begin{tabular}{|c|c|c|c|c|c|}
\hline
$\rho$ &  $q_\Phi $ & $\lam_{\bar{x}}$ & $\lam_{\bar{y}}$  & $SU(5)$ rep.  \\
\hline
$a_p^+$  & $-x$ & $\lam_+$ & -$q_{S} \frac{\lam_+}{\lam_+ -  q_P}$  & $\br{5}$ \\
\hline
$a_p^-$ &  $x$ & $\lam_-$ & $q_{S} \frac{\lam_-}{\lam_- + q_P}$   & $\r{5}$ \\
\hline
$b_q^+$ &  $y$ & -$q_{S} \frac{\lam_+}{\lam_+ + q_P}$ & $\lam_+$ &  $\r{10}$ \\ 
\hline
$b_q^-$ &  $-y$ & $q_{S} \frac{\lam_-}{\lam_- - q_P}$ & $\lam_-$ &  $\br{10}$ \\ 
\hline
$c_r^+$ &  $x-y$ & $\frac{q_S \lam_+-m^4}{\lam_++q_P-q_S}$ & -$\lam_+$- $\frac{q_S \lam_+-m^4}{\lam_++q_P-q_S}$ & $\br{5}$ \\
\hline
$c_r^-$ &  $-(x-y)$ &  -$\frac{q_S\lam_-+m^4}{\lam_--q_P+q_S} $ &  -$\lam_-$ + $\frac{q_S\lam_-+m^4}{\lam_--q_P+q_S} $ &  $\r{5}$ \\
\hline
\end{tabular}
\end{center}
\caption{\small Wavefunction parameters.}
\label{t3}
\end{table}
assuming for simplicity that $M_{xy} = 0$ (see appendix \ref{ap:wave} for the general expressions). For the sectors $a_p^+$, $b_q^+$ and $c_r^+$, $\lam_+$ is defined as the lowest (negative) eigenvalue of the flux matrix
\be
\mathbf m_{\rho}=\left (\begin{array}{ccc}
-q_P&q_S&im^2q_x\\
q_S&q_P&im^2q_y\\
-im^2q_x&-im^2q_y&0 \end{array}\right )
\ee
and one can check that the three lower entries of the vector in (\ref{gensolpsi}) are the corresponding eigenvector of this matrix. The same definition applies to $\lam_-$ for the sectors $a_p^-$, $b_q^-$ and $c_r^-$.\footnote{Although they have a similar definition, $\lam_{\pm}$ have different values. Indeed, since $\mathbf m_{\a^+} = - \mathbf m_{\a^-}$ we have that $-\lam_-(\a^-)$ is the highest positive eigenvalue of $\mathbf m_{\a^+}$.} In general $\pm\lam_{\pm}$ satisfy a complicated cubic equation discussed in appendix \ref{ap:wave}, which depends on the flux densities $q_S$ and $q_P$. Since these two quantities contain the hypercharge flux, $\lam_\pm$ will be different for each of the subsectors $a_p^\pm$, $b_q^\pm$ and $c_r^\pm$. Indeed, it is precisely in the value of the flux densities $q_P$ and $q_S$ that the wavefunctions within the same $SU(5)$ multiplet but with different hypercharge differ. 

Although by working in  the holomorphic gauge it is not easy to see which wavefunctions converge locally, by going to a real gauge we can see that if we impose condition (\ref{condify}) the locally convergent zero modes lie in the sectors $a_n^+$, $b_p^+$ and $c_r^+$, see appendix \ref{ap:wave}. In the following we will consider this choice of signs for the fluxes, and so we will concentrate in the wavefunctions for these sectors, that in table \ref{dictMSSM} are identified with the MSSM chiral multiplets. In particular, in section \ref{s:npwave} we will compute how these zero modes are modified in the presence of non-perturbative effects.

\begin{table}[htb] 
\renewcommand{\arraystretch}{1.25}
 \begin{center}
    \begin{tabular}{ | c | c | c | c | c | c |}
    \hline
    Sector & Chiral mult. & $SU(3) \times SU(2)$ & $q_Y$ & $q_x$ & $q_y$  \\ \hline
    $a_1^+$ & $D_R$ & $3(\bar{\mathbf 3},\mathbf 1)$ & $-\frac{1}{3}$& $-1$ & $0$  \\ \hline
    $a_2^+$ & $L$ & $3(\mathbf 1,\mathbf 2)$ & $\frac{1}{2}$ & $-1$ & $0$  \\ \hline
    $b_1^+$ &$U_R$ & $3(\bar{\mathbf 3},\mathbf 1)$  & $\frac{2}{3}$ & $0$ & $1$ \\ \hline
    $b_2^+$ & $Q_L$ & $3(\mathbf 3,\mathbf 2)$ & $-\frac{1}{6}$ & $0$ & $1$ \\ \hline
    $b_3^+$ &$E_R$ & $3(\mathbf 1,\mathbf 1)$ & $-1$ & $0$ & $1$\\ \hline
    $c_1^+$ & $D_d$ &$(\mathbf{ \bar 3},\mathbf 1)$ & $-\frac{1}{3}$ & $1$ & $-1$ \\ \hline
    $c_2^+$ & $H_d$ & $(\mathbf 1,\mathbf 2)$ & $\frac{1}{2}$ & $1$ & $-1$ \\ \hline
    \end{tabular}
\end{center}
\caption{Dictionary $SO(12)$-MSSM}
\label{dictMSSM}
\end{table}

Finally, one may solve the wavefunctions for the bulk sector $(X,Y)^{\pm}$, which is only sensitive to the presence of the hypercharge flux. Although the global properties of $\langle F_Y \rangle$ can be chosen so that no chiral matter arises from this sector \cite{bhv2}, there will always be massive modes which we can be identified with the $X^\pm$, $Y^\pm$ bosons of 4d $SU(5)$ GUTs. As analyzed in appendix \ref{ap:wave} such massive modes will have a Gaussian profile, a fact that can be used to suppress operators mediating proton decay \cite{nosusy}. 

\subsection{Yukawa couplings}
\label{sec:treeYukawas}

Let us compute the Yukawa couplings between the chiral zero modes of this model, before any non-perturbative effect is taken into account. As in \cite{afim} such couplings arise from the trilinear term in (\ref{supo7}), 
which in terms of the wavefunctions above can be written as
\be
Y_{\rho\sig\tau}^{ijk}\, =\, m_* f_{\rho\sig\tau} \int_{S}\, {\rm det\, } (\vec{\psi}_\rho^{\, i} , \vec{\psi}_\sig^{\, j}, \vec{\psi}_\tau^{\, k}) 
\, {\rm d }{\rm vol}_S
\label{yukawa7}
\ee
where $f_{\rho\sig\tau} = -i \tr\, ([E_\rho, E_\sig] E_\tau)$, ${\rm d vol}_S = 2 \om^2 = \, {\rm d} x \wedge {\rm d}y \wedge {\rm d}  \bar x \wedge {\rm d} \bar y$ and
the vectors $\vec{\psi}_\rho^{\, i}$ are given by the three lower entries of $\Psi$. From the last subsection and the results of appendix \ref{ap:wave} we have that these vectors read
\be
\vec{\psi}_{a^+_p}^i
\, =\,
\left(\!\!
\begin{array}{c}\vspace*{.1cm}
-\displaystyle{\frac{i\lam_{a_p}}{m^2}} \\ \zeta_{a_p}  \displaystyle{\frac{i\lam_{a_p}}{m^2}}  \\ 1
\end{array}
\!\! \right)\chi_{a_p}^i
\quad  \quad
\vec{\psi}_{b^+_q}^j
\, =\,
\left(\!\!
\begin{array}{c}\vspace*{.1cm}
-\zeta_{b_q} \displaystyle{\frac{i\lam_{b_q}}{m^2}} \\ \displaystyle{\frac{i\lam_{b_q}}{m^2}}  \\ 1
\end{array}
\!\! \right)\chi_{b_q}^j
\quad  \quad
\vec{\psi}_{c^+_r}
\, =\,
\left(\!\!
\begin{array}{c}
\dfrac{i \zeta_{c_r}}{m^2} \\[3mm] \dfrac{i(\zeta_{c_r}-\lam_{c_r})}{m^2}  \\ 1
\end{array}
\!\! \right)\chi_{c_r}
\label{simvecs}
\ee
where the $\lam$'s and $\zeta$'s are real constants defined in appendix \ref{ap:wave}. The scalar wavefunctions $\chi$ are given by
\bea\nonumber
\chi_{a_p}^i\, =\, m_* e^{ \lam_{a_p}  x(\bar{x} - \zeta_{a_p} \bar{y})}  f_i(y+ \zeta_{a_p}x) & \quad  \quad&
\chi_{b_q}^j\, =\, m_* e^{ \lam_{b_q}  y(\bar{y} - \zeta_{b_q}\bar x)}  g_j(x + \zeta_{b_q}y) \\
\chi_{c_r}\, =\,m_*  \g_{c_r}  e^{(x-y)\left (\zeta_{c_r} \bar x - (\lam_{c_r} - \zeta_{c_r})\bar y\right )} 
\label{simchis}
\eea
and following \cite{hv08} the holomorphic factor can be chosen as
\be
f_i = \g^{i}_{a_p}  m_*^{3-i} (y+\zeta_{a_p}x)^{3-i} \quad \quad 
g_j =  \g^{j}_{b_q}  m_*^{3-j} (x+ \zeta_{b_q}y)^{3-j} \quad \quad i,j=1,2,3
\label{simhol}
\ee
with the normalization factors $\g_{a_p}^i$, $\g_{b_q}^j$ and $\g_{c_r}$ to be fixed later.

We then see that the structure of wavefunctions and Yukawas is quite similar to the one in the $U(3)$ toy model 
analyzed in \cite{afim}. One difference is the more involved sector structure, which as illustrated in table \ref{dictMSSM} 
is necessary to accommodate the MSSM chiral spectrum. Notice also that, due to the extra components of $\langle F \rangle$ 
that we have introduced, the holomorphic factors in the wavefunctions not only depend on the complex coordinate along 
the matter curve, but also on the transverse one. This however does not affect the general result of \cite{cchv09}, 
in the sense that the Yukawa matrices are of rank one. 
Indeed, substituting in (\ref{yukawa7}) shows that the integrand is the product of $f_i g_j$ times an exponential
whose argument is invariant under a diagonal $U(1)$ rotation of $x$ and $y$. Since ${\rm d vol}_S$ is also invariant
under such rotation, the integral can be non-vanishing only when $f_i$ and $g_j$ are constants, which happens
for $i=j=3$. Computing the integral yields the only non-zero coupling
\be
Y^{33}_{a^+_p b^+_q c^+_r} = \pi^2 \left(\frac{m_*}{m}\right)^4 \g^3_{a_p} \g^3_{b_q} \g_{c_r} f_{a^+_p b^+_q c^+_r}
\label{y33pert}
\ee
where $f_{a^+_p b^+_q c^+_r}$ are the structure constants of $SO(12)$ in the fundamental representation.
Except for the normalization factors this coupling does not depend on the fluxes.



\section{Non-perturbative effects in local models}
 \label{s:nplocal}

\subsection{The non-perturbative superpotential}

As can be seen from the previous section, the $SO(12)$ model yields the right particle content but an oversimplified flavor structure. In particular, given the zero mode wavefunctions above only one generation of down-type quarks and leptons will develop non-vanishing Yukawa couplings. This feature has been shown to be general for F-theory models where all $D$-type Yukawa couplings arise from a single triple intersection, and a similar statement holds for $U$-type Yukawa couplings in points of $E_6$ enhancement \cite{cchv09}.

Following \cite{mm09}, one may solve the above rank-one Yukawa problem by considering the contribution of non-perturbative effects to Yukawa couplings. Indeed, as shown in there, the presence of an Euclidean D3-brane instanton wrapping a 4-cycle $S_{\rm np}$ of the three-fold base $B$ may induce a non-perturbative correction to the tree-level 7-brane superpotential (\ref{supo7}).\footnote{A similar effect is sourced by a gaugino condensate on a 7-brane wrapping $S_{\rm np}$, see\cite{mm09,bdkkm10}, but for simplicity in the following we will focus on the case of D3-brane instantons.} Such non-perturbative correction will not only depend on the 7-brane 4-cycle $S$, but also on the 4-cycle $S_{\rm np}$ that the D3-instanton is wrapping, and which is characterized by a holomorphic divisor function $h$: $S_{\rm np} = \{(x,y,z) \in B | h(x,y,z)=0\}$. Indeed, by repeating the computations of \cite{mm09,afim} (see also appendix \ref{ap:supo}) one obtains a full superpotential for the 7-brane on $S$ of the form
\be
W\, =\, m_*^4\left[ \int_S \tr (\Phi_{xy} F) \wedge dx \wedge dy + \frac{\eps}{2} \sum_n \int_S \theta_n \,  \str \left( \Phi_{xy}^n F \wedge F\right)\right]
\label{supo}
\ee
where the first contribution is nothing but the tree-level superpotential (\ref{supo7}) while the second is the non-perturbative correction $W_{\rm np}$ created by the non-perturbative effect. Here $(x,y)$ are local complex coordinates on the 4-cycle $S$ (which is locally described as $S = \{z=0\}$) and $\theta_n$  stand for the $n^{\rm th}$-derivative of ${\rm log\, } h$ along the holomorphic coordinate $z$ transverse to $S$. Finally, $\eps$ is a suppression factor that measures the strength of the non-perturbative effect. More precisely we have that
\bea
\eps & = & \CA\, e^{-T_{\rm np}} h_0^{N_{\rm D3}} \\
\theta_n & = & \frac{g_s^{-n/2} \mu^3}{(2\pi)^{2+ \frac{3n}{2}} m_*^{4+2n}}\, [\p_z^n\, {\rm log\, } (h/h_0)]_{z=0} 
\label{deftheta}
\eea
with $T_{\rm np} \, =\, {\rm Vol\, } (S_{\rm np}) + i \int_{S_{\rm np}} C_4$ the complexified K\"ahler modulus corresponding to $S_{\rm np}$, $h_0 = \int_S h$ the mean value of $h$ in $S$ and $N_{\rm D3} = (8\pi^2)^{-1} \int_S \tr (F \wedge F)\in \IN$ the total D3-brane charge induced on $S$ by the presence of the magnetic field $F$. In addition, $\mu \sim m_*$ is the fundamental scale for the non-perturbative effect, and $\CA$ a holomorphic function of the closed string moduli fields which will not play any role in the following.

Notice that the analysis of \cite{afim} did not consider the general non-perturbative superpotential $W_{\rm np}$ above, but rather the particular case where only $\theta_1$ (denoted $\theta$ therein) was non-vanishing. The reason for such Ansatz was the assumption taken in \cite{mm09} that the two 4-cycles $S$ and $S_{\rm np}$ do not intersect, which implies that $h|_S \equiv h_0$ and so $\theta_0$ vanishes.\footnote{This can be understood as follows: if the intersection of two divisors $\Sigma_1 \cap \Sigma_2$ is homologically trivial, then the restriction of the line bundle $\CL_{\Sigma_1}$ into $\Sigma_2$ is trivial, and vice versa. Hence, as the divisor function $h_1$ of $\Sigma_1$ is a section of $\CL_{\Sigma_1}$, we can always take $h_1|_{\Sigma_2}$ as a constant section of the trivial bundle $\CL_1|_{\Sigma_2}$.} Taking $\theta_1$ as an arbitrary holomorphic function on $S$ and neglecting all those terms with higher suppression on the scale $m_*^{-2}$ the Ansatz of \cite{afim} follows. Now, while these are valid assumptions for a large class of local F-theory models (including the $U(3)$ model of \cite{afim}) we will see that they need to be reconsidered for the $SO(12)$ model of the previous section. As a result, the computation of zero modes and Yukawas at the non-perturbative level will have to be revisited to include the effects of the more general superpotential (\ref{supo}).

In the present context, the assumption $S \cap S_{\rm np}=0$ is well-motivated if the 4-cycle $S$ localizes the MSSM degrees of freedom of the F-theory compactification. Indeed, for $S \cap S_{\rm np} \neq 0$ D3-instanton zero modes charged under the MSSM gauge group may arise at the intersection of the two divisors. These extra zero modes would in principle invalidate the analysis of \cite{mm09} which, similarly to \cite{bdkmmm06} and \cite{ag06}, assumes that all instanton modes charged under the gauge group of interest are massive. However, it can still happen that $S \cap S_{\rm np} \neq 0$ if $S$ is wrapped by a 7-brane that does not localize the MSSM gauge group. 
In fact, from the results of \cite{bcm11} we know that $[S \cap S_{\rm np}] \neq 0$ must be true for at least some 7-brane 4-cycle $S$, since otherwise a D3-instanton wrapping $S_{\rm np}$ would not have the right structure of zero modes to generate a superpotential. This result is in perfect agreement with the intuition of D3-instanton generated superpotentials in type IIB orientifold compactifications. Indeed, if our F-theory compactification has a type IIB limit, then by taking it the 4-cycle $S$ such that $S \cap S_{\rm np}\neq0$ will become the 4-cycle of an O7-plane, while the D3-instanton zero modes localized at $S \cap S_{\rm np}$ will correspond to the universal neutral zero modes of an $O(1)$ D3-brane instanton \cite{cgh11}.

\begin{figure}[ht]
\begin{center}
\begin{tabular}{lcr}
\includegraphics[width=7.375cm]{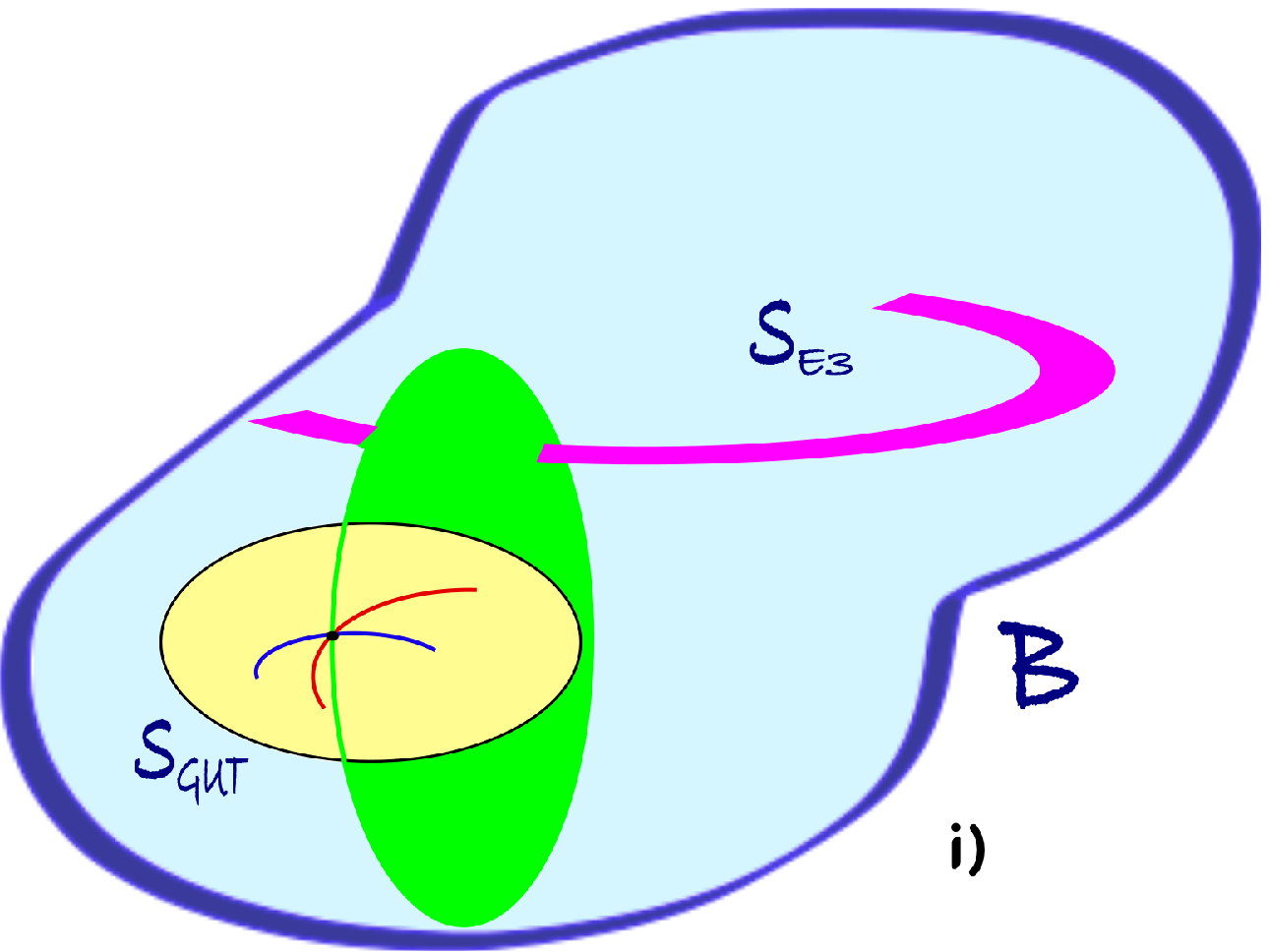}
& \quad &
\includegraphics[width=7.375cm]{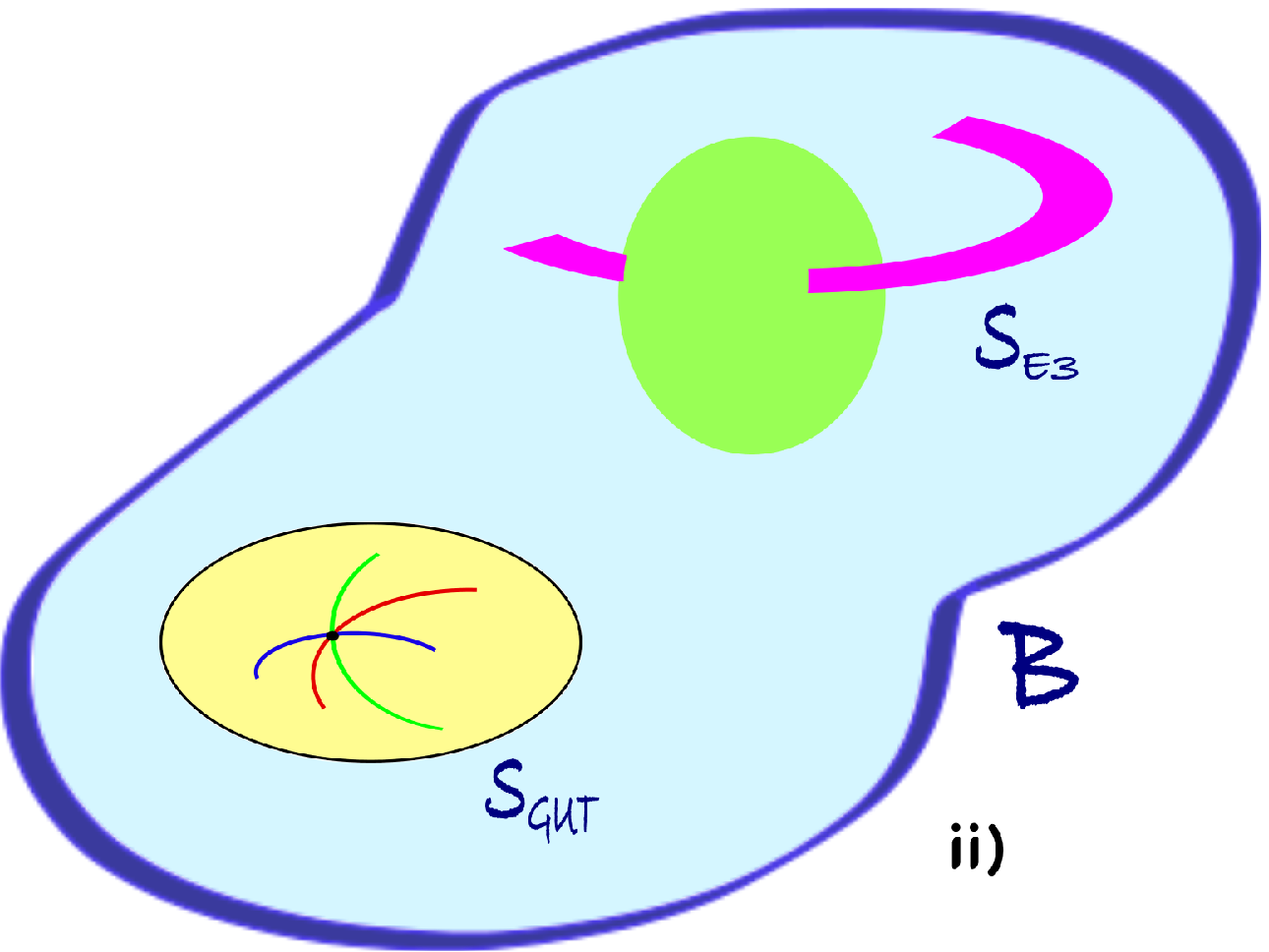}
\end{tabular}
\end{center}
\caption{\small{Two possible scenarios for non-perturbative corrections to Yukawa couplings. In figure $i)$ the D3-brane instanton intersects a 7-brane that in turn intersects the 4-cycle $S_{GUT} = S_{MSSM}$. In figure $ii)$ the D3-instanton intersects only 7-brane divisors fully disconnected from $S_{GUT}$.}}
\label{twoposs}
\end{figure}

To summarize, even if the D3-instanton generating the non-perturbative superpotential $W_{\rm np}$ within (\ref{supo}) does not intersect the 4-cycle $S_{\rm MSSM}$ that localizes the MSSM degrees of freedom, it must intersect some other 7-branes and this, in a potential type IIB limit, should correspond to a transverse intersection of the D3-instanton with an O7-plane. Given this, one may consider two different scenarios:

\begin{itemize}

\item[{\em i)}] The D3-instanton intersects a 4-cycle $S'$ which in turn intersects the 4-cycle $S_{\rm MSSM}$ in a matter curve $\Sigma$. That is, $[S_{\rm D3} \cap S'] \neq 0 \neq [S' \cap S_{\rm MSSM}]$ but $[S_{\rm D3} \cap S_{\rm MSSM}] = 0$.

\item[{\em ii)}] The D3-instanton intersects a 4-cycle $S'$ that does not intersect $S_{\rm MSSM}$ at all. That is, $[S_{\rm D3} \cap S'] \neq 0$ and  $[S' \cap S_{\rm MSSM}]=0=[S_{\rm D3} \cap S_{\rm MSSM}]$.

\end{itemize}

Let us consider both scenarios in the context of the $SO(12)$ model of the previous section, taking advantage that it can also be understood as a local type IIB orientifold model. Indeed, if we interpret the $SO(12)$ model in terms of D7-branes at angles, we obtain a configuration of the form

\begin{table}[htb] 
\renewcommand{\arraystretch}{1.25}
\begin{center}
\begin{tabular}{ccl}
Brane & Stack & \quad Divisor \\
\hline
O7-plane &  & $S$ : $z\, =\, 0$\\
1 D7-brane & $\a$ & $S_\a$ : $z \,= \, \sig\, (\oh y - x)$ \\
1 D7-brane & $\a^*$ & $S_{\a^*}$ : $z \,= \, -\sig\, (\oh y - x)$ \\
5 D7-branes & $\b$ & $S_\b$ : $z \,= \, \oh \sig\, y$ \\
5 D7-branes & $\b^*$ & $S_{\b^*}$ : $z \,= \, -\oh \sig\, y$ \\
\end{tabular}
\end{center}
\label{curvesso12}
\end{table}
\noindent
where we have ignored the world volume flux $F$ on each stack. Here we have defined the adimensional quantity $\sigma = (m/m_{st})^2$, with $m_{st}^{-2} = 2\pi \alpha'$ being the string scale in the orientifold limit, that determines the angle of intersection between D7-branes. The D7-branes $(\a,\a^*)$ are mapped to each other by the orientifold action, same for $(\b,\b^*)$, and so we only get a unitary gauge group from each pair. Namely, we obtain a gauge group $U(5) \times U(1)_\a \simeq SU(5) \times U(1)_\a \times U(1)_\b$, as expected. Finally, we can also match the matter spectrum obtained in the previous section by realizing that the sectors $a$, $b$, $c$ arise from the matter curves $\Sigma_a = S_\a \cap S_\b$, $\Sigma_b = S_\b \cap S_{\b^*}$ and $\Sigma_c = S_\a \cap S_\b^*$, and then including the effect of the world volume flux. 

In this type IIB version of the $SO(12)$ model it is clear how to realize the scenario $i)$ above. First we should require that the 4-cycle $S_{\rm np}$ wrapped by the D3-instanton intersects the O7-plane 4-cycle $S$, and then that it does not intersect $S_\a$, $S_\b$ or their orientifold images. Notice that by intersecting the 4-cycle $S$ we do not generate any extra D3-instanton zero mode. On the contrary, this nontrivial intersection is necessary for  $S_{\rm np}$ to host a $O(1)$ D3-instanton and so to get rid of unwanted zero modes. 

Hence, in this scenario we have that the D3-instanton intersects the O7-plane in a non-trivial 2-cycle $\Sigma_{\rm E3} = S \cap S_{\rm np}$, which by assumption is far away from the Yukawa point $S \cap S_\a \cap S_\b$ at $x=y=z=0$. By the discussion above, this means that the pull back $h|_S$ of the divisor function $h$ will not be a constant, nor will be the function $\theta_0 \propto [{\rm log} (h/h_0)]_{z=0}$ that enters into the superpotential (\ref{supo}) if $S$ refers to the 4-cycle of the O7-plane. Now the important point is that, for the $SO(12)$ model under discussion, the 4-cycle $S$ in (\ref{supo}) indeed corresponds to the embedding of the O7-plane, rather than to $S_{\rm MSSM}$.\footnote{Note that for $\langle \Phi \rangle = 0$ we recover an $SO(12)$ gauge group localized at the 4-cycle $S$, namely a stack of 12 D7-branes on top of the O7-plane locus, while for $\langle \Phi \rangle \neq 0$ the gauge group is broken because these D7-branes are wrapping different 4-cycles, one of them being $S_{\rm MSSM}$. Hence $S_{\rm MSSM} \neq S$.} As a result, for the $SO(12)$ model and within the scenario $i)$ above, $\theta_0(x,y)$ will be a non-constant holomorphic function on $S$. 

On the other hand, if we consider the scenario $ii)$ for the $SO(12)$ model we have that $S \cap S_{\rm np} = 0$, and so in this case $\theta_0 = 0$ just like in the Ansatz of \cite{afim}. In fact, as we will see momentarily, it happens for the $SO(12)$ model that the contribution to (\ref{supo}) coming from $\theta_1$ also vanishes, and so for this scenario the first non-vanishing contribution to the non-perturbative Yukawa couplings would come from $\theta_2$, which is already rather suppressed in the scale $m$. Thus, for the purpose of generating realistic Yukawa couplings, the previous scenario seems more promising.

We then see that the two scenarios $i)$ and $ii)$, which have a clear geometrical meaning in terms of a global model (cf. fig. \ref{twoposs}), can be characterized in terms of $\theta_0$ in our local description. Such function should be taken constant for scenario $ii)$, while it will be given by a certain holomorphic function for scenario $i)$. Finally, this result will also hold away from the type IIB orientifold description of the $SO(12)$ model, the only difference being that we should replace the O7-plane by a more complicated system of $(p,q)$ 7-branes.

The discussion above shows how the geometry of our F-theory compactification can constrain the superpotential (\ref{supo}), and one may find further geometrical constraints on these non-perturbative corrections. Indeed, let us go again to the type IIB orientifold description of our local $SO(12)$ model. There the non-perturbative piece of the superpotential (\ref{supo}) is generated by a $O(1)$ D3-instanton, whose embedding $S_{\rm np} = \{h=0\}$ determines the holomorphic functions $\theta_n$ via the divisor function $h$. Now, the 4-cycle $S_{\rm np}$ will host a $O(1)$ D3-instanton only if it is invariant under the orientifold action, which close to our Yukawa point is given by $z \mapsto -z$. This means that in the vicinity of this point we have to impose $h(z) = h(-z)$, and from the definition (\ref{deftheta}) it follows that $\theta_n = 0$ for $n$ odd. In particular the function $\theta_1$, in which the Ansatz used in \cite{afim} was based, does vanish for the $SO(12)$ model, and so in principle the whole analysis needs to be revisited.

Again, this kind of constraint is not only true in the type IIB orientifold limit, and it holds for general local F-theory models. In fact, one can see directly from the superpotential (\ref{supo}) that the terms that correspond to $\theta_n$ with $n$ odd vanish automatically for $SO(N)$ groups, in agreement with our type IIB intuition. Indeed, such terms are proportional to symmetrized traces of the form $\str(\mathfrak{t}_1\mathfrak{t}_2\dots\mathfrak{t}_{n+2})$, with $\mathfrak{t}_i$ generators of $SO(N)$ in the vector representation. As these $\mathfrak{t}_i$ are antisymmetric matrices, the symmetrized trace will vanish identically for any $n$ which is odd. In particular, the symmetric tensor $d_{abc} = \str(\mathfrak{t}_a \mathfrak{t}_b \mathfrak{t}_c)$ which corresponds to the correction of $\theta_1$ and played a key role in \cite{afim}  will also vanish identically. As already pointed out in \cite{afim}, the fact that $d_{abc} =0$ is not only true for $SO(N)$, but also for other groups of interest in F-theory GUT model building like $E_6$, $E_7$ and $E_8$. Hence, an analysis of non-perturbatively generated Yukawas that not only involves $\theta_1\neq 0$ but the more general superpotential (\ref{supo}) is in order. 

\subsection{The corrected equations of motion}
\label{ss:eom}

As a consequence of the new superpotential (\ref{supo}), the equations of motion for the 7-brane fields $\Phi$ and $A$ are modified, affecting a local model in two different ways. On the one hand the background values $\langle \Phi \rangle$ and $\langle A \rangle$ will be shifted to a new value. On the other hand the fluctuation fields  $(a_{\bar{m}}, \psi_{\bar{m}})$ and $(\varphi_{xy}, \chi_{xy})$ will satisfy different zero mode equations, and so the wavefunctions obtained in the previous section will receive a correction that accounts for the non-perturbative effect. In the following we will discuss the new equations that the 7-brane fields need to satisfy, solving for the background to first order in the non-perturbative parameter $\eps$. We will also analyze the new zero mode equations for the bosonic fluctuations  $(a_{\bar m}, \varphi_{xy})$, which by supersymmetry satisfy the same equations as the fermionic modes $(\psi_{\bar m}, \chi_{xy})$. The next section will be devoted to apply this latter set of equations in order to obtain the corrected zero mode wavefunctions for the $SO(12)$ model of section \ref{s:so12}.

The equations of motion that follow from the superpotential (\ref{supo}) are
\begin{subequations}
\label{Fterm7np}
\begin{align}
\label{Fterm7Anp}
D_{\bar{m}} \Phi_{xy} + \eps \sum_n \left( D_{y} [{\rm S} (\theta_n \Phi_{xy}^n F_{\bar{m}x})]  + D_{x} [{\rm S} (\theta_n \Phi_{xy}^n F_{y\bar{m}})] +
D_{\bar{m}} [{\rm S} (\theta_n \Phi_{xy}^n F_{xy})] \right)  & =  0\\
\label{Fterm7phinp}
F \wedge dx \wedge dy + \frac{\eps}{2} \sum_n n\, \theta_n\, {\rm S} (\Phi_{xy}^{n-1} F \wedge F ) & =  0
\end{align}
\end{subequations}
where S stands for the symmetrization of all the elements within the brackets, and $D_m$ are the covariant derivatives along the GUT 4-cycle defined below eq.(\ref{matrixDirac}). Notice that if we take all $\theta_n =0$ except for $\theta_1$ we recover eqs.(3.40) of \cite{afim}, as expected.

In addition, the D-term (\ref{FI7}) gives the additional equation 
\be
\omega \wedge F + \frac{1}{2} [\Phi, \bar{\Phi}]\, =\, 0
\label{Dterm7}
\ee
where in the local coordinate system that we are working with the K\"ahler form is taken to be
\be
\omega = \frac{i}{2}\left( dx\wedge d\bar{x} +  dy\wedge d\bar{y}\right).
\label{kahlerform}
\ee
Notice that, unlike the F-term equations (\ref{Fterm7np}), the D-term equations do not receive any correction due to the non-perturbative effects. This may seem surprising, but it can be derived directly from the results of \cite{dm10}, as we discuss in appendix \ref{ap:supo}.

While the system of differential equations (\ref{Fterm7np}) and  (\ref{Dterm7}) is quite difficult to solve, one may follow \cite{afim} and perform a perturbative expansion in the small parameter $\eps$
\be
\Phi_{xy}\, =\, \Phi^{(0)}_{xy} + \eps\, \Phi^{(1)}_{xy} +\eps^2\, \Phi^{(2)}_{xy} + \dots \quad \quad \quad A_{\bar{m}}\, =\, A_{\bar{m}}^{(0)} + \eps\, A_{\bar{m}}^{(1)} + \eps^2\, A_{\bar{m}}^{(2)} + \dots
\label{expeps}
\ee
and then solve for the above equations order by order in $\eps$. Indeed, to zeroth order in this expansion (\ref{Fterm7np}) reduce to
\be
(D_{\bar{m}} \Phi_{xy})^{(0)}\, =\, F_{\bar{x}\bar{y}}^{(0)} \, = \,  0
\label{Fterm7np0}
\ee
which are indeed the 7-brane F-term equations in the absence of any non-perturbative effect. To first order we obtain
\begin{subequations}
\label{Fterm7np1}
\begin{align}
\label{Fterm7Anp1} \nonumber
 (D_{\bar{m}} \Phi_{xy})^{(1)} \,=\, & - \sum_{n=0} (\p_y \theta_n) {\rm S} [\Phi_{xy}^{n} F_{\bar{m}x}]^{(0)}  -  (\p_x \theta_n) {\rm S} [\Phi_{xy}^{n} F_{y\bar{m}}]^{(0)}  \\
& - \sum_{n=1} n\, \theta_n  \left( {\rm S} [\Phi_{xy}^{n-1} D_{y} \Phi_{xy}  F_{\bar{m}x}] + {\rm S} [\Phi_{xy}^{n-1} D_{x} \Phi_{xy}  F_{y\bar{m}}]   \right)^{(0)} \\
\label{Fterm7phinp1}
F^{(1)} \wedge dx \wedge dy\,  =\, & - \frac{1}{2} \sum_n n\, \theta_n\, {\rm S} [ \Phi_{xy}^{n-1} F \wedge F]^{(0)}  
\end{align}
\end{subequations}
where we have used the Bianchi identity $D_{[m} F_{np]} = 0$ and we have defined
\be
(D_{\bar{m}} \Phi_{xy})^{(1)} \,=\, D_{\bar{m}}^{(0)} \Phi^{(1)}_{xy} -i [A_{\bar{m}}^{(1)}, \Phi^{(0)}]
\ee
%


Let us now provide a solution to these $\CO(\eps)$ equations at the level of the background, generalizing the results in eqs.(3.45) of \cite{afim}, where they were solved for the particular case where  
$\theta_n = \d_{n1} \theta$.\footnote{As pointed out in \cite{afim} these solutions for the background are directly 
related to a Seiberg-Witten map \cite{sw99} acting on the 7-brane fields $A$ and $\Phi$. Hence, a simple way to find the solutions for 
(\ref{Fterm7np1}) is to generalize the SW map in \cite{afim}, eqs.(4.18) to the present case. A natural generalization is given by
\begin{subequations}
\label{ansatz}
\begin{align}
\hat{A}_{\bar{m}} \, =\, A_{\bar{m}} + \tilde{A}_{\bar{m}} \, =\ & A_{\bar{m}} - \frac{\eps}{2} \sum_n n\, \theta_{n}^{ij}\,  {\rm S}  \left[\Phi_{xy}^{n-1} A_i (\p_jA_{\bar{m}} + F_{j\bar{m}}) \right] + \CO (\eps^2)
\label{ansatzA} \\
\hat{\Phi}_{xy}\, = \, \Phi_{xy} + \tilde{\Phi}_{xy} \, = \ & \Phi_{xy} -  \frac{\eps}{2} \sum_n {\rm S} \left[ A_i (\p_j + D_j) (\theta_n^{ij} \Phi_{xy}^n) \right] + \CO(\eps^2)
\label{ansatzPhi}
\end{align}
\end{subequations}
where $i,j = x,y$ and $\theta_n^{xy} = - \theta_n^{yx} = - \theta_n$.
At the level of the background and in the holomorphic gauge $\langle A_{\bar{m}}\rangle^{(0)} = 0$, we can identify $\langle \tilde{A}_{\bar{m}} \rangle = - \eps\langle A_{\bar{m}}\rangle^{(1)}$ and $\langle \tilde{\Phi}_{xy} \rangle = - \eps\langle \Phi_{xy}\rangle^{(1)}$, obtaining eqs.(\ref{solAbkg}) and (\ref{solPhibkg}).\label{SWfoot}}  
This more general solution is given by
\be
\langle A_{\bar{m}}\rangle^{(1)}\, =\, - \oh \sum_{n=1} n\, \theta_{n}\,  {\rm S}  \left[\langle \Phi_{xy}\rangle^{n-1}( \langle A_x\rangle \langle F_{y\bar{m}} \rangle - \langle A_y \rangle \langle F_{x\bar{m}}\rangle) \right]^{(0)}
\label{solAbkg}
\ee
which indeed solves (\ref{Fterm7phinp1}) at the level of the background. Note that this result is true even if the matrices in (\ref{solAbkg}) do not commute. 
Similarly, we have that
\bea
\label{solPhibkg}
\langle \Phi_{xy} \rangle^{(1)} & =  & \sum_{n=0} \left[ (\p_x \theta_n)  {\rm S} [\Phi_{xy}^{n} \langle A_y \rangle]^{(0)}  -  (\p_y \theta_n)   {\rm S} [\Phi_{xy}^{n} \langle A_x \rangle]^{(0)}\right] \\  \nonumber
& + & \oh  \sum_{n=1} n\, \theta_n \big( {\rm S} \left[ \langle A_y\rangle \langle \Phi_{xy}\rangle^{n-1} (\p_x + D_x) \Phi_{xy} \right] 
 -{\rm S} \left[ \langle A_x\rangle \langle \Phi_{xy}\rangle^{n-1} (\p_y + D_y) \langle \Phi_{xy} \right]  \big)^{(0)}
\eea
satisfies eq.(\ref{Fterm7Anp1}) at the level of the background. Indeed, by applying $D_{\bar{m}}^{(0)} = \p_{\bar{m}}$ to the rhs of (\ref{solPhibkg}) and using that $\th_n$ and $\langle\Phi_{xy}\rangle^{(0)}$ are holomorphic we obtain the rhs of (\ref{Fterm7Anp1}) plus the term
\bea
\nonumber
&& \frac{i}{2} \sum_n n\, \theta^{ij}_n \left({\rm  S} 
\left[  
\langle \Phi_{xy}\rangle^{n-1} \langle F_{\bar{m}i} \rangle [\langle A_{j}\rangle,  \langle \Phi_{xy}\rangle ]
\right] 
- {\rm  S} \left[
\langle \Phi_{xy}\rangle^{n-1} \langle A_{i} \rangle [\langle F_{\bar{m}j} \rangle,  
\langle \Phi_{xy}\rangle]
\right] 
\right)^{(0)} \\
&& =\, \frac{i}{2} \sum_n n\, \theta^{ij}_n 
\left[ 
{\rm  S} \left[
\langle \Phi_{xy}\rangle^{n-1} \langle A_i\rangle \langle F_{j\bar{m}} \rangle 
\right], 
 \langle \Phi_{xy}\rangle 
\right]^{(0)}\, =\, i 
\left[ \langle A_{\bar{m}} \rangle^{(1)} , \langle \Phi_{xy} \rangle^{(0)}\right] 
\eea
as needed from eq.(\ref{Fterm7Anp1}). Here $i,j = x,y$ and $\theta_n^{xy} = - \theta_n^{yx} = - \theta_n$.

The next step is to derive the zero mode equations for the bosonic fluctuations $(\varphi_{xy}, a_{\bar{m}})$  defined by
\be
\Phi_{xy}\, =\, \langle \Phi_{xy} \rangle + \varphi_{xy} \quad \quad A_{\bar{m}}\, =\, \langle A_{\bar{m}} \rangle + a_{\bar{m}}
\label{fluct2}
\ee
and which also follow an expansion of the form (\ref{expeps}). 
Expanding the $\CO(\eps^0)$ equation (\ref{Fterm7np0}) to linear order in fluctuations we obtain the 
equations of motion for $(\varphi_{xy}^{(0)}, a_{\bar{m}}^{(0)})$. One can check that together with 
the zeroth order D-term equation, they amount to the system of equations (\ref{Dirac9}), 
where now $\Psi\, =\, (0, a_{\bar{x}}^{(0)}, a_{\bar{y}}^{(0)}, \varphi_{xy}^{(0)})$. 
Hence at zeroth order in $\eps$ the wavefunctions for the bosonic fluctuations match the fermionic 
wavefunctions in absence of non-perturbative effects, as expected from supersymmetry. 
For the $SO(12)$ model, the zeroth order wavefunctions for each sector are then given by (\ref{simvecs}). 

Let us now consider the equations for  $(\varphi_{xy}^{(1)}, a_{\bar{m}}^{(1)})$, that arise from expanding (\ref{Fterm7np}) to first order in $\eps$. Just like in \cite{afim}, in order to write down the zero mode equation to order $\CO(\eps)$ we need to take into account the corrections to the background (\ref{solAbkg}) and (\ref{solPhibkg}).
Indeed, let us first consider the particular case where $\theta_0 \neq 0$ but $\theta_n = 0$ for all $n>0$.  Then, at the linear order in fluctuations and in the holomorphic gauge, eqs.(\ref{Fterm7np}) read
\begin{subequations}
\label{Ftermth0}
\begin{align}
D_{\bar{x}} a_{\bar{y}} - D_{\bar{y}} a_{\bar{x}} \, = \ & 0\\
D_{\bar{m}} \vphi_{xy} + i [\langle \Phi_{xy} \rangle, a_{\bar{m}}] \, = \ & \eps\left[(\p_y\th_0) D_x a_{\bar{m}} -  (\p_x\th_0) D_y a_{\bar{m}}\right]
\end{align}
\end{subequations}
In addition, the corrected background solutions reduce to
\bea
\langle A_{\bar{m}} \rangle & = & 0 + \CO(\eps^2)\\
\langle \Phi_{xy} \rangle & = & \langle\Phi_{xy}\rangle^{(0)} + \eps \left(\langle A_y\rangle^{(0)} \p_x\th_0  - \langle A_x\rangle^{(0)} \p_y\th_0  \right)  + \CO(\eps^2)
\eea
and so, by plugging the latter into (\ref{Ftermth0}) we obtain
\bea
\p_{\bar{x}} a_{\bar{y}} - \p_{\bar{y}} a_{\bar{x}} & = & 0 + \CO(\eps^2)
\label{ft01}\\
\p_{\bar{m}} \vphi_{xy}+ i [\langle \Phi_{xy} \rangle^{(0)}, a_{\bar{m}}] & = & \eps\left[(\p_y\th_0) \p_x a_{\bar{m}} -  (\p_x\th_0) \p_y a_{\bar{m}}\right] + \CO(\eps^2)
\label{ft02}
\eea
which could have also been obtained by linearizing (\ref{Fterm7np0}) and (\ref{Fterm7np1}) in fluctuations.
Notice that these F-term equations are independent of the worldvolume flux, and that the cubic coupling that arises from (\ref{supo}) for only $\theta_0 \neq 0$ is also flux-independent. Hence, following the arguments of \cite{cchv09} we have that the holomorphic piece of the Yukawa couplings should not depend on these fluxes. Moreover, it is easy to see that shifting $\theta_0$ by a constant will not modify nor the wavefunctions nor the Yukawa couplings. This can be traced back to the fact that a constant $\theta_0$ will only add a constant piece to the superpotential (\ref{supo}). A detailed analysis of the F-term zero mode equations for this case as well as a computation of the holomorphic Yukawas in terms of a residue formula is given in appendix \ref{ap:res}.

Similarly, the zero mode equations for $(\varphi_{xy}^{(1)}, a_{\bar{m}}^{(1)})$ with general $\th_n$ can be obtained by linearizing eqs.(\ref{Fterm7np1}) in fluctuations. Let us do so for the case where only $\theta_0$ and $\theta_2$ are non-vanishing, which will be the case analyzed in the next section. From (\ref{Fterm7Anp1}) we have
\bea\label{FtermA02}
\p_{\bar{m}} \vphi_{xy}^{(1)}+ i [\langle \Phi_{xy} \rangle^{(0)}, a_{\bar{m}}^{(1)}] & = & (\p_y\th_0) \p_x a_{\bar{m}}^{(0)} -  (\p_x\th_0) \p_y a_{\bar{m}}^{(0)} \\ \nonumber
& + & (\p_y\th_2) {\rm S} [ \langle \Phi_{xy}  \rangle^2  D_x a_{\bar{m}}]^{(0)}  -  (\p_x \th_2)   {\rm S} [  \langle \Phi_{xy}  \rangle^2 D_{y} a_{\bar{m}}]^{(0)} \\ \nonumber
& + & i  (\p_y\th_2)  [ \langle \Phi_{xy}  \rangle^2   \langle A_x \rangle, a_{\bar{m}}]^{(0)}  -  i (\p_x \th_2)   [  \langle \Phi_{xy}  \rangle^2 \langle A_y \rangle, a_{\bar{m}}]^{(0)} \\ \nonumber
& + &  2\,\th_2 \,{\rm S} [ \langle \Phi_{xy}  \rangle \left( \p_y  \langle \Phi_{xy}  \rangle  D_x a_{\bar{m}} -  \p_x  \langle \Phi_{xy}  \rangle D_{y} a_{\bar{m}} \right)]^{(0)} 
\\ \nonumber
& + & 2 i\, \th_2 \left[\langle \Phi_{xy} \rangle \left( \p_y  \langle \Phi_{xy} \rangle \langle A_x \rangle - \p_x   \langle \Phi_{xy} \rangle \langle A_y \rangle \right), a_{\bar{m}}\right]^{(0)} 
 \\ \nonumber
 & + & (\p_y\th_2) {\rm S} [ \langle \Phi_{xy} \rangle \langle F_{x\bar{m}} \rangle \varphi_{xy}  ]^{(0)}  -  (\p_x \th_2)   {\rm S} [  \langle \Phi_{xy}  \langle F_{y\bar{m}} \rangle \varphi_{xy} ]^{(0)} \\ \nonumber
 & + &  2\, \th_2\, {\rm S} [ \langle \Phi_{xy}  \rangle \left( D_y \varphi_{xy}  \langle F_{x\bar{m}}  \rangle -  D_x   \vphi_{xy} \langle F_{y\bar{m}}  \rangle \right)]^{(0)} 
 \\ \nonumber
 & + &  2\, \th_2\,  {\rm S} [ \vphi_{xy} \left( \p_y \langle \Phi_{xy}  \rangle \langle F_{x\bar{m}}  \rangle -  \p_x  \langle \Phi_{xy}  \rangle \langle F_{y\bar{m}}  \rangle \right)]^{(0)}  \\ \nonumber
 & + &  i \th_2 \left[\langle \Phi_{xy} \rangle \left(  \langle F_{x\bar{m}} \rangle \langle A_y \rangle  -  \langle F_{y\bar{m}} \rangle \langle A_x \rangle \right), \vphi_{xy}\right]^{(0)} 
\eea
where we have used the assumption that $\langle A_{\bar{m}} \rangle$ and $ \langle \Phi_{xy}  \rangle$ commute. 
From (\ref{Fterm7phinp1}) we obtain 
\bea\label{Ftermphi02}
\p_{\bar{x}} a_{\bar{y}}^{(1)} - \p_{\bar{y}} a_{\bar{x}}^{(1)} & = &  2 \theta_2\, {\rm S} \left[ \vphi_{xy} \left(\langle F_{x\bar{x}} \rangle \langle F_{y\bar{y}}\rangle - \langle F_{x\bar{y}} \rangle \langle F_{y\bar{x}}\rangle \right)  \right]^{(0)} \\ \nonumber
& + &   2 \theta_2\, {\rm S} \left[ \langle \Phi_{xy} \rangle \left(\langle F_{x\bar{x}} \rangle  D_{y} a_{\bar{y}} -  \langle F_{y\bar{x}}\rangle D_x a_{\bar{y}}  + \langle F_{y\bar{y}} \rangle  D_{x} a_{\bar{x}} -  \langle F_{x\bar{y}}\rangle D_y a_{\bar{x}}  \right)  \right]^{(0)}  \\ \nonumber
& + & i \theta_2\,  \left[ \langle \Phi_{xy}\rangle ( \langle F_{x\bar{x}}\rangle \langle A_y \rangle - \langle F_{y\bar{x}} \rangle \langle A_x\rangle), a_{\bar{y}} \right]^{(0)} \\ \nonumber 
& + &  i \theta_2\,  \left[ \langle \Phi_{xy}\rangle ( \langle F_{y\bar{y}}\rangle \langle A_x \rangle - \langle F_{x\bar{y}} \rangle \langle A_y\rangle), a_{\bar{x}} \right]^{(0)}
\eea
which for non-vanishing fluxes are rather involved equations. Nevertheless, by means of the Seiberg-Witten map of footnote \ref{SWfoot} one can simultaneously get rid of all the flux dependence within these two F-term equations and in the cubic coupling that arises from (\ref{supo}) with $\theta_2 \neq 0$, see appendix \ref{ap:res} for details. Hence, according to the results of \cite{cchv09} we have that the holomorphic piece of the Yukawa couplings should not depend on the worldvolume flux $\langle F \rangle$, a fact that we will use in the next two sections. 

Finally, the fluctuations must satisfy the D-term equation that can be derived from 
expanding (\ref{Dterm7}) to linear order in fluctuations, namely
\be
\omega \wedge \p_{\langle A \rangle} a - \frac{1}{2} [\langle \bar{\Phi} \rangle, \varphi]   = 0
\label{Dterm}
\ee
where $a = a_{\bar{x}} d\bar{x} + a_{\bar{y}} d\bar{y}$ and $\vphi = \vphi_{xy} dx \wedge dy$. 
As mentioned above, at zeroth order in $\eps$ we obtain part of the set of equations (\ref{Dirac9}) for 
$(\varphi_{xy}^{(0)}, a_{\bar{m}}^{(0)})$. 
At first order in $\eps$ we obtain
\be
D_{x} a_{\bar{x}}^{(1)} + D_{y} a_{\bar{y}}^{(1)} - i  [\langle \bar{\Phi}_{\bar{x}\bar{y}} \rangle^{(0)}, \vphi_{xy}^{(1)}]\,=\, i \left[ \langle A_x \rangle^{(1)}, a_{\bar{x}}^{(0)} \right] +  i \left[ \langle A_y \rangle^{(1)}, a_{\bar{y}}^{(0)} \right] +  i \left[ \langle \bar{\Phi}_{\bar{x}\bar{y}} \rangle^{(1)}, \vphi_{xy}^{(0)} \right] 
\label{Dterm02}
\ee
where $\langle A_m \rangle^{(1)}$ and $\langle \bar{\Phi}_{\bar{x}\bar{y}} \rangle^{(1)}$ are the Hermitian conjugates 
of (\ref{solAbkg}) and (\ref{solPhibkg}), respectively.


\section{Non-perturbative zero modes at the {SO(12)} point}
\label{s:npwave}

Let us now apply the non-perturbative scheme of the previous section to the Yukawa point of $SO(12)$ enhancement. More precisely, we would like to consider the superpotential (\ref{supo}) for the local $SO(12)$ model described in section \ref{s:so12}, and see how this new superpotential affects the wavefunction profile for the matter fields and the Yukawa couplings. The aim of this section is to solve for the zero mode wavefunctions in the presence of the non-perturbative piece of the superpotential, leaving the computation of Yukawa couplings for the next section. 

Before attempting to solve for such wavefunctions, it is useful to recall some results obtained in the previous section, which impose some constraints on the superpotential (\ref{supo}) for the model at hand. Indeed, because the 7-brane fields $\Phi$ and $A$ take values in the algebra of $SO(12)$, all those non-perturbative contributions in (\ref{supo}) with $n$ = odd vanish identically. Hence, we should consider those terms that involve the holomorphic functions $\th_0$, $\th_2$, $\dots$ etc. 

As also discussed in the previous section, the new zero mode wavefunctions can be written as a perturbative expansion in the small parameter $\eps$ that measures the strength of the non-perturbative effect. More precisely we have that
\be
\vec{\psi}_\rho \, =\, \vec{\psi}_\rho^{(0)} + \eps \left(\vec{\psi}_{\rho,\th_0}^{(1)} + \vec{\psi}_{\rho,\th_2}^{(1)} + \dots \right) + \CO(\eps^2)
\label{npzmexp}
\ee
where $\vec{\psi}_\rho^{(0)}$ are the tree-level wavefunctions (\ref{simvecs})-(\ref{simhol}) for the sector $\rho=a_p^+,b_q^+,c_r^+$, $\vec{\psi}_{\rho, \th_0}^{(1)}$ is the $\CO(\eps)$ correction to this sector when all $\theta_{2n}$ vanish except $\theta_0$, and the same for $\vec{\psi}_{\rho, \th_2}^{(1)}$. If we consider both $\theta_0$ and $\theta_2$ present at the same time, the first order correction to the zeroth order wavefunctions must satisfy the F and D-term equations (\ref{FtermA02}), (\ref{Ftermphi02}) and (\ref{Dterm02}). Notice that $\th_0$ and $\th_2$ appear linearly in the rhs of these equations, which justifies that we can write the first order correction as the sum $\vec{\psi}_{\rho,\th_0}^{(1)} + \vec{\psi}_{\rho,\th_2}^{(1)}$. This statement generalizes to the case where more $\th_{n}$ are present in the non-perturbative piece of the superpotential. However, as we will see in the next section, for the purpose of computing Yukawa couplings considering only $\theta_0$ and $\theta_2$ is enough.

In the remaining of this section we will solve for the first order corrections $\vec{\psi}_{\rho,\th_n}^{(1)}$ to the wavefunctions (\ref{simvecs})-(\ref{simhol}), first for the case where $n=0$ and then for $n=2$. As it is clear from eqs.(\ref{FtermA02})-(\ref{Dterm02}) the corrections for the case $n=2$ are more difficult to obtain but, as mentioned before, one can do a field redefinition that removes the worldvolume flux dependence. Hence, in section \ref{ss:n2} we will determine the corrections $\vec{\psi}_{\rho,\th_2}^{(1)}$ with fluxes turned off. Still, the computation for the $n=2$ case is slightly more technical and the reader not interested in such details may focus on the simpler $n=0$ case in order to get the idea of the computation, and then proceed to the computation of Yukawa couplings in the next section. 

\subsection{Zero modes for $n=0$}
\label{ss:n0}

Let us then consider the wavefunction corrections for the case where $\theta_0 \neq 0$ but $\theta_n = 0$ for all $n>0$. Notice that in \cite{afim} $\th_0$ was also present but assumed to be constant, and shown that Yukawa couplings are independent of it. As discussed in the previous section we may now relax this condition and take $\th_0$ to be non-constant and holomorphic on $x,y$. For simplicity let us take it to be
\be
\th_0\, =\, im^2\left(\th_{00} + x\, \th_{0x}   + y\, \th_{0y}\right) 
\label{thetaeq}
\ee
where $\th_{00}$, $\th_{0x}$, $\th_{0y}$ are complex constants, and the factor $im^2$ has been added for later convenience. We will now see that the corrected wavefunctions do depend on $\th_{0x}$ and $\th_{0y}$, and in the next section that they also enter into the corrected Yukawa couplings.

The corrected equations of motion for the fluctuations were derived in section \ref{s:nplocal}, and they can be conveniently rewritten 
by using the notation
\be
\vec{\psi}_{(\rho)}
\, =\, \vec{\psi}_{\!\! \rho} E_\rho
\, =\,
\left(
\begin{array}{c}
a_{\rho \bar x} \\ a_{\rho \bar{y}} \\ \vphi_{\rho xy}
\end{array}
\right) \, E_\rho
\label{defvec}
\ee
Recall that because of supersymmetry the bosonic fluctuations $(a_{\rho \bar m}, \varphi_{\rho xy})$ pair up with 
fermionic fluctuations $(\psi_{\rho \bar m}, \chi_{\rho})$ analyzed in section \ref{ss:pzm}, and so in the absence of 
non-perturbative effects the components of $\vec{\psi}_{\!\! \rho}$ match those of the vectors in eq.~(\ref{simvecs}).  
Using the above notation we can express the corrected F-term equations (\ref{ft01}) and (\ref{ft02}) as
\bea
\p_{\bar{x}} a_{\rho\bar{y}} - \p_{\bar{y}} a_{\rho\bar{x}} & = & 0 \label{ft03}\\
\p_{\bar{m}} \vphi_{\rho xy} + i m^2 q_\Phi(\rho) a_{\rho\bar{m}} & = & i\eps m^2 \left[\th_{0y} \p_x a_{\rho\bar{m}} -  \th_{0x} \p_y a_{\rho\bar{m}}\right] + \CO(\eps^2)
\label{ft04}
\eea
which, in the holomorphic gauge that we are considering, do not depend on the worldvolume flux densities $M_{xy}$, $q_P$ and $q_S$. By the results of \cite{cchv09,fi09} this implies that the holomorphic Yukawa couplings cannot depend on these quantities either. We will verify this explicitly by means of a residue computation performed in appendix \ref{ap:res}.

On the other hand, the D-term (\ref{Dterm02}) translates into
\bea
& & \hspace*{-1cm}\left\{\partial_x + \bar x[q_P(\rho) - M_{xy} q_F(\rho)] - \bar y q_S(\rho) \right\} a_{\rho \bar x}  \nonumber \\
& &  + \hspace{3mm}  \left\{\partial_y  - \bar y[q_P(\rho) + M_{xy} q_F(\rho)] - \bar x q_S(\rho) \right\} a_{\rho \bar y} - i m^2 \bar{q}_\Phi(\rho) \varphi_{\rho xy} \qquad 
\label{dsr}  \\
&  = & 
i \eps m^2 \bar\th_{0x}\left\{y[M_{xy} q_F(\rho) + q_P(\rho)] + xq_S(\rho) \right\} \varphi_{\rho xy} \nonumber \\
& \hspace*{1cm} & - \hspace{3mm}  
i \eps m^2  \bar\th_{0y}\left\{x [M_{xy} q_F(\rho) - q_P(\rho)] + y q_S(\rho) \right\} \varphi_{\rho xy}
\nonumber
\eea
where the specific values of the charges $q_\Phi$, $q_F$, $q_P$, and $q_S$, for each sector $\rho$ are given in tables \ref{t1} and \ref{t2}. 
Notice that as usual the D-term equation depends on the flux densities, and in particular on the hypercharge fluxes contained in $q_P$ and $q_S$. Hence, just like in \cite{afim}, the holomorphic  Yukawa couplings will not depend on the hypercharge but the physical Yukawas will, as we show in the next section. 

As already mentioned, the zero modes to zeroth order in $\eps$ are given by eq.~(\ref{simvecs}).
To find the corrected zero modes the strategy is to start with an Ansatz motivated by the zeroth order solutions and then proceed perturbatively in $\eps$. Notice that the zeroth order solutions consist of a fixed vector $\vec{v}_\rho$ given by (\ref{ap:vec}) multiplying a scalar wavefunction $\chi_\rho= \varphi_{\rho xy}$, that has a simple dependence on the complex variables $\lam_{\bar{x}} y + \lam_{\bar{y}} x$ and $\lam_{\bar{x}} x - \lam_{\bar{y}} y$. We find that the first order solutions are also of this form, but now with a corrected scalar wavefunction, namely
\be
\varphi_{\rho xy} = \vphi_{\rho xy}^{(0)} + \eps \vphi_{\rho xy}^{(1)} + \CO(\eps^2)
\label{chirexp}
\ee
where $\vphi_{\rho xy}^{(0)}$ are given by the scalar wavefunctions $\chi_\rho$ in eq.~(\ref{simchis}). In the following we report the results for the correction $\vphi_{\rho xy}^{(1)}$ in the different sectors, dropping the subscripts $xy$ to avoid cluttering the equations. While we will keep the worldvolume flux dependence in the zero mode equations, for simplicity we will set $M_{xy} = 0$. Our solutions are however easily generalized for non-vanishing $M_{xy}$.

\subsubsection*{Sector $a^+$}

Let us first define the complex variables
\be
u_a\, =\, x - \zeta_a y \quad \quad \quad \quad v_a\, =\, y + \zeta_a x
\ee
that come from a rescaling of $\lam_{\bar{x}} y + \lam_{\bar{y}} x$ and $\lam_{\bar{x}} x - \lam_{\bar{y}} y$ for this sector. To simplify notation we have suppressed the subindex $p=1,2$, for each sector $\rho = a_p^+$, that labels elements of the ${\bf \bar{5}}$ representation with different hypercharge, and therefore with different values of $\lam_{a}$ and $\zeta_a$. Notice that the variables $u_a$ and $v_a$ are different for each of these subsectors. 

The correction to the wavefunctions $\varphi_{a^+}^i$ can be conveniently written in terms of $(u_a, v_a)$. Concretely,
\be
\varphi_{a^+}^{i (1)} = m_*e^{-  q_\Phi \lam_a \bar{u}_a} \left( A_i(v_a) + \Upsilon_{a^+}^{i}\right)
\label{chia11}
\ee
where $q_\Phi(a^+) = - x$ for these sectors, $A_i$ is a holomorphic function of $v_a$ and 
\be
\Upsilon_{a^+}^{i}  =  \oh \th_{0y} \lam_a^2 \bar u_a^2  f_i(v_a) +  
\lam_a \bar{u}_a (\zeta_a \th_{0y} -  \th_{0x}) f_i^\prime(v_a)
+\left[\frac{\a_1}{2} x^2 + \a_2 x v_a\right]f_i(v_a)
\label{achinc}
\ee
The last two terms are in fact only necessary to fulfill the D-term equation (\ref{dsr}).
In particular, we obtain that the coefficients $\a_1$ and $\a_2$ must take the following values
\bea
\a_1 & = & -\frac{m^4}{\lam_a} 
\left\{\bar\th_{0x}\left(q_S^a - \zeta_a q_P^a \right) + \bar\th_{0y} \left(q_P^a + \zeta_a q_S^a \right) \right\}
\nonumber \\[2mm]
\a_2 & = & -\frac{m^4}{\lam_a}  
\left\{\bar\th_{0x} q_P^a -  \bar\th_{0y}q_S^a  \right\}
\label{A12def}
\eea

To sum up, taking into account the zeroth order solution in eq.(\ref{simchis}) the final result for 
$\varphi_{a^+}^i$ can be cast as
\be
 \varphi_{a^+}^i\, =\, m_*e^{- q_\Phi \lam_a \bar u_a }\left(f_i(v_a) + \eps A_i(v_a) 
+ \eps \Upsilon_{a^+}^{i}\right) + \CO(\eps^2)
\label{chia12}
\ee
Naively the functions $A_i$ remain unfixed by the above equations of motion. This is because we could think of them as a $\CO(\eps)$ correction to the holomorphic functions $f_i$, which are also not fixed. However, given the choice (\ref{simhol}) of $f_i$, the $A_i$ are fixed as follows. Notice that the F-term equation (\ref{ft03}) implies \mbox{$a_{\rho \bar m} = \partial_{\bar m} \xi_\rho$}, where $\xi_\rho$ is a regular function \cite{cchv09}. As shown in appendix \ref{ap:res}, this function can be found by integrating (\ref{ft04}). Imposing that $\xi_\rho$ is regular at the loci $q_\Phi(\rho)=0$ where the zero modes are localized implies nontrivial constraints for the wavefunctions. In particular, for the  $a^+$ sector requiring $\xi_{a^+}^i$ to be free of poles at $x=0$ fixes the $A_i$, which read
\be
A_2=A_3=0 \quad ; \quad
A_1 = \g_a^1 a_0 \quad ; \quad a_0=m_*^2\zeta_a (\zeta_a\th_{0y} - 2\th_{0x})
\label{fixa00}
\ee
Interestingly, this form of $A_1$ guarantees that the Yukawa couplings computed via overlap of zero modes
will be flux independent up to normalization factors, as we comment on the next section.

\subsubsection*{Sector $b^+$}

The sectors $b^+$ and $a^+$ are quite similar, so let us first define the variables
\be
u_b\, =\, y - \zeta_b x \quad \quad \quad \quad v_b\, =\, x + \zeta_b y
\ee
that again are different for each subsector $\rho = b_q^+$, $q=1,2,3$. As in (\ref{chia11}) we obtain 
\be
\chi_{b^+}^{j (1)} = m_*e^{q_\Phi \lam_b \bar u_b} \left(B_j(v_b) + \Upsilon_{b}^{j} \right)
\label{chib11}
\ee
where now $q_\Phi(b^+)=y$, $B_j$ is a holomorphic function of $v_b$ and
\bea
\Upsilon_{b}^{i} & = & \oh \th_{0x} \lam_b^2 \bar u_b^2  g_j(v_b) +  
\lam_b \bar{u}_b (\zeta_b \th_{0x} -  \th_{0y}) g_j^\prime(v_b)
+\left[\frac{\b_1}{2} y^2 + \b_2 y v_b\right]g_j(v_b)
\label{bchinc}
\eea
Again, the terms proportional to $\b_1$, $\b_2$ are required to satisfy the D-term equation. 
These coefficients can be obtained from (\ref{A12def}) by replacing $\bar\th_{0x} \leftrightarrow \bar\th_{0y}$,
$q_P^a \to -q_P^b$, $q_S^a \to q_S^b$, and consistently $\lam_a \to \lam_b$, $\zeta_a \to \zeta_b$.  

Including the zeroth order solution from eq.(\ref{simchis}) yields the full result
\be
\vphi_{b^+}^j = m_*e^{q_\Phi \lam_b \bar u_b}\left(g_j(v_b) + \eps B_j(v_b) + \eps \Upsilon_{b}^{j}\right) + \CO(\eps^2)
\label{chib12}
\ee
The functions $B_i$ are determined demanding that $\xi_{b^+}^j$ be free of singularities as explained
in appendix \ref{ap:res}. In this way we find
\be
B_2=B_3=0 \quad ; \quad
B_1 = \g_b^1 b_0 \quad ; \quad b_0=m_*^2\zeta_b (\zeta_b\th_{0x} - 2\th_{0y})
\label{fixb00}
\ee

\subsubsection*{Sector $c^+$}

In this case it is helpful to introduce the variables
\be
u_c\, =\, x - \tau_c y \quad \quad \quad \quad v_c\, =\, y + \tau_c x
\ee
with $\tau_c=(\lam_c-\zeta_c)/\zeta_c$. The quantities $\lam_c$ and $\zeta_c$, defined in appendix \ref{ap:wave},
actually depend on the subsector $c_r^+$, $r=1,2$. The correction to the wavefunction is found to be 
\be 
\varphi_{c^+}^{(1)} = m_*\g_c e^{q_\Phi \zeta_c \bar u_c} \left(C(v_c) + \Upsilon_{c^+}^{(1)}\right)
\label{chic11}
\ee
where now $q_\Phi=(x-y)$, $C$ is a function of $v_c$, and
\be
\Upsilon_{c^+}^{(1)} =
-\frac{1}{2}\zeta_c^2 {\bar u_c}^2 (\th_{0x} + \th_{0y}) 
+ \frac{\d_1}{2} (x-y)^2 + \d_2 (x-y)v_c   
\label{cchinc}
\ee
The constants $\d_1$ and $\d_2$ are given by
\bea
\d_1 & = & \frac{m^4}{\zeta_c(1+\tau_c)^2} 
\left\{\bar\th_{0x}\left(q_S^c - \tau_c q_P^c \right) + \bar\th_{0y} \left(q_P^c + \tau_c q_S^c \right) \right\}
\nonumber \\[2mm]
\d_2 & = & \frac{m^4}{\zeta_c(1+\tau_c)^2}  
\left\{\bar\th_{0x}\left(q_P^c + q_S^c \right) + \bar\th_{0y} \left(q_P^c - q_S^c \right) \right\}
\label{C12def}
\eea
The holomorphic terms in $\Upsilon_{c^+}^{(1)}$, which depend on $\bar\th_{0x}$ and $\bar\th_{0y}$ through $\d_1$ and $\d_2$,
are needed to satisfy the corrected D-term equation. In appendix \ref{ap:res} we show that $C=0$.

\subsection{Normalization and mixings of corrected zero modes}
\label{sec:mix}

Given the corrected zero mode wavefunction one must compute their normalization factors, since it is through these factors that the physical Yukawa couplings depend on worldvolume fluxes. The computation of normalization factors for perturbative zero modes is carried out in appendix \ref{ap:wave}, where the following norms are computed explicitly
\bea
K_\rho^{ij}\, =\, \langle \vec{\psi}_{\rho \, i}^{\rm real} | \vec{\psi}_{\rho \, j}^{\rm real} \rangle  & = &
  m_*^2 \int_S \tr \,( \vec{\psi}_{\rho \, i}^{\rm real} \cdot \vec{\psi}_{\rho\, j}^{\, \dag\, {\rm real}})\, {\rm d vol}_S \, = \,  2 \, ||\vec{v}_{\rho}||^2  X_\rho^{ij}
 \eea
 with $\vec{v}_\rho$ defined in (\ref{ap:vec}), and the scalar wavefunction metrics
 \bea
X_\rho^{ij}& =& m_*^2 \left (\chi_{\rho}^{i\,  \text{real}}, \chi_{\rho}^{j \, \text{real}} \right)\, = \,
m_*^2 \int_{S}\, \left(\chi_{\rho}^{i \,  \text{real}}\right)^* \chi_{\rho}^{j\,  \text{real}} \, d{\text{vol}}_S
\label{chimix}
\eea
are calculated by extending the local patch to $\mathbb C^2$. Here the superscript `real' stands for the zero mode wavefunction expressed in a real gauge rather than in the holomorphic gauge that we have used in the previous sections. Following \cite{fi1} one may switch from wavefunctions in the holomorphic to the real gauge by multiplying them by an appropriate sector dependent prefactor. For instance, in the sector $a^+$ we have
\be
\chi_{a^+}^{i\, \text{real}}\,=\, e^{\frac{q_P^a}{2}(|x|^2 -|y|^2) - q_S^a\, {\rm Re } (x\bar{y})}
\chi_{a^+}^i
\ee
where $\chi_{a^+}^i$ is the zero mode computed in the holomorphic gauge. As we are now dealing with non-perturbative zero modes, we should take $\chi_{a^+}^i$ to be the corrected scalar wavefunction $\varphi_{a^+}^i$ given in (\ref{chia12}).

For the perturbative zero modes analyzed in appendix \ref{ap:wave}, $X_\rho^{ij} =0$ for $i\not=j$, since the integrand needs to be invariant under the $U(1)$ diagonal rotation $(x,y) \raw e^{i\alpha} (x,y)$. However, this no longer needs to be true at order $\eps$. Indeed, let us consider the sector $a^+$ and the corrected zero modes in this sector when only $\theta_0$ is present. Due to the correction $\Upsilon_{a^+}^{i}$ given in (\ref{achinc}), one can in principle have  non-diagonal metrics. In fact, to order $\eps$ one finds that only $X_{a^+}^{31}$ and its conjugate are different from zero.

Substituting the corrected zero modes and computing the Gaussian integrals we find
\be
K_{a^+}^{ij} \,=\, \frac{2\pi^2 m_*^4}{\Delta_a q_P^a} ||\vec{v}_{a}||^2 \g_a^{i *} \g_a^j \,{\bf x}^{ij}_{a} + \CO(\eps^2)
\ee
where $\Delta_a=-(2\lam_a +  q_P^a (1 + \zeta_a^2))$,  $||\vec{v}_{a}||$ is given by (\ref{norma}) and the matrix ${\bf x}_a$ is 
\be
{\bf x}_{a} =  
\left(\!
\begin{array}{ccc}
2\frac{m_*^4}{(q_P^a)^2} & 0 & \eps \left[  m_*^2 (2r_a \bar{\theta}_{0x} + r_a^2 \bar{\theta}_{0y})  \right]   \\
0 & \frac{m_*^2}{q_P^a} & 0\\
\eps \left[ m_*^2 (2r_a \theta_{0x} + r_a^2 \theta_{0y}) \right]  & 0 & 1
\end{array}
\! \right)
\label{matx}
\ee
\be
r_a\, =\, - \frac{q_S^a}{q_P^a}
\label{defaxy}
\ee
The quantity $r_a$ is the quotient between the off-diagonal and diagonal worldvolume fluxes felt by this sector. Hence, when off-diagonal fluxes are turned on $K_{a^+}^{31}$ is non-zero. Finally, notice that the diagonal terms $K_{a^+}^{ii}$ do not get corrections to order $\eps$. 

As the metric for the zero-modes is non-trivial, the Yukawas computed from them do not yet correspond to the physical couplings. From the 4d effective theory viewpoint, to obtain physical Yukawas one performs a field redefinition such that the 4d chiral fields have canonical kinetic terms. The higher dimensional counterpart of this field redefinition is to take a linear combination of zero modes such that the matrix $K_{a^+}$ becomes the identity. In the absence of non-perturbative effects $\eps=0$ and such redefinitions are rather easy to perform, since $K_{a^+}$ is diagonal and one only needs to choose the normalization factors as in (\ref{normga}) to have $K_{a^+} = \uno_3$. The same applies to order $\CO(\eps)$ whenever $q_S^a=0$. 

In general we will have 4d fields $\Phi^m_\rho$ whose metric $K_\rho$ has off-diagonal terms, and so a wavefunction rescaling is not enough to have canonically normalized fields. However, as the field metric $K_\rho$ must be Hermitian and definite positive, we can always write it as 
\be
K_\rho\, =\, P_\rho^\dag P_\rho
\label{KPP}
\ee
and describe the fields with canonical kinetic terms and the physical Yukawa couplings as
\be
{\Phi}^i_{{\rm phys}}\, =\, (P_\rho)^{i}_{\, m} {\Phi}^m \quad \quad Y_{ijk}^{\rm phys} \, =\, (P_{\rho}^{-1})_i^{\, m}  (P_{\rho'}^{-1})_j^{\, n}  (P^{-1}_{\rho''})_k^{\, p}\, Y_{mnp}
\label{changeP}
\ee
where $Y_{mnp}$ stand for the initial set of Yukawa couplings. Finally, the set of internal wavefunctions ${\psi}_m$ associated to these fields will transform under this change of basis as
\be
{\psi}_i^{\rm phys}\, =\, (P^{-1\, t})_i^{\, m}  {\psi}_m
\ee

In the case of $K_{a^+}$ one can easily find a matrix $P_{a^+}$ such that (\ref{KPP}) is satisfied
\be
P_{a^+}\,=\, 
\frac{\sqrt{2}\pi m_*^2  ||\vec{v}_a||}{\sqrt{\Delta_a q_P^a}}\, 
\left(\!
\begin{array}{ccc}
\sqrt{2} \frac{m_*^2}{q_P^a} && \eps  \bar{\mu}_a \\ & \frac{m_*}{(q_P^a)^{1/2}} \\ & & 1
\end{array}
\! \right)
\left(\!
\begin{array}{ccc}
\g_a^1 \\ & \g_a^2 \\ & & \g_a^3
\end{array}
\! \right)
+ \CO(\eps^2)
\label{totalP}
\ee
\be
\mu_a\, =\, \frac{q_P^a}{\sqrt{2}} ( 2 r_a \theta_{0x} + r_a^2 \theta_{0y})
\ee
Upon choosing the normalization factors $\g_a^i$ as in (\ref{normga}) this matrix simplifies to
\be
P_{a^+}\, \xrightarrow{\g_a^i = (\ref{normga})} 
\left(\!
\begin{array}{ccc}
1 && \eps \bar{\mu}_a \\ & 1 \\ & & 1
\end{array}
\! \right)
+ \CO(\eps^2)
\label{newP}
\ee
Hence, in order to have canonically normalized fields we not only need to choose appropriate normalization factors, but also perform a rotation among the families of this sector. One can express such rotation in terms of the holomorphic representatives $f_i$ that describe each of our families as
\be
\tilde{f}_1\, =\, f_1 \quad \quad \tilde{f}_2\, =\, f_2 \quad \quad \tilde{f}_3\, =\, f_3 - \eps \bar{\mu}_a f_1
\label{changea}
\ee
$\tilde{f}_i$ being the holomorphic representatives that describe canonically normalized 4d fields. 

The analysis of the metrics in the $b^+$ sector proceeds along the same lines. Again considering the case where only
$\theta_0 \neq 0$, the non-zero off-diagonal entries of the mixing matrix amount to $X_{b^+}^{31}$ and its conjugate $X_{b^+}^{13}$. Similarly to the sector $a^+$ there are no $\CO(\eps)$ corrections to the diagonal terms $X_{b^+}^{ii}$, and obtaining canonically normalized fields amounts to appropriately choosing the normalization factors $\g_b^j$ and performing a redefinition of chiral representatives. More precisely, we must choose $\g_b^j$ as in (\ref{normgb}) and perform the redefinition
\be
\tilde{g}_1\, =\, g_1 \quad \quad \tilde{g}_2\, =\, g_2 \quad \quad \tilde{g}_3\, =\, g_3 - \eps \bar{\mu}_b g_1
\ee
where now
\be
\mu_b= \frac{q_P^b}{\sqrt{2}} (2r_b \theta_{0y} + r_b^2 \theta_{0x}) \quad  \quad r_b = \frac{q_S^b}{q_P^b}
\label{changeb}
\ee

To summarize, in the presence of non-diagonal fluxes of the form (\ref{fluxcso12}) non-trivial corrections appear at 
first order in $\eps$ for the wavefunction metrics. The corrections appear in the off-diagonal metrics $K_\rho^{ij}$, $i\not=j$ 
that vanish at zeroth order in $\eps$. One can perform a change of basis of the families of zero modes in order to
set the off-diagonal entries of $K_\rho$ to vanish, and choose the appropriate wavefunction factors $\g_\rho^i$ to have canonically normalized 4d kinetic terms. Notice that such factors are the ones found in appendix \ref{ap:wave}, namely the normalization factors at the perturbative level, and that the only effect of $\CO(\eps)$ corrections to the field metrics is encoded in the redefinitions (\ref{changea}) and (\ref{changeb}). These redefinitions have however no effect for the physical Yukawa couplings, at least at the level of approximation that we are working. 

Indeed, the above redefinitions are only nontrivial in the cases $f_3 \raw \tilde{f}_3$ and $g_3 \raw \tilde{g}_3$, and amount to say that in this new basis the third families of both sectors $a^+$ and $b^+$ have a contamination of $\CO(\eps)$ from the first family. In principle this contamination will modify the Yukawa couplings $Y_{3j}$ and $Y_{i3}$. However, as the Yukawa couplings involving the first family are already of order $\epsilon$, the modification will be $\CO(\eps^2)$. The same result is obtained by directly applying (\ref{changeP}) to the Yukawa couplings computed in the next section. 

We then find that, although non-perturbative effects modify the metrics for the 4d matter fields, this modification can be neglected at the level of approximation that we are working, at least for the purpose of computing fermion masses. The whole effect of wavefunction normalization is already captured by tree-level wavefunctions, and more precisely by the computations carried out in appendix \ref{ap:wave}. This fact is quite relevant in the present scheme, because holomorphic Yukawas do not depend on worldvolume fluxes. Hence, the only place where hypercharge flux will enter into the expression for the physical Yukawa couplings will be via the normalization factors of perturbative zero modes.  

While the above results have been obtained for the superpotential (\ref{supo}) with only $\theta_0 \neq 0$, they hold more generally. Indeed, one can check that a similar result is obtained for the $U(3)$ model of \cite{afim}, that corresponds to the case $n=1$, and one expects the same for the case $n=2$. That is, one expects $\CO(\eps)$ corrections to the off-diagonal metric elements, which can nevertheless be absorbed into a redefinition of the third family and so do not affect the expressions for the physical Yukawa couplings. In particular, the physical Yukawas are expected to depend only on the $\CO(\eps^0)$ normalization factors $\gamma_\rho^i$ computed in appendix \ref{ap:wave}, where the full hypercharge flux dependence will be contained.

\subsection{Zero modes for $n=2$}
\label{ss:n2}

Let us now turn to compute the corrections $\vec{\psi}_{\rho,\th_2}^{(1)}$ defined in (\ref{npzmexp}). To this end we set $\theta_0 = 0$, while for simplicity we take $\theta_2 \not= 0$ to be constant. Note that the F-term equations given in (\ref{FtermA02}) and (\ref{Ftermphi02}) are rather unwieldy when fluxes are present, but simplify considerably when fluxes are turned off. On the other hand, just like in \cite{afim} and in the case $n=0$ above, the holomorphic Yukawa couplings for the case $n=2$ are expected to be independent of worldvolume fluxes. We show that this is the case  in appendix  \ref{ap:res}, by means of the Seiberg-Witten map of footnote \ref{SWfoot}. Indeed, it is easy to see that the SW map (\ref{ansatz}) defines certain variables $\hat{a}_{\bar{m}}$ and $\hat{\vphi}_{xy}$ for the 7-brane fluctuations. When we express the F-term zero mode equations and the superpotential trilinear couplings in terms of such hatted variables, we find that they are flux-independent.\footnote{More precisely, the F-term equations and trilinear couplings for ($\hat{a}_{\bar{m}}$, $\hat{\vphi}_{xy}$) are those for $({a}_{\bar{m}}$, ${\vphi}_{xy}$) after setting $\langle F \rangle =0$.} We can then compute the holomorphic Yukawa couplings via a residue formula as in \cite{cchv09}, showing explicitly their flux independence. 

Since we know that holomorphic Yukawas are flux independent, in order to compute them we may solve for such wavefunctions  by considering the zero mode equations in the absence of worldvolume fluxes. Of course, we are ultimately interested in computing the physical Yukawa couplings, which do depend on fluxes. However, as we have argued in the last subsection, this flux dependence is expected to arise {\em only} via the normalization factors for the tree-level zero modes. Hence, one can simplify the computation of Yukawas by switching off worldvolume fluxes when computing corrected wavefunctions and their overlap integrals. The flux dependence is taken into account by including the normalization factors $\g_\rho^i$ calculated in appendix \ref{ap:wave}. This last step will be carried in the next section, where the Yukawa couplings of the $SO(12)$ model will be computed.

To proceed let us consider the zero mode equations corrected up to $\CO(\eps^2)$ and in the absence of worldvolume fluxes. Using (\ref{s3b}) one finds that the F-term (\ref{FtermA02}) reduces to
\be
\bar\p_{\bar m} \varphi_{\rho xy} + i m^2 q_\Phi(\rho) a_{\rho \bar m} + 
2\eps \theta_2 m^4 \left[t_x(\rho) \partial_y a_{\rho \bar m} 
- t_y(\rho)\partial_x a_{\rho \bar m} \right]= 0
\label{fnpsr}
\ee
where we have defined the quantities
\be
t_m(\rho) = x s_{mx}(\rho) + y s_{my}(\rho) \quad ; \quad m=x,y
\label{tmrho}
\ee
These $t_m(\rho)$ can be regarded as the components of a 1-form. The values of \mbox{$s_{mn}=s_{nm}$} given in table \ref{t1}
imply that $\p t(\rho)=0$.
{}From (\ref{Ftermphi02}) we simply obtain \mbox{$\p_{\bar{x}} a_{\rho\bar{y}} = \p_{\bar{y}} a_{\rho\bar{x}}$}.
The D-term equation is given by
\be
\partial_x a_{\rho \bar x}   +  \partial_y a_{\rho \bar y} - i m^2 \bar{q}_\Phi(\rho) \varphi_{\rho xy} = 0
\label{dsrnf}
\ee
In the following we present the solutions in the relevant sectors, again dropping subscripts $xy$ in $\varphi_{\rho xy}$.

\subsubsection*{Sector $a^+$}
\label{ss:sa}

We obtain a localized solution with $a_{a^+\bar y}=0$ and
\be
a_{a^+\bar x}=  -\frac{i\lam_a m_*}{m^2} e^{\lam_a  |x|^2} H_a(y,x,\bar x) \quad ; \quad
\varphi_{a^+}= m_* e^{\lam_a  |x|^2} \left(H_a + \frac{\lam_a}{m^4 \bar x} \p_x H_a\right)
\label{ansA}
\ee 
The constant $\lam_a$ is such that the D-term equation (\ref{dsrnf}) is satisfied for any $H_a$. 
We further demand that $\varphi_{a^+} \to m_* f_i(y)$ when $x \to 0$. Substituting into the F-term (\ref{fnpsr})
and solving to first order in $\eps$ we obtain
\beqa
H_a & \!\!\!\! = \!\!\!\! & f_i(y) + 2\eps\th_2\left(\bar x \a -
\frac{i\lam_a^2 m^2}{3(\lam_a^2 + m^4)}f_i^\prime(y)\right)  \label{hasol}\\[3mm]
\a & \!\!\!\! = \!\!\!\! & -\frac{i\lam_a^2 m^2}{4(\lam_a^2 + 2m^4)}
(\lam_a - m^4 x\bar x )f_i(y) -\frac{i\lam_a^2 m^2}{8} \bar x y f_i(y)
-\frac{i\lam_a m^2}{12}\left(\! 3y - \frac{4m^4}{(\lam_a^2+ m^4)}x \! \right)f_i^\prime(y) 
\nonumber
\eeqa
The F-term equation is satisfied to $\co(\eps)$ for any $\lam_a$. Hence, we expect the $\lam_a$ dependence to drop out completely in the computation of holomorphic Yukawa couplings.

\subsubsection*{Sector $b^+$}
\label{ss:sb}

There are localized zero modes with $a_{b^+\bar x}=0$ and
\be
a_{b^+\bar y}=  \frac{i\lam_b m_*}{m^2} e^{\lam_b  |y|^2} H_b(x,y,\bar y) \quad ; \quad
\varphi_{b^+}=  m_* e^{\lam_b  |y|^2} \left(H_b + \frac{\lam_b}{m^4 \bar y} \p_y H_b\right)
\label{ansB}
\ee 
Imposing that the D-term equation (\ref{dsrnf}) is satisfied for any $H_b$ determines $\lam_b$.
It is also required that $\varphi_{b^+} \to m_* g_j(x)$ when $y \to 0$. Inserting into the F-term (\ref{fnpsr})
and solving to first order in $\eps$ leads to
\beqa
H_b & \!\! = \!\! & g_j(x) + 2\eps\th_2\left(\bar y \b -
\frac{i\lam_b^2 m^2}{12(\lam_b^2 + m^4)}g_j^\prime(x)\right)  \label{hbsol}\\[3mm]
\b & \!\! = \!\! & \frac{i\lam_b m^6}{12(\lam_b^2+ m^4)} y g_j^\prime(x) 
\nonumber
\eeqa
The F-term equation is verified for generic $\lam_b$.

\subsubsection*{Sector $c^+$}
\label{ss:sc}

The wavefunctions in this sector have the form
\beqa
a_{c^+\bar x} & = & -a_{c^+\bar y}=  \frac{i\lam_c m_*}{m^2} e^{\lam_c  |x-y|^2} H_c(x,\bar x, y,\bar y) \label{ansC} \\[3mm]
\varphi_{c^+} & = &  m_* e^{\lam_c  |x-y|^2} \! \left(\! H_c + \frac{\lam_c}{m^4}\frac{\p_x H_c - \p_y H_c}{\bar x-\bar y} \! \right)
\nonumber
\eeqa 
Requiring that the D-term (\ref{dsrnf}) is satisfied for any $H_c$ fixes $\lam_c$. When $x \to y$
it must be that $\varphi_{c^+} \to m_* \g_c$. {} Solving the F-term (\ref{fnpsr})
to first order in $\eps$ we find
\beqa
H_c & \!\! = \!\! & \gamma_c\left[1 + 2\eps\th_2 (\bar x - \bar y) \nu \right]
\label{hcsol}\\[3mm]
\nu & \!\! = \!\! & -\frac{i\lam_c^3 m^2}{8(\lam_c^2+ m^4)} + 
\frac{i\lam_c m^2}{16(\lam_c^2+ m^4)}(\bar x - \bar y) \left[(\lam_c^2 + 2m^4) x + \lam_c^2 y\right] 
\nonumber
\eeqa
The F-term equation leaves $\lam_c$ free.


\section{Yukawa couplings}
\label{s:yc}

Given the non-perturbative zero mode wavefunctions one may proceed to compute the non-perturbative corrections to the Yukawa couplings. For this one must consider the cubic terms in fluctuations that arise from the full superpotential (\ref{supo}), insert the values for the background fields and zero modes and, finally, compute the overlapping integrals that contribute up to $\CO(\eps)$. 

As before, we consider the case where, from all the possible holomorphic functions $\theta_n$ present in (\ref{supo}), only $\theta_0$ and $\theta_2$ are non-vanishing. It is then clear that the cubic terms arising from this superpotential have the structure
\be
W^{\rm cubic} = W_0^{\rm cubic} + W_1^{\rm cubic}
\label{wtotex}
\ee
where $W_0^{\rm cubic}$ arises from the tree-level term (\ref{supo7}) and $W_1^{\rm cubic}$ from the sum  of the two terms in (\ref{supo}) proportional to $\theta_0$ and $\theta_2$.

Let us analyze these cubic terms in some detail. We have that the tree-level coupling reads
\be
W_0^{\rm cubic}= - i m_*^4 \int_S \tr \, \left(\varphi \wedge a \wedge a \right)
\label{w03}
\ee
and leads to the expression (\ref{yukawa7}). If we just insert the tree-level wavefunctions $\vec{\psi}_\rho^{(0)}$ into (\ref{yukawa7}) we recover the tree-level Yukawa couplings that we have already computed in section \ref{sec:treeYukawas}, whose single non-zero coupling is given by (\ref{y33pert}). If we instead insert the corrected wavefunctions (\ref{npzmexp}) there will be in addition two sets of $\CO(\eps)$ integrals, that will contain two tree-level zero modes and one $\CO(\eps)$ correction $\vec{\psi}_{\rho, \th_n}^{(1)}$, with $n=0,2$. The Yukawa structure that arises from (\ref{w03}) is then
\be
Y_0\, =\, Y_0^{(0)} + Y_{0,0}^{(1)} + Y_{0,2}^{(1)} + \CO(\eps^2)
\ee
where $Y_0^{(0)}$ is the tree-level Yukawa and $Y_{0,n}^{(1)}$ an $\CO(\eps)$ correction linear in $\vec{\psi}_{\rho, \th_n}^{(1)}$.

The corrected couplings induced by $W_1$ depend explicitly on the background and read
\be
W_1^{\rm cubic} =  m_*^4\, \eps \! \int_S \theta_2 \,  \str \left( \langle \Phi_{xy} \rangle \varphi_{xy} \, Da \wedge Da 
+ \varphi_{xy}^2 \, Da \wedge \langle F \rangle  \right
)
\label{w1t}
\ee
where $D_ma = \p_m a - i[\langle A_m \rangle, a]$. The non-perturbative piece depending on $\theta_0$ does not yield a term trilinear in fluctuations, as explained in detail in appendix \ref{ap:res}. The Yukawa structure that arises from this term is simply
\be
Y_1\, =\, Y_{1}^{(0)} + \CO(\eps^2)
\ee
where $(0)$ indicates that we must only insert tree-level wavefunctions into (\ref{w1t}). 

In order to compute the couplings arising from $W_1^{\rm cubic}$ it is useful to derive a more explicit expression for the integral (\ref{w1t}), by applying some specific features of the $SO(12)$ model. Recall that quarks and leptons arise from the sectors $a^+_p$ and $b^+_q$ and come in families indexed by $i$ and $j$ respectively. They couple to a Higgs in the sector $c^+$ when we insert the zero modes $\vec{\psi}_{\! \rho} E_\rho$, $\rho=a_p^+,b_q^+,c^+$ in the symmetrized traces of the superpotential (\ref{supo}). To simplify the notation in this section we drop 
the subscripts that identify the subsectors with different hypercharge.

Let us now consider an specific 7-brane background. Namely, we will consider $\langle \Phi_{xy} \rangle$ given in (\ref{pvev}) and a flux of the form (\ref{f1vev}), which we will write as $\langle F \rangle = \mff\,  Q_F$. It is then useful to define the quantities
\beqa
\theta_{a^+b^+c^+} & = & \theta_2 \str\left(  \langle \Phi_{xy} \rangle E_{a^+} E_{b^+} E_{c^+} \right) = 
\theta_2 m^2 (x d_{xa^+b^+c^+} + y d_{ya^+b^+c^+})
\label{tabc} \\
\kappa_{a^+b^+c^+} & = & \theta_2 \str\left(Q_F E_{a^+} E_{b^+} E_{c^+} \right) = 
-\theta_2 (d_{xa^+b^+c^+} + d_{ya^+b^+c^+}) 
\label{kabc} 
\eeqa
where the $d_{ma^+b^+c^+}$ are the traces
\be
d_{ma^+b^+c^+} = \str\left(Q_m  E_{a^+} E_{b^+} E_{c^+} \right) \quad ; \quad m=x,y
\label{dmabc}
\ee
with the generators in the fundamental representation. The correction due to $W_1^{\rm cubic}$ is then found to be
\beqa
\left(Y_1\right)_{a^+b^+c^+}^{ij} \!\! & = & \!\!\! 2 m_* \,\eps \! 
\int_{S} \theta_{a^+b^+c^+} \, Da_{a^+}^i \wedge Da_{b^+}^j  \varphi_{c^+ xy} \ + \ 
\kappa_{a^+b^+c^+} \,\varphi_{a^+ xy}^i  \varphi_{b^+ xy}^j  Da_{c^+} \wedge \mff \nonumber \\
& + & \ {\rm cyclic \ permutations \ in \ } a^+, \, b^+, \, c^+
\label{yuk1}
\eeqa 
It is straightforward to generalize this formula for a more general flux of the form (\ref{totalflux}). This will however not be necessary for our purposes, since in section \ref{ss:y2} we will compute the corrected couplings for the case $\langle F \rangle = 0$ which already captures the correction to the holomorphic Yukawa couplings. In appendix \ref{ap:res} we will explain how the formula is applied when fluxes are present.

To summarize, we have that the corrected Yukawa couplings can be expressed as
\be
Y_{a^+b^+c^+}^{ij} = \left(Y_0^{(0)}\right)_{\! a^+b^+c^+}^{ij} + \left(Y_{0,0}^{(1)}\right)_{\! a^+b^+c^+}^{ij} 
+ \left[\left(Y_{0,2}^{(1)}\right)_{\! a^+b^+c^+}^{ij}  + \left(Y_1^{(0)}\right)_{\! a^+b^+c^+}^{ij}\right] +  \co(\eps^2)
\label{yexp} 
\ee
where both $Y_0^{(0)}$ and $Y_{0,n}^{(1)}$ come from (\ref{yukawa7}) after inserting 
the corrected wavefunctions (\ref{npzmexp}), and respectively extracting the zeroth and first order contributions in $\eps$. Similarly, $Y_1^{(0)}$ is calculated substituting the uncorrected wavefunctions in (\ref{yuk1}). Notice that $Y_{0,0}^{(1)}$ represent the $\CO(\eps)$ corrections to the Yukawas due to $\theta_0$, that is the corrections that we would obtain if $\theta_2$ vanished, and which we compute in section \ref{ss:y0}. Similarly, the sum in brackets corresponds to the corrections due to $\theta_2$, which are computed in section \ref{ss:y2}. Finally we will consider that $\theta_0$ is lineal in $x$ and $y$ and that $\theta_2$ is constant, as this is enough to generate all the possible  corrections to the Yukawa couplings below $\CO(\eps^2)$. With these assumptions the  coupling $Y_{a^+b^+c^+}^{33}$ remains uncorrected to $\CO(\eps^2)$ and it is given by (\ref{y33pert}).

\subsection{Couplings due to $\theta_0$}
\label{ss:y0}

The corrected Yukawas implied by $\theta_0$ are easy to compute by substituting the zero modes of section \ref{ss:n0} in (\ref{yukawa7}) and evaluating the integrals. The only non-vanishing couplings turn out to be
\begin{subequations}
\label{th0resu}
\begin{align}
Y_{a^+b^+c^+}^{22} \, = \ & \frac{ \eps \pi^2  m_*^6}{m^4} 
\, f_{a^+b^+c^+} \g_a^2 \g_b^2 \g_c (\theta_{0x}+\theta_{0y}) \label{y22}\\[2mm]
Y_{a^+b^+c^+}^{31}  \, =\ & \frac{\eps \pi^2 m_*^6}{ m^4} \, f_{a^+b^+c^+} \g_a^3 \g_b^1 \g_c \theta_{0y} 
\label{y31} \\[2mm]
Y_{a^+b^+c^+}^{13}  \, =\ & \frac{\eps \pi^2 m_*^6}{ m^4} \, f_{a^+b^+c^+} \g_a^1 \g_b^3 \g_c \theta_{0x}  
\label{y13}
\end{align}
\end{subequations}
To obtain these couplings we have used the functions $A_i$ and $B_j$ given in (\ref{fixa00}) and (\ref{fixb00}) respectively.
In particular, the constants $a_0$ and $b_0$ are such that the couplings
$Y^{13}$ and $Y^{31}$ turn out to be flux independent, up to normalization factors. For instance, before
substituting the value of $a_0$ we find that
\be
Y_{a^+b^+c^+}^{13}  =  \frac{\eps \pi^2 m_*^4}{ m^4} \, f_{a^+b^+c^+} \g_a^1 \g_b^3 \g_c
\left[(1+2\zeta_a)  m_*^2 \theta_{0x}- \zeta_a^2 m_*^2 \theta_{0y} + a_0 \right]  
\label{y13full}
\ee 
We see that taking $a_0 =m_*^2(\zeta_a^2 \th_{0y} -2 \zeta_a  \th_{0x})$ indeed leads to (\ref{y13}). 
In appendix \ref{ap:res} we show how the couplings (\ref{th0resu}) can also be deduced from a residue formula.

To conclude, we have derived a set of Yukawa couplings that are flux independent up to the $\CO(\eps^0)$ normalization factors 
$\g_\rho^i$, similarly to the structure obtained in \cite{afim}.

\subsection{Couplings due to $\theta_2$}
\label{ss:y2}

In this case we need to substitute the wavefunctions of section \ref{ss:n2} in (\ref{yukawa7}), and also in (\ref{yuk1}). Since we have computed the zero modes with fluxes switched off, the corrections to the Yukawa couplings are simply given by\footnote{A parameter $\theta_1$ in the non-perturbative superpotential (\ref{supo}) gives rise to a similar coupling but with $\theta_{a^+b^+c^+}$ replaced by $\theta_1 \str \left(E_{a^+} E_{b^+} E_{c^+}\right)$, which vanishes for $SO(12)$ \cite{afim}.}
\beqa
\left(Y_1\right)_{a^+b^+c^+}^{ij} \!\! & = & \!\!\! -2 m_* \,\eps \! 
\int_{S} \!  {\rm d vol}_S \, \theta_{a^+b^+c^+} 
\big(\p_x a_{a^+\bar x}^i \,\p_y a_{b^+ \bar y}^j 
-\p_y a_{a^+\bar x}^i \,\p_x a_{b^+ \bar y}^j 
 -  \p_x a_{a^+\bar y}^i \,\p_y a_{b^+ \bar x}^j 
\nonumber\\[2mm]  
& + & \!\!\!  \p_y a_{a^+\bar y}^i \,\p_x a_{b^+ \bar x}^j \big) \, \varphi_{c^+ xy} \ 
 + \ {\rm cyclic \ permutations \ in \ } a^+, \, b^+, \, c^+
\label{yuk1nf}
\eeqa 

The final outcome for the couplings follows by adding the different contributions that depend on $\theta_2$, as shown in (\ref{yexp}). In the $SO(12)$ model that we are discussing such sum simplifies by virtue of the properties
\be
d_{x a^+b^+c^+} = -\frac{i}6 f_{a^+b^+c^+} \quad ; \quad
d_{y a^+b^+c^+} = \frac{i}{12} f_{a^+b^+c^+}
\label{dres}
\ee
Evaluating the various integrals involving the explicit wavefunctions gives the non-zero couplings
\begin{subequations}
\label{th2resu}
\begin{align}
Y_{a^+b^+c^+}^{23} \, =\ & \phantom{-}\frac{\eps \pi^2 m_*^5}{3 m^2} 
\, f_{a^+b^+c^+} \g_a^2 \g_b^3 \g_c \theta_{20} 
\label{y23} \\[2mm]
Y_{a^+b^+c^+}^{32}  \, =\ & -\frac{\eps \pi^2 m_*^5}{6 m^2} \, f_{a^+b^+c^+} \g_a^3 \g_b^2 \g_c \theta_{20}  
\label{y32}
\end{align}
\end{subequations}
where $\theta_{20}=-i\theta_2$. Notice that these couplings are completely independent of the parameters $\lambda_a$,
$\lambda_b$, and $\lambda_c$, which are not determined by the F-terms.

We expect that with fluxes turned on the couplings will have the same structure (\ref{th2resu}), namely a flux independent 
core multiplied by  flux-dependent normalization factors $\g_\rho^i$.
Had we computed the wavefunctions and Yukawa couplings for $\langle F \rangle \neq 0$, we would have probably arrived to an 
expression similar to (\ref{y13full}) in which the flux dependence in the core actually drops out after some parameters take 
their prescribed values.  
To sustain this assumption, in appendix \ref{ap:res} we compute the holomorphic couplings in presence of fluxes. 
To this purpose we will derive and apply a residue formula which is explicitly independent of fluxes. 
The couplings determined in this way completely agree with the results in (\ref{th2resu}).

\section{Quark-lepton mass hierarchies and hypercharge flux}
\label{s:pheno}

Gathering the results of eqs.(\ref{y33pert}),(6.11),(6.15) one obtains that the physical Yukawa mass matrix has a structure of the form
\be
Y_{D/L} \  = \ 
2\pi^2 \varrho^{-2} \g_c\left(
\begin{array}{ccc}
\CO(\e^2)   &   \CO(\e^2)   &   \eps  \varrho^{-1} { \g_{a}^{1}}{\g_{b}^{3}} {\tilde \th}_{0x}\\
\CO(\e^2)   &  \eps \varrho^{-1}  { \g_{a}^{2} \g_{b}^{2} }  ({\tilde \theta}_{0x}+{\tilde \theta}_{0y})
&  {\eps}  \varrho \g_{a}^{2}  \g_b^3    {\tilde {\theta}}_{20} \\
\eps \varrho^{-1} { \g_{b}^{1}}{\g_{a}^{3}}{\tilde \th}_{0y}
& - \frac {\eps}{2} \varrho   \g_{b}^{2} \g_{a}^{3}  {\tilde \theta}_{20}  &  \g_a^3\g_b^3
\end{array}
\right) 
\label{yukasmm}
\ee
where the coefficients 
\beq
{\tilde \theta}_{0x}=m^2\theta_{0x} \quad \quad
{\tilde \theta}_{0y}=m^2\theta_{0y} \quad \quad 
{\tilde \theta}_{20}=\frac {1}{3}m_*^3\theta_{20} 
\eeq
are adimensional, and we have defined the quotient of scales
\be
\varrho \, = \, \left( \frac{m}{m_*} \right)^2 \, =\,  (2\pi)^{3/2} g_s^{1/2} \sigma
\label{varrho}
\ee
The second expression for $\varrho$ can be derived in the type IIB orientifold limit of the $SO(12)$ model. 
There, as explained in section \ref{s:nplocal}, $\sigma = (m/m_{st})^2$ describes the intersection slope of 
the 7-branes in units of the string scale $m_{st}^{-2} = 2\pi \alpha'$. In this limit we also obtain the relation 
$m_{st}^4 = g_s (2\pi)^3 m_*^4$, and combining both results the second equality of (\ref{varrho}) follows. 

As already explained, the holomorphic Yukawa couplings are independent from fluxes, this dependence only
appearing in the normalization factors $\gamma_{a,b,c}^i$.
In particular the dependence on hypercharge fluxes is the only possible source of distinction between 
D-quark and charged lepton Yukawas  which are equal before this flux is turned on.
 In what follows we will be using the
uncorrected normalization factors as computed in appendix A, since these corrections would only
induce terms of order $\epsilon^2$ in the physical Yukawa couplings. The same happens with the mixing
discussed in section \ref{sec:mix}, which only affects the physical Yukawa couplings at higher order.

Our expressions for Yukawa couplings apply at the unification-string scale, presumably of order $10^{16}$ GeV, 
so that in order to compare with experimental fermion masses one needs to run the data up to the unification scale.
An updated two-loop analysis for this running within the MSSM  has been performed in ref.\cite{Ross:2007az}
 from which we will take the data 
below. 
Here we will only discuss  fermion masses for charged leptons and D-quarks, which are
the ones relevant for the $SO(12)$ case studied here.  Table \ref{massas} shows  the relevant fermion mass ratios
evaluated at the unification scale for various values of tan$\beta$ (the ratio of the two Higgs vevs in the MSSM).
We also show for reference the Yukawa couplings of the $\tau$ lepton and $b$ and $t$ quarks.
Recall that $m_{\tau,b}=Y_{\tau,b}V \cos\beta $, $m_t=Y_tV \sin\beta $, with $V=\sqrt{V_u^2+V_d^2}\simeq 174$ GeV.
\begin{table}[htb] 
\renewcommand{\arraystretch}{1.25}
\begin{center}
\begin{tabular}{|c||c|c|c|}
\hline
tan$\beta$  &  10&   38  &  50 \\
\hline\hline
$m_d/m_s$ &   $5.1\pm 0.7\times 10^{-2}$   &  $5.1\pm 0.7\times 10^{-2}$  & $5.1\pm 0.7\times 10^{-2}$  \\
\hline
$m_s/m_b$ &    $1.9\pm 0.2\times 10^{-2}$   &  $1.7\pm 0.2\times 10^{-2}$  & $1.6\pm 0.2\times 10^{-2}$  \\
\hline\hline
$m_e/m_\mu$  &   $4.8\pm 0.2\times 10^{-3}$   &  $4.8\pm 0.2\times 10^{-3}$  & $4.8\pm 0.2\times 10^{-3}$   \\
\hline
$m_\mu/m_\tau$  &    $5.9\pm 0.2\times 10^{-2}$   &  $5.4\pm 0.2\times 10^{-2}$  & $5.0\pm 0.2\times 10^{-2}$  \\
\hline\hline
$m_b/m_\tau$  &    $0.73\pm0.03 $   &  $0.73\pm0.03 $ &    $0.73\pm0.04 $  \\
\hline\hline
$Y_\tau $  &    $0.070\pm0.003 $   &  $0.32\pm0.02 $ &    $0.51\pm0.04 $ \\
\hline
$Y_b $  &    $0.051\pm0.002 $   &  $0.23\pm0.01 $ &    $0.37\pm0.02 $ \\
\hline
$Y_t $  &    $0.48\pm0.02 $   &  $0.49\pm0.02 $ &    $0.51\pm0.04 $ \\
\hline
\end{tabular}
\end{center}
\caption{\small Running mass ratios  of leptons and D-quarks at the unification scale from ref.\cite{Ross:2007az}.
The Yukawa couplings $Y_{\tau,b,t}$ at the unification scale are also shown.}
\label{massas}
\end{table}
As emphasized e.g.  in ref.\cite{Ross:2007az},
the Yukawa couplings obtained at the GUT scale seem to depart from the predictions 
from the minimal $SU(5)$  GUT, which yield $Y_{\tau,\mu,e}=Y_{b,s,d}$. This happens not only, as is well known,  for the first two generations 
but also for the third for which one has a sizable departure from unification, see table \ref{massas}.  
On the other hand, for large tan$\beta$, which is going to be our case as we will see momentarily, there are additional
large threshold corrections to $Y_b$  from the low-energy SUSY thresholds. In particular the leading such corrections
in the MSSM are given by
\beq
\delta Y_b \ =\ -\frac {g_3^2}{12\pi^2 } \frac {\mu M_3}{m_{\tilde b}} \tan\beta \ -\ 
\frac {Y_t^2}{32\pi^2 } \frac {\mu A_t}{m_{\tilde t}} \tan\beta 
\label{thresholdios}
\eeq
where $M_3$,$m_{\tilde b}$,$m_{\tilde t}$ are the gluino, sbottom and stop masses and $\mu$, $A_t$ are the Higgsino mass 
and top trilinear  parameter. These corrections may easily be of order 20\% (see 
e.g.\cite{Ross:2007az},\cite{Elor:2012ig}) and the sign depends on the relative signs of the
soft terms. So from table \ref{massas} and taking into account these corrections 
we will take  for the third generation ratio
\beq
\frac {Y_\tau}{Y_b} \ =\ 1.37 \pm 0.1 \pm 0.2
\label{ratiotaub}
\eeq
at the unification scale. The low energy threshold corrections may render $b/\tau$ unification 
\cite{Ross:2007az,Elor:2012ig}  but only for particular choices of parameters, particularly of signs.

Since we do not know the corrections of order $\epsilon^2$ in the matrix (\ref{yukasmm}) we will not attempt to 
describe the first generation masses but will only require  that one of the eigenvalues should be much smaller than the 
other two. We will thus only try to describe the hierarchies between the third and second generation. 
 From the table one  gets for the mass ratios
\beq
\frac {m_\mu }{m_\tau } \ =\ 4.8-6.1\times 10^{-2}  \ \ ,\ \ 
\frac {m_s }{m_b } \ =\ (1.4-2.1) \times 10^{-2} 
\label{jerarquias}
\eeq
for tan$\beta=10-50$. One then has for the (lepton/quark) ratio
\beq
\frac { {m_\mu }/{m_\tau }} { {m_s }/{m_b } } \ \simeq \   3.3\pm 1
\label{ratioratio}
\eeq
We would  like to see whether such hierarchies arise in our scheme. 

\subsection{The third generation Yukawa couplings}

In order to compute the physical Yukawa couplings we need to compute the normalization factors $\g_{a,b,c}^i$. In principle 
this requires full knowledge of the matter wavefunctions along the matter curves. If however we assume that the relevant
wavefunctions are localized close to the (unique)  intersection point of the three matter curves, one can use the local 
convergent expressions for these normalization factors given in appendix A. 
As we said,  we will be using the
uncorrected normalization factors  shown in that appendix, since the corrections would only
induce terms of order $\epsilon^2$ in the physical Yukawa couplings.
Then the leading contribution to the third generation Yukawa couplings is given by the 33 entry of the above matrix. 
Using (\ref{y33pert}) together with (\ref{normga}-\ref{normgc}) we obtain
\bea\nonumber
& &\hspace{-.8cm} Y_{b,\tau}=(4\pi g_s)^{1/2} \,\sigma 
\left(\! -\frac{q_P^{a_{1,2}}(2\lam_{a_{1,2}}+q_P^{a_{1,2}}(1+\zeta_{a_{1,2}}^2))}{m^4+\lam_{a_{1,2}}^2(1+\zeta_{a_{1,2}}^2)}
\right)^{\!\!\! 1/2} \!\!
\left(\!-\frac{q_P^{b_{2,3}}(-2\lam_{b_{2,3}}+q_P^{b_{2,3}}(1+\zeta_{b_{2,3}}^2))}{m^4+\lam_{b_{2,3}}^2(1+\zeta_{b_{2,3}}^2)}
\right)^{\!\!\! 1/2}\\
& &\hspace{.8cm}\left(\!-\frac{(2\zeta_c+q_P^c)(q_P^c+2\zeta_c-2\lam_c)+(q_S^c+\lam_c)^2}{m^4+\zeta_c^2+(\zeta_c-\lam_c)^2}
\right)^{\!\!\! 1/2}
\label{gorda}
\eea
where the $q_P$, $\lambda$'s and $\zeta$'s are defined in appendix \ref{ap:wave} and all fluxes, as in previous sections,  have been
taken constant in the vicinity of the intersection point.
 
To estimate the value of the couplings we assume that the fluxes are such that $q_S^a\simeq 0$, $q_S^b \simeq 0$,
and $q_S^c \ll q_P^c$, as we will indeed find in our numerical fits. Using the results in appendix  \ref{ap:wave},
with $M_{xy}=0$, we then find the approximate result
\be
Y_{b/\tau} \simeq \left ( 8\, \pi g_s \sigma^2 \frac{q_P^{a_{1,2}}\,q_P^{b_{2,3}}\,q_S^c}{\lam_a \lam_b \lam_c}\right )^{1/2} 
\label{yukaproxb}
\ee
The eigenvalues $\lambda$ are expected to be generically of the order of the charges $q_P \simeq M$.
Taking $q_S^c \ll q_P^c$ then shows that $Y_{b/\tau}$ is proportional to $\sigma g_s^{1/2}$
with a small coefficient $\CO(1)$. The intersection slope $\sigma$ is assumed to be small whereas $g_s$ is constrained
as we now explain. Flux quantization requires \mbox{$\int_{\Sigma_2} \langle F \rangle  \simeq 2\pi$}, 
so that taking $V_{\Sigma_2}\simeq V_S^{1/2}$ implies $M \simeq N_Y\simeq {\tilde N}_Y \simeq (2\pi)/V_S^{1/2}$. 
Next, the volume of the 7-brane surface $S$ enters in
the perturbative equality for the unification coupling (see e.g.\cite{thebook})
\be
\alpha_G \ =\  \frac{2\pi^2g_s}{m_{st}^4V_S} 
\ee
Setting $\a_G \simeq 1/24$ leads to the estimate for the fluxes 
\be
\frac {M}{m_{st}^2}\ =\ \left( \frac {2 \alpha_G}{g_s } \right)^{1/2}\ \simeq \   \frac {0.29}{g_s^{1/2}}  
\label{estimflux}
\ee
Having diluted fluxes imposes $M < m_{st}^2$. We then conclude that $g_s^{1/2}$ cannot be arbitrarily small.
The conditions of small intersection slope and fluxes are needed to justify the effective description of the 7-brane
theory.

Although (\ref{yukaproxb}) is just an approximation, it indicates  that in the present scheme with the wavefunctions 
localized at the matter curve intersection point the third generation Yukawa couplings are 
large, of the same order of the Yukawa coupling of the top quark which we know
on phenomenological grounds is of that order (see table \ref{massas}).  So e.g. in a MSSM scheme
one expects a large  $\tan \beta \simeq m_t/m_b\simeq 20-50$.
In the computations below we show that the above qualitative statements 
remain true in our F-theory scheme.

\subsection{Hierarchies of fermion masses. }

We have discussed above the Yukawa couplings for the third generation.
The corrections discussed in previous chapters are in principle able to generate 
Yukawa couplings and masses also for the first two generations.
We would like to explore now to what extent the above results could be able to describe the
observed structure of hierarchical lepton and D-quark masses.
We will be interested now on the relative hierarchies among different D-quarks 
or different charged leptons, so that we will study the matrix
\be
\frac{Y}{Y^{33}} = 
\left(
\begin{array}{ccc}
\CO(\e^2)   &   \CO(\e^2)   &   \eps \varrho^{-1}  \frac { \g_{a}^{1}}{\g_{a}^{3}} {\tilde  \th}_{0x}\\
\CO(\e^2)   &  \eps \varrho^{-1}   \frac { \g_{a}^{2} \g_{b}^{2} } {\g_{a}^{3}\g_{b}^{3}}  ({\tilde \theta}_{0x}+{\tilde \theta}_{0y})
&    {\eps} \varrho \frac {  \g_{a}^{2}}{\g_{a}^{3}}  {\tilde \theta}_{20} \\
\eps \varrho^{-1}  \frac { \g_{b}^{1}}{\g_{b}^{3}}  {\tilde \th}_{0y}
& - \frac {\eps}{2} \varrho \frac {  \g_{b}^{2}}{\g_{b}^{3}} {\tilde  \theta}_{20}  &  1
\end{array}
\right) 
\label{yukasmmr}
\ee
where we have divided the matrix (\ref{yukasmm}) by the largest $Y_{33}$ entry.
It is easy to check that this matrix has eigenvalues
\bea\nonumber
\lam_1&=&1+\CO(\eps^2)\\\nonumber
\lam_2&=& \eps \varrho^{-1} \frac { \g_{a}^{2} \g_{b}^{2} } {\g_{a}^{3}\g_{b}^{3}}  ({\tilde \theta}_{0x}+{\tilde \theta}_{0y})+\CO(\eps^2)\\\nonumber
\lam_3&=&\CO(\eps^2).
\eea
This is interesting since we automatically get a hierarchy of masses of order $(1,\epsilon, \epsilon^2)$ from the start, without any
further assumption.  Note that if we had $\theta_0=0$ and we were left only with the corrections from $\theta_2$, the matrix would be 
still rank one up to order $\epsilon^2$ and the non-perturbative corrections would be unable to create the desired hierarchies.
So in order to obtain hierarchies it turns out to be crucial the presence of a non-constant $\theta_0$ as studied in the previous sections.

\begin{figure}
\center\includegraphics[width=7.5cm]{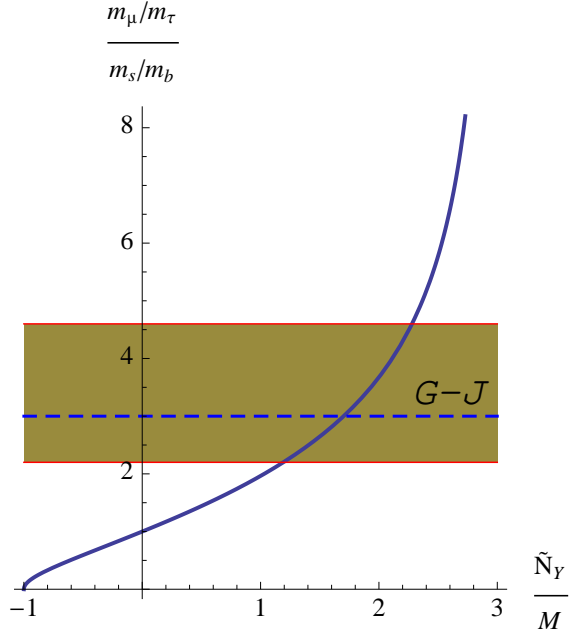}
\caption{$\frac{m_\mu/m_\tau}{m_s/m_b}$ at the unification scale as a function of $\tilde N_Y/M$.
The band shows the region consistent with  the running of the low energy data up to the unification scale taken from ref.\cite{Ross:2007az}.
The dashed line is the Georgi-Jarlskog value \cite{GJ}.}
\label{hierarnew}
\end{figure}

Identifying the first and second eigenvalues with the third and second generations one gets to leading order in $\epsilon$ 
at the unification scale.
\bea\label{eq:md}
\frac{m_s}{m_b}&=&\left (-\frac{q_P^{a_1}q_P^{b_2}}{m_*^4} \right )^{1/2}\eps\,\varrho^{-1} (\tilde \th_{0x}+\tilde \th_{0y})\\
\frac{m_\mu}{m_\tau}&=&\left (-\frac{q_P^{a_2}q_P^{b_3}}{m_*^4} \right )^{1/2}\eps\,\varrho^{-1}(\tilde \th_{0x}+\tilde \th_{0y})\label{eq:ml}
\eea
where we have used ${\g_a^2}/{\g_a^3}=\left ( {q_P^a}/{m_*^2} \right )^{1/2}$ and ${\g_b^2}/{\g_b^3}=\left (- {q_P^b}/{m_*^2} \right )^{1/2}$.
Using the expressions above together with the charges in table \ref{t2}  we get for the ratio of ratios
\be
\frac{m_\mu/m_\tau}{m_s/m_b}=\left ( \frac{(M+\frac{1}{2}\tilde N_Y)(M+\tilde N_Y)}{(M-\frac{1}{3}\tilde N_Y)(M+\frac{1}{6}\tilde N_Y)} \right )^{1/2}.
\ee
This ratio is interesting because the dependence on the non-perturbative correction and extra flux-dependent factors cancel
out yielding a result which only depends on the ratio ${\tilde N}_Y/M$.
Experimentally one has from table \ref{massas} that 
 $\frac{m_\mu/m_\tau}{m_s/m_b}=3.3\pm1$ so that one has the constraint
\be
\left ( \frac{(1+\frac{\tilde N_Y}{2M})(1+\frac{\tilde N_Y}{M})}{(1-\frac{\tilde N_Y}{3M})(1+\frac{\tilde N_Y}{6M})} \right )^{1/2}=3.3\pm1 \ .
\ee
This condition is displayed in figure \ref{hierarnew}.  One observes that agreement with experiment may be obtained with a ratio of
fluxes ${\tilde N}_Y/M= 1.8\pm 0.6$, independently from the value of the rest of the parameters. Note that 
 conditions for  consistent local chirality are $M_y+q_Y^{a^+}\tilde N_Y>0$ and $M_x+q_Y^{b^+}\tilde N_Y<0$ which are satisfied as long as $-1<\tilde N_Y/M < 3$.
To reproduce the particular D-quark and charged lepton hierarchies 
we can substitute the last result in (\ref{eq:md}) and (\ref{eq:ml}) which can be rewritten as
\bea\label{eq:md2}
\frac{m_s}{m_b}&=& \left[\left(1-\frac{\tilde N_Y}{3M}\right)\left(1+\frac{\tilde N_Y}{6M}\right)\right]^{1/2}  \,\frac{M}{m_*^2}\eps\,\varrho^{-1}(\tilde \th_{0x}+\tilde \th_{0y})\\
\frac{m_\mu}{m_\tau}&=&  \left[\left(1+\frac{\tilde N_Y}{2M}\right)\left(1+\frac{\tilde N_Y}{M}\right)\right]^{1/2} \,\frac{M}{m_*^2}\eps\,\varrho^{-1}(\tilde \th_{0x}+\tilde \th_{0y})\label{eq:ml2}
\eea
From table \ref{massas} we see that $\frac{m_\mu}{m_\tau}=5.4\pm0.6 \times 10^{-2}$ so that we get an estimate
\be
\frac{M}{m^2}\,\eps\,(\tilde \th_{0x}+\tilde \th_{0y})\simeq (2.3 \pm 0.2) \times 10^{-2}
\label{pertpar}
\ee
where we have used the central value ${\tilde N}_Y/M=1.8$ and the definition of $\varrho$. 
Thus taking ${\tilde N}_Y/M\simeq 1.8$ and the non-perturbative correction of the size given by eq.(\ref{pertpar}) 
one obtains values consistent with the experimental $m_\mu/m_\tau$ and $m_s/m_b$ ratios.
Note that the number in eq.(\ref{pertpar}) is quite small, consistent with a non-perturbative origin.

\subsection{ $b-\tau$ (non)-unification}

\begin{figure}[ht]
\begin{center}
\begin{tabular}{lcr}
\includegraphics[scale=1.3]{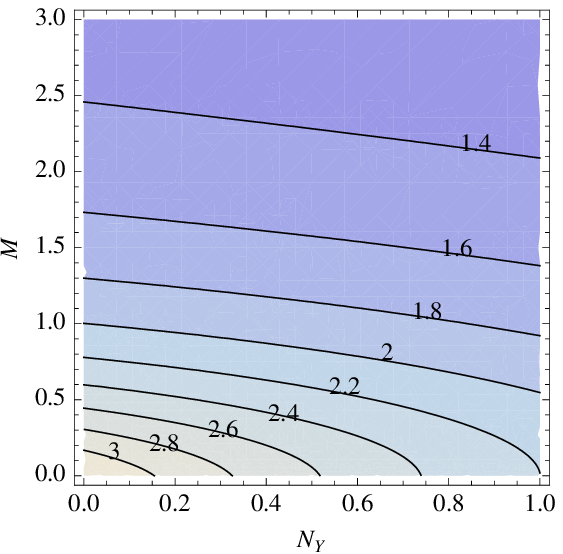}
& \quad &
\includegraphics[scale=1.3]{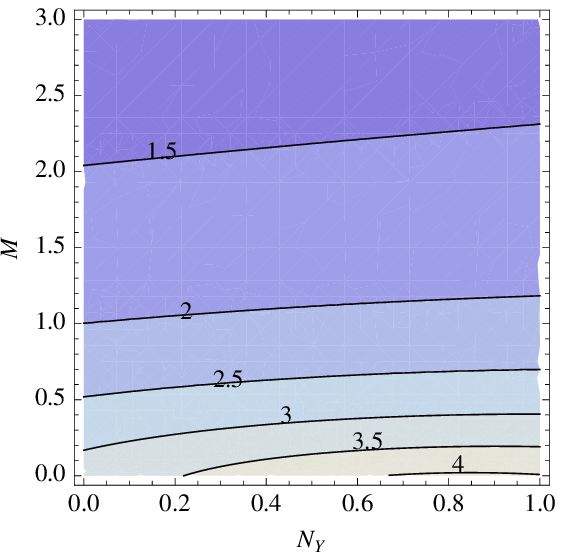} \\
\includegraphics[scale=1.3]{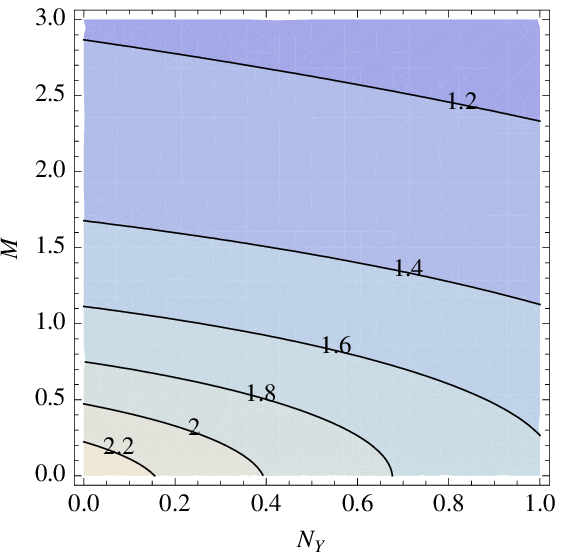}  & \quad &
\includegraphics[scale=1.3]{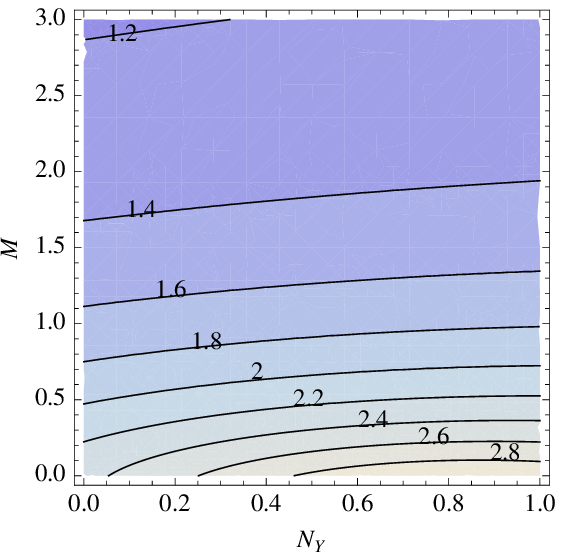} 
\end{tabular}
\end{center}
\caption{\small{ $Y_\tau/Y_b$ as a function of $M,N_Y$ for
$N_a =-1(1)$  left(right). Flux choices are consistent with $\frac{m_\mu/m_\tau}{m_s/m_b}= 3(2.2)$
for upper(lower) plots. }}
\label{ytyb2}
\end{figure}

Hypercharge fluxes also violate the equality of the $\tau$ and $b$-quark Yukawas at unification. Indeed, 
using eq.(\ref{y33pert}) one gets
\be\label{eq:cy}
\frac{Y_\tau}{Y_b}=\frac{\g_{a_2}^3\g_{b_3}^3}{\g_{a_1}^3\g_{b_2}^3} \ .
\ee
with the expressions for the $\gamma$'s given in eqs.(A.65-A.67). 
We would like to see now whether one can obtain the result  $\frac{Y_\tau}{Y_b}=1.37\pm0.1\pm 0.2$ discussed above for some choice of fluxes consistent with the equations for the Yukawa hierarchies of second to third generation discussed above. Recall that the latter require ${\tilde N}_Y/M\simeq 1.2 - 2.4$.  We have as free flux parameters $M,N_Y, N_a$, since $N_b$ is determined by eq.(\ref{cond}) which sets $N_Y=3(N_a-N_b)$. In figure \ref{ytyb2} we show the value obtained for $Y_\tau/Y_b$ as a function of the fluxes  $M$ and $N_Y$ and for values $N_a=\pm 1$, which is sufficient to show the general behavior. This is done for ${\tilde N}_Y/M=1.8$ (upper plots) and 1.3 (lower plots), which correspond in turn to $\frac{m_\mu/m_\tau}{m_s/m_b}=3,2.2$ respectively, yielding consistent  2nd to 3rd generation mass  hierarchies. A first conclusion is that correctly yielding the latter hierarchy requires in turn $Y_\tau/Y_b>1$, as observed. One can easily have flux choices with $\frac{Y_\tau}{Y_b}=1.37\pm0.1\pm 0.2$, particularly for $N_a=-1$ and ${\tilde N}_Y/M=1.3$  (lower left figure).

%
An example of consistent  parameter choices is
 (the fluxes are given in $m^2$ units)
\beq
(M,N_a,N_Y,{\tilde N}_Y; \  {\tilde \epsilon}) \ =\ (2,0,0.1,3.6;\   7.5\times 10^{-4})
\eeq
where ${\tilde \epsilon }=g_s^{1/2}\eps\,(\tilde \th_{0x}+\tilde \th_{0y})$.
One obtains 
\be
\left (\frac{Y_{\tau}}{Y_b},\,\frac{m_s}{m_b},\,\frac{m_\mu}{m_\tau},\,Y_b\right )= ( 1.38,\,1.7\times10^{-2}, \,5.4\times10^{-2},\,0.66 g_s^{1/2}\sigma)
\ee
in very good agreement with the experimental results in table \ref{massas}. 
Note  that the value for $Y_b$ at the unification scale are consistent with those in table \ref{massas} for $g_s \sigma^2 \simeq 0.1$
which   suggest indeed a large value 
of tan$\beta$ in the MSSM context. This is generally the case for all examples able to appropriately
describe the mass ratios.
Let us finally comment that the condition of diluted fluxes corresponds to  e.g. $M/m_{st}^2=(\sigma M)/m^2 <  \CO(1)$,
and the same for the rest of the fluxes. This may be achieved 
in the above example and also for the examples in figure \ref{ytyb2} by considering an appropriately small value for $\sigma$.

One can repeat the analysis in this chapter for the case of the non-SUSY Standard Model remaining below the
string scale. Indeed, although we made our discussion in terms of superpotentials, the Yukawa coupling sector
remains essentially unchanged in the presence of terms breaking SUSY below the string scale. 
 In the case of the SM, extrapolating the low-energy masses up to the unification scale the ratios of the second to
third generation masses remain similar to those shown in eq.(\ref{jerarquias}), see e.g. ref.\cite{bora}. 
Concerning the  $b/\tau$ ratio one gets  around the unification scale  $Y_\tau/Y_b=1.73$, with no relevant
 low-energy thresholds giving additional contributions.  The flux analysis above would equally apply to this 
 non-SUSY case and there are wide ranges of fluxes consistent with the fermion hierarchies and 
 $\tau/b$ ratio.  The required fluxes tend to be however more diluted in this non-SUSY case.

We conclude that instanton effects are able to generate the observed second to third generation hierarchies of charged leptons and D-quarks, via the superpotential deformation (\ref{suponpintro}). The relevant non-perturbative correction corresponds to the term proportional to $\theta_0$ in (\ref{suponpintro}), while the correction from $\theta_2$ only affects the first generation.
The hierarchies  between second and third generations can then be easily understood in this scheme. One can reproduce the values $(Y_\mu/Y_\tau)/( Y_s/Y_b)= 3.3\pm 1.0$  at the unification scale, consistent with the low-energy data and,  at the same time, the $b/\tau$ ratio with Yukawa couplings  $Y_\tau/Y_b (m_{st})\simeq 1.37\pm 0.1\pm 0.2$, as obtained from the RGE in the MSSM.  These two attractive features are purely due to the hypercharge flux which explicitly breaks the underlying $SU(5)$ symmetry, which otherwise predicts equal masses for D-quarks and leptons of each generation at the unification scale. 
 The Yukawa couplings of the third generation are large, corresponding to large 
values of tan$\beta$ (for not too small $g_s \sigma^2$)   within the context of the MSSM.  
 In order to compute the masses for the first generation we would need
to know the corrections to the Yukawa couplings at order $\epsilon^2$ which, although feasible, is a more involved task.

\section{Conclusions and outlook}

In this paper we have analyzed the generation and structure of Yukawa couplings in local F-theory models of $SU(5)$ unification, taking into account the presence of non-perturbative effects. More precisely, we have shown how non-perturbative effects can induce non-trivial Yukawas for the two lighter families of quarks and leptons, and naturally reproduce a hierarchical mass structure among the three families. In this sense, our work can be thought of as a continuation of \cite{afim}, in which the same kind of approach was applied to a toy model of unification. While most of our previous conclusions remain true, there are several important new features that are crucial to understand the physics of Yukawa couplings in the present context of $SU(5)$ unification. 

First, in a realistic F-theory GUT model the non-perturbative effects on 7-brane Yukawa couplings are non-trivially constrained. Recall that in \cite{mm09,afim} one could understand the effect of distant instantons on Yukawa couplings by simply adding the term $W_{\rm np} = \eps \int \theta_1 \tr (\Phi F \wedge F)$ to the 7-brane tree-level superpotential. However, such term identically vanishes when evaluated at Yukawa points of $SO(2N)$ or $E_{6,7,8}$ enhancement, which are precisely the cases of interest for F-theory GUT models with chiral matter localized at matter curves. As pointed out in \cite{afim} this does not mean that non-perturbative effects are trivial, but rather that one needs to consider the more general superpotential (\ref{suponpintro}) and repeat the analysis with those terms that do not vanish. In the scheme of \cite{mm09,afim} the less suppressed of such terms is $\eps \int \theta_2 \tr (\Phi^2 F \wedge F)$, which we have indeed included in our analysis. We have however seen that the hierarchical structure that this term generates for the masses of the three fermion families is of the form $(\CO(\eps^2), \CO(\eps^2), \CO(1))$, which is rather unsatisfactory from a phenomenological viewpoint. 

A more interesting result is obtained when the term $\eps \int \theta_0 \tr (F \wedge F)$ is included in the analysis. This term is in fact the least suppressed of the series (\ref{suponpintro}) but becomes irrelevant for $\theta_0$ constant, which was the case for the scenario analyzed in \cite{mm09,afim}. However, in the context of realistic F-theory GUT models we may have a non-constant holomorphic $\theta_0$. Indeed, we have analyzed an $SU(5)$ F-theory model in the vicinity of a Yukawa point of $SO(12)$ enhancement, where the couplings $10\times {\bar 5}\times {\bar 5}$ are generated. This model admits a natural type IIB realization, and there one can argue that $O(1)$ D3-brane instantons can induce a non-constant $\theta_0$. We have analyzed the $10\times {\bar 5}\times {\bar 5}$ Yukawa structure with $W_{\rm np} = \eps \int \theta_0 \tr (F \wedge F) +\eps \int \theta_2 \tr (\Phi^2 F \wedge F)$, and obtained a fermion mass hierarchy of the form $(\CO(\eps^2), \CO(\eps), \CO(1))$ in a much better agreement with observation. It would be interesting to extend this analysis to Yukawa points of $E_{6,7,8}$ enhancement, for which we expect a similar result.

Second, in the model of $SO(12)$ enhancement one can analyze in detail the effect of the hypercharge flux $F_Y$ in the Yukawa couplings. As emphasized throughout the text the hypercharge flux is the only ingredient that breaks the $SU(5)$ gauge symmetry of the model and, as a result, it should be able to explain the observed differences between, e.g., D-quarks and charged lepton masses. One can indeed check explicitly that due to the presence of $F_Y$, all the SM particles have different internal wavefunctions. Hence, in principle the same $SU(5)$ coupling $10\times {\bar 5}\times {\bar 5}$ can lead to different SM Yukawas. It is however not clear that such differences in the wavefunctions is sufficient to understand the experimental data, as Yukawa couplings turn out to depend on the hypercharge flux in a rather subtle manner. 

Indeed, it was shown in \cite{cchv09} that holomorphic Yukawa couplings do not depend on worldvolume fluxes, including the hypercharge flux. In \cite{afim} this result was verified in the presence of non-perturbative corrections of the sort $W_{\rm np} = \eps \int \theta_1 \tr (\Phi F \wedge F)$, with the help of the non-commutative formalism of \cite{cchv09}. In this work we have extended the above results and shown that the same statement is true for the non-perturbative correction $W_{\rm np} = \eps \int \theta_0 \tr (F \wedge F) +\eps \int \theta_2 \tr (\Phi^2 F \wedge F)$ relevant for F-theory GUTs. Unlike before, this time there is no obvious interpretation of $W_{\rm tree} + W_{\rm np}$ in terms of non-commutative geometry, but we have nevertheless derived a residue formula for the holomorphic Yukawas in which the worldvolume flux independence is manifest. 
It would be interesting to study whether there is also a non-commutative structure behind \mbox{$W_{\rm tree} + W_{\rm np}$}
in this case.

As a result, at the holomorphic level and within each family, all the Yukawas that arise from $10\times {\bar 5}\times {\bar 5}$ have the same value. 
The only way in which one can reproduce the  pattern of D-quarks and charged lepton masses observed in nature is via the wavefunction normalization factors that take us from holomorphic to physical Yukawa couplings. We have computed such normalisation factors for the local $SO(12)$ model under study up to $\CO(\eps^2)$ corrections. Such computation is reliable when the matter wavefunctions are sufficiently peaked around the Yukawa point, which is a typical requirement for having reasonable Yukawas for the heaviest family. In fact, our analysis suggests that the Yukawa of the bottom is of the same order of magnitude as the top-quark Yukawa, which points to a large  $\tan \beta$ within the context of the MSSM. Regarding the mass hierarchies within the third and second family, thanks to the hypercharge flux dependence on the normalisation factors one is able to reproduce  reasonable  ratios (\ref{ratiotaub}) and (\ref{ratioratio}) at the unification scale. This is a clear improvement compared to the classical 4d field theory models of $SU(5)$ unification, in which such hierarchies can only be achieved by means of a complicated Higgs system
and substantial threshold corrections.

\bigskip

\bigskip

\centerline{\bf \large Acknowledgments}

\bigskip

We thank  P.G.~C\'amara, I.~Garc\'{\i}a-Etxebarria, L.~Martucci,  E.~Palti and A.~Uranga  for useful discussions. 
A.F. thanks the IFT-UAM/CSIC, the AEI-Potsdam, as well as the AS ICTP, for hospitality and support during progress of this work, and the Alexander von Humboldt Foundation for financial assistance.
This work has been partially supported by the grants FPA 2009-09017, FPA 2009-07908, Consolider-CPAN (CSD2007-00042) from the MICINN, HEPHACOS-S2009/ESP1473 from the C.A. de Madrid, the contract ``UNILHC" PITN-GA-2009-237920 of the European Commission, the REA grant agreement PCIG10-GA-2011-304023 from the People Programme of FP7 (Marie Curie Action) and the grant  SEV-2012-0249 of the ÒCentro de Excelencia Severo OchoaÓ Programme. F.M. is supported by the Ram\'on y Cajal programme through the grant RYC-2009-05096. D.R. is supported through the FPU grant AP2010-5687.

\clearpage

\appendix


\section{{SO(12)} wavefunctions at tree-level}
\label{ap:wave}

In this appendix we solve in for the zero mode wavefunctions of the $SO(12)$ model of the main text. In particular, we will solve for the zero modes of the tree-level superpotential (\ref{supo7}) and D-term (\ref{FI7}). First we will find such zero modes in the holomorphic gauge (\ref{aholo}), and show that they can be expressed as (\ref{gensolpsi}). We will then express them in a physical real gauge and find the normalization factors $\g_{a_p}^i$, $\g_{b_q}^j$ and $\g_{c_r}$ that correspond to canonical kinetic terms in the 4d effective action. 

\subsection{Zero modes in the holomorphic gauge}

As discussed in \cite{afim} and in the main text, in the absence of non-perturbative effects the 7-brane bosonic $(a_{\bar{m}}, \varphi_{xy})$ and fermionic $(\psi_{\bar{m}}, \chi_{xy})$ zero modes satisfy a Dirac-like equation 
\be
{\bf D_A} \Psi\, =\, 0
\label{ap:Dirac9}
\ee
where $\Psi$ is a 4-vector of wavefunctions and ${\bf D_A}$ a $4\times 4$ matrix of operators, both defined in (\ref{matrixDirac}). Massive modes satisfy instead 
\be
{\bf D_A}^\dag {\bf D_A}\, \Psi\, =\, |m_\rho|^2 \Psi
\label{eigenferm}
\ee
where ${\bf D_A}^\dag$ is given by 
\be
{\bf D_A}^\dag\, =\, 
\left(
\begin{array}{cccc}
0 & {D}_{\bar{x}} & {D}_{\bar{y}} & {D}_{\bar{z}} \\
-{D}_{\bar{x}} & 0 & -{D}_{z} & {D}_{y} \\
-{D}_{\bar{y}} & {D}_{z} & 0 & -{D}_{x} \\
-{D}_{\bar{z}} & -{D}_{y} & {D}_{x} & 0
\end{array}
\right)
\label{Ddag}
\ee
In practice, in order to solve for the zero modes it is useful to consider eq.(\ref{eigenferm}) with $m_\rho =0$. That is because the operator $\mathbf{{D_A}^{\dagger} D_A}$ takes the form 
\begin{equation}
\mathbf{{D_A}^{\dagger} D_A}=-\Delta \mathbb{I}_4+ \mathbf{M} 
\label{DDform}
\end{equation}
where $\Delta$ is a Laplacian on the GUT 4-cycle $S$, and $\mathbf{M}$ is a matrix whose coefficients are 7-brane worldvolume fluxes and intersection slopes. For constant fluxes $\mathbf{M}$ and $\Delta$ commute, and so the first step to solve for the eigenmodes (\ref{eigenferm})  is to diagonalize $\mathbf{M}$. In the following we will solve for the $SO(12)$ model zero modes for each sector of table \ref{t2}. We will be sketchy,  and we refer  to \cite{afim,cdp11} for more details on this class of computations.


\subsubsection*{Sector a}

 For the sectors $a^\pm_p$ the matrix $\mathbf{M}$ in (\ref{DDform}) takes the form
\begin{equation}
\mathbf{M}_{a_{p}^{\pm}}=\pm \left( \begin{array}{cccc}
M_{xy} & 0 & 0 & 0 \\
0 & -q_P^a & q_S^{a} & -im^2 \\
0 & q_{S}^{a} & q_P^a & 0 \\
0 & im^2 & 0 & 0 \end{array} \right)
= \pm \left( \begin{array}{cc}
M_{xy} & 0 \\
0 & \mathbf{m}_{a_p^+} - M_{xy} \mathbb{I}_3
 \end{array} \right)
\end{equation}
where the entries $q^{a}_S$, $q^a_P$ take the values 
\begin{table}[htb] 
\renewcommand{\arraystretch}{1.25}
\begin{center}
\begin{tabular}{|c|c|c|}
\hline
& $q_S^a$ & $q_P^a$\\
\hline
$a_1^\pm$ & $-N_a-\frac{1}{3}N_Y$ & $M -\frac{1}{3} \tilde{N}_Y$  \\
\hline
 $a_2^\pm$ & $-N_a+\frac{1}{2}N_Y$ & $M + \frac{1}{2} \tilde{N}_Y$  \\
 \hline
\end{tabular}
\end{center}
\end{table}

\noindent
and we have defined $M$ and $M_{xy}$ as in (\ref{DterM}). 
This flux matrix is diagonalized by the following unitary matrix
\begin{equation}
\mathbf{J}_{a_{p}}=  \left( \begin{array}{cccc}
1 & 0 & 0 & 0 \\
0&\lam_1^a \ca_1 & -\lam_2^a \ca_2 &  -i\lam_3^a \ca_3 \\
0&-\lam_1^a \zeta_1^a \ca_1 & \lam_2^a \zeta_2^a \ca_2 &  i\lam_3^a \zeta_3^a  \ca_3 \\
0&i m^2 \ca_1 & -i m^2 \ca_2  &  m^2\ca_3 \\
 \end{array} \right)
\end{equation}
where $\lambda_i^a$ are the three  eigenvalues of $\mathbf{m}_{a_p^+}$, chosen so that $\lambda_1^a<\lambda_2^a< \lambda_3^a$, and satisfying the cubic equation\footnote{In terms of the quantities $\lam_+$, $\lam_-$ defined below table \ref{t3} we have that $\lam_+ (a_p^+) = \lam_1^a$ and $\lam_- (a_p^-) = \lam_3^a$.}
\begin{equation}
(\lambda_i^a)^3-2M_{xy}(\lambda_i^a)^2 -  [m^4+ (q_P^a)^2 + (q_S^{a})^2 - M_{xy}^2]\lambda_i^a+m^4 (q_P^a + M_{xy})=0
\label{cubica}
\end{equation}
Finally, we have defined $\ca_i = \left ((\lam_i^a)^2 + (\lam_i^a \zeta_i^a)^2 + m^4\right )^{-1/2}$ and

\be
\zeta_{i}^a =-\frac{q_S^a}{\lam_i^a - q_P^a -M_{xy}}
\ee

It is easy to see that the lowest eigenvalue of $\mathbf{M}$ for the sectors $a_p^+$ is given by $\lam_1^a - M_{xy}$, and by $M_{xy} - \lam_3^a$ for the sectors $a_p^-$. The potential fermionic zero modes for any of these sectors should then be proportional to the corresponding eigenvectors of $\mathbf{M}$. Indeed, considering the sectors $a_p^+$, the Ansatz for a fermionic zero mode would be
\begin{equation}
\Psi_{a^+_{p}}=
\mathbf{J}_{a_{p}}
 \left( \begin{array}{c}
0  \\
1  \\
0  \\
0  \end{array} \right) \chi_{a^+_{p}}^\prime\, E_{a_p^+}
=
\left( \begin{array}{c}
0  \\
- \frac{i\lam_1^a}{m^2}  \\
 - \frac{i q_S^a \lam_1^a}{(\lam_1^a - q_P^a -M_{xy})m^2}  \\
1  \end{array} \right) \chi_{a^+_{p}}\, E_{a_p^+}
\label{totalwfsola}
\end{equation}
where $\chi_{a^+_{p}}^\prime$ and $\chi_{a^+_{p}}$ are scalar wavefunctions that differ by a constant factor. With this Ansatz eq.(\ref{ap:Dirac9}) translates to
\begin{equation}\label{eqa}
\mathcal D_x\chi_{a^+_{p}}=\mathcal D_{\bar y}\chi_{a^+_{p}}=\mathcal D_{\bar z}\chi_{a^+_{p}}=0
\end{equation}
where the differential operators $\mathcal D_\a$ correspond to the entries of the Dirac operator $\mathcal D_{\mathbf A}= \mathbf{J}^t_{a_{p}} \mathbf{D_A} \mathbf{J}_{a_{p}} $ in the basis rotated by $\mathbf{J}_{a_{p}}$. Namely, we have that
\be
\left(\begin{array}{c}
\cd_x \\
\cd_y \\
\cd_z \\
\end{array}\right) =
\left(\begin{array}{c}
\ca_1\left[\lam_1^a D_x - \lam_1^a \zeta_1^a D_y + im^2 D_z\right] \\
\ca_2\left[-\lam_2^a D_x + \lam_2^a \zeta_2^a D_y - im^2 D_z\right]  \\
\ca_3\left[-i\lam_3^a D_x +i \lam_3^a \zeta_3^a D_y + m^2 D_z\right] \\
\end{array}\right)
\ee

\be
\left(\begin{array}{c}
\cd_{\bar x} \\
\cd_{\bar y} \\
\cd_{\bar z} \\
\end{array}\right) =
\left(\begin{array}{c}
\ca_1\left[\lam_1^a D_{\bar x} - \lam_1^a \zeta_1^a D_{\bar y} - im^2 D_{\bar z}\right] \\
\ca_2\left[-\lam_2^a D_{\bar x} + \lam_2^a \zeta_2^a D_{\bar y} + im^2 D_{\bar z}\right]  \\
\ca_3\left[i\lam_3^a D_{\bar x} - i \lam_3^a \zeta_3^a D_{\bar y} + m^2 D_{\bar z}\right]  \\
\end{array}\right)
\ee

The general solution to eqs.(\ref{eqa}) in the holomorphic gauge (\ref{aholo}) is then given by
\begin{equation}
\chi_{a^+_{p}}^{\text{hol}}=e^{\lambda_1^a x \left(\bar{x} +\frac{q_S^{a}}{\lam_1^a - q_P^a -M_{xy}}\bar y \right)}f_{a^+_{p}}\left ( y-\frac{q_S^{a}}{\lam_1^a - q_P^a -M_{xy}}x \right )
\label{scalarwfa}
\end{equation} 
where $f_{a^+_{p}}$, $p =1,2$ are holomorphic functions to be determined by boundary conditions. Taking $\lam_{a_p}\, =\, \lam_1^a$ and $\zeta_{a_p} = \zeta_1^a$  we obtain the zero mode wavefunctions
\begin{equation}
\Psi_{a^+_{p}}^{\rm hol}\, =\,
 \left( \begin{array}{c}
0  \\
- \frac{i\lam_{a_p}}{m^2}  \\
\zeta_{a_p} \frac{i\lam_{a_p}}{m^2}  \\
1  \end{array} \right) e^{ \lam_{a_p}  x(\bar{x} - \zeta_{a_p} \bar{y})}  f_{a_p^+} (y+ \zeta_{a_p}x) E_{a_p^+}
\label{finalwfa}
\end{equation}
that is indeed of the form (\ref{gensolpsi}), with $\lam_+ = \lam_{a_p}$.


\subsubsection*{Sector b}

For the sectors $b_q^{\pm}$, $q=1,2,3$, the flux matrix $\mathbf{M}$ takes the form
\begin{equation}
\mathbf{M}_{b_q^{\pm}}=\pm \left( \begin{array}{cccc}
-M_{xy} & 0 & 0 & 0 \\
0 & -q_P^b & q_S^{b} & 0 \\
0 & q_{S}^{b} & q_P^b & im^2 \\
0 & 0 & -im^2 & 0 \end{array} \right)
= \pm \left( \begin{array}{cc}
- M_{xy} & 0 \\
0 & \mathbf{m}_{b_q^+} + M_{xy} \mathbb{I}_3
 \end{array} \right)
\end{equation}
where $q^{b}_S$, $q_P^b$ take the following values 
\begin{table}[htb] 
\renewcommand{\arraystretch}{1.25}
\begin{center}
\begin{tabular}{|c|c|c|}
\hline
& $q_S^b$ & $q_P^b$\\
\hline
$b_1^\pm$ & $N_b+\frac{2}{3}N_Y$ & $- M +\frac{2}{3} \tilde{N}_Y$  \\
\hline
 $b_2^\pm$ & $N_b-\frac{1}{6}N_Y$ & $- M - \frac{1}{6} \tilde{N}_Y$  \\
 \hline
 $b_3^\pm$ & $N_b-N_Y$ & $ - M - \tilde{N}_Y$  \\
 \hline
\end{tabular}
\end{center}
\end{table}

\noindent
It is diagonalized by the unitary matrix
\begin{equation}
\mathbf{J}_{b_q}=  \left( \begin{array}{cccc}
1 & 0 & 0 & 0 \\
0&\lam_1^b \ca_1 & -\lam_2^b \ca_2 &  -i\lam_3^a \ca_3 \\
0&-\lam_1^b \zeta_1^b \ca_1 & \lam_2^a \zeta_2^a \ca_2 &  i\lam_3^a \zeta_3^a  \ca_3 \\
0&i m^2 \ca_1 & -i m^2 \ca_2  &  m^2\ca_3 \\
 \end{array} \right)
\end{equation}
where $\lambda_1^b<\lambda_2^b<\lambda_3^b$ are the eigenvalues of $\mathbf{m}_{b_q^+}$ and so they satisfy the equation
\begin{equation}
(\lambda^b_i)^3+2M_{xy}(\lambda^b_i)^2 - [m^4+ (q_P^{b})^2 + (q_S^{b})^2- M_{xy}^2]\lambda^b_i - m^4( q_P^b + M_{xy}) =0
\label{cubicb}
\end{equation}
Finally, we have defined $\ca_i = \left ((\lam_i^b)^2 + (\lam_i^a \zeta_i^b)^2 + m^4\right )^{-1/2}$ and

\be
\zeta_{i}^b=-\frac{q_S^b}{\lam_1^b +q_P^b+ M_{xy}}
\ee

As before, the appropriate Ansatz for fermionic zero modes in the sectors $b^+_q$ is
\begin{equation}
\Psi_{b^+_q}= 
\mathbf{J}_{b_{q}}
\left( \begin{array}{c}
0  \\
1  \\
0  \\
0  \end{array} \right) \chi_{b^+_q}  E_{b^+_q}
\end{equation}
Therefore, the wavefunctions $\chi_{b^+_q}$ must satisfy the differential equations
\begin{equation}\label{eqb}
\mathcal D_x\chi_{b^+_{q}}=\mathcal D_{\bar y}\chi_{b^+_{q}}=\mathcal D_{\bar z}\chi_{b^+_{q}}=0
\end{equation}
where $\mathcal D_{\mathbf A} = \mathbf{J}^t_{b_q} \mathbf{D_A} \mathbf{J}_{b_q} $ is the Dirac operator in the rotated basis:
\be
\left(\begin{array}{c}
\cd_x \\
\cd_y \\
\cd_z \\
\end{array}\right) =
\left(\begin{array}{c}
\ca_1\left[\lam_1^b D_x - \lam_1^b \zeta_1^b D_y + im^2 D_z\right] \\
\ca_2\left[-\lam_2^b D_x + \lam_2^b \zeta_2^b D_y - im^2 D_z\right]  \\
\ca_3\left[-i\lam_3^b D_x +i \lam_3^b \zeta_3^b D_y + m^2 D_z\right] \\
\end{array}\right)
\ee

\be
\left(\begin{array}{c}
\cd_{\bar x} \\
\cd_{\bar y} \\
\cd_{\bar z} \\
\end{array}\right) =
\left(\begin{array}{c}
\ca_1\left[\lam_1^b D_{\bar x} - \lam_1^b \zeta_1^b D_{\bar y} - im^2 D_{\bar z}\right] \\
\ca_2\left[-\lam_2^b D_{\bar x} + \lam_2^b \zeta_2^b D_{\bar y} + im^2 D_{\bar z}\right]  \\
\ca_3\left[i\lam_3^b D_{\bar x} - i \lam_3^b \zeta_3^b D_{\bar y} + m^2 D_{\bar z}\right]  \\
\end{array}\right)
\ee
It is easy to check that the zero mode equations in the sector $b_q^+$ are identical to the sector $a_p^+$ provided that the following replacements are performed
\begin{equation}
\begin{array}{ccc}
x\leftrightarrow y & \lambda_i^a\leftrightarrow \lambda_i^b & \zeta_i^a \leftrightarrow \zeta_i^b \\
q_S^{a}\leftrightarrow q_S^{b}&q_P^{a}\leftrightarrow -q_P^{b} & M_{xy} \leftrightarrow - M_{xy}
\end{array}
\end{equation}
Thus, the solution of (\ref{eqb}) in the holomorphic gauge is
\begin{equation}
\chi_{b^+_q}^{\text{hol}}=e^{\lambda_1^b y \left(\bar y + \frac{q_S^{b}}{\lam_1^b +q_P^b+ M_{xy}}\bar x\right)} f_{b^+_q}\left ( x-\frac{q_S^{b}}{\lam_1^b +q_P^b+ M_{xy}}y \right )
\end{equation}
where $f_{b^+_q}$, $q=1,2,3$ are arbitrary holomorphic functions. Taking $\lam_{b_q}\, =\, \lam_1^b$ and $\zeta_{b_q} = \zeta_1^b$ we obtain the zero mode wavefunctions
\begin{equation}
\Psi_{b^+_q}^{\rm hol}\, =\,
 \left( \begin{array}{c}
0  \\
- \zeta_{b_q} \frac{i\lam_{b_q}}{m^2}  \\
 \frac{i\lam_{b_q}}{m^2}  \\
1  \end{array} \right) e^{ \lam_{b_q}  y(\bar{y} - \zeta_{b_q} \bar{x})}  f_{b_q^+} (x+ \zeta_{b_q}y) E_{b^+_q}
\end{equation}
that is indeed of the form (\ref{gensolpsi}), with $\lam_+ = \lam_{b_q}$.


\subsubsection*{Sector c}

For this sector the flux matrix reads
\begin{equation}
\mathbf{M}_{c_{r}^{\pm}}=\pm \left( \begin{array}{cccc}
0 & 0 & 0 & 0 \\
0 & -q_P^c & q_S^{c} & im^2 \\
0 & q_{S}^{c} & q_P^c & -im^2 \\
0 & -im^2 & im^2 & 0 \end{array} \right)
= \pm \left( \begin{array}{cc}
0 & 0 \\
0 & \mathbf{m}_{c_r^+} 
 \end{array} \right)
\end{equation}
with $q^{c}_S$, $q_P^c$ now taking the following values 
\begin{table}[htb] 
\renewcommand{\arraystretch}{1.25}
\begin{center}
\begin{tabular}{|c|c|c|}
\hline
& $q_S^c$ & $q_P^c$\\
\hline
$c_1^\pm$ & $N_a-N_b-\frac{1}{3}N_Y$ & $-\frac{1}{3} \tilde{N}_Y$  \\
\hline
 $c_2^\pm$ & $N_a-N_b+\frac{1}{2}N_Y$ & $\frac{1}{2} \tilde{N}_Y$  \\
 \hline
\end{tabular}
\end{center}
\end{table}

\noindent
The unitary matrix diagonalizing $\mathbf{M}_{c_{r}^{\pm}}$ is
\begin{equation}
\mathbf{J}_{c_r}=  \left( \begin{array}{cccc}
1 & 0 & 0 & 0 \\
0 & \zeta_1^c \ca_1 & \zeta_2^c \ca_2 & \zeta_2^c \ca_2 \\
0 & (\zeta_1^c+\lam^c_1)\ca_1 & (\zeta_2^c+\lam^c_2)\ca_2 & (\zeta_3^c+\lam^c_3)\ca_3 \\
0 & im^2\ca_1 & im^2\ca_2 & im^2\ca_3 \end{array} \right)
\end{equation}
with $\lam_1^c<\lam_2^c<\lam_3^c$ being the solutions to
\begin{equation}
(\lam_i^c)^3-\left[ 2m^4+(q_P^c)^2 + (q_S^c)^2\right] \lam_i^c+2 m^4 q_S^c = 0
\label{cubic}
\end{equation}
and where we have defined $\ca_i^c= \left ( (\zeta_i^c)^2+(\lambda_i^c -\zeta_i^c)^2+ m^4 \right )^{-1/2}$ and 
\be
\zeta^c_i= - \frac{q_S^c \lam^c_i-m^4}{\lam^c_i+q_P^c-q_S^c} =
\frac{\lam^c_i(\lam^c_i -q_P^c-q_S^c)}{2(\lam^c_i -q_S^c)}
\ee
In the last equality we used (\ref{cubic}).

As discussed in appendix \ref{ap:flux}, for $q_S^{c}\neq 0$ this sector yields a chiral spectrum. Let us focus on the case where $q_S^c > 0$ and look for zero mode solutions in the sector $c_r^+$. The appropriate Ansatz is 
\begin{equation}
\Psi_{c^+_r}= 
\mathbf{J}_{c_r}
\left( \begin{array}{c}
0  \\
1  \\
0  \\
0  \end{array} \right) \chi_{c^+_r} E_{c^+_r}
\end{equation}
with the wavefunctions $\chi_{c^+_{r}}$ satisfying
\begin{equation}\label{eqc}
\mathcal D_x\chi_{c^+_r}=\mathcal D_{\bar y}\chi_{c^+_r}=\mathcal D_{\bar z}\chi_{c^+_r}=0
\end{equation}
and where $\mathcal D_{\mathbf A}$ is the Dirac operator in the new basis, namely
\begin{equation}
\left( \begin{array}{c}
\mathcal D_x  \\
\mathcal D_y  \\
\mathcal D_z   \end{array} \right)=\left( \begin{array}{c}
  \ca_1\left (-\zeta_1^cD_x+ (\lam_1^c - \zeta_1^c)D_y+im^2D_z \right )\\
  \ca_2\left (-\zeta_2^cD_x+ (\lam_2^c - \zeta_2^c)D_y+im^2D_z \right )\\
  \ca_3\left (-\zeta_3^cD_x+ (\lam_3^c - \zeta_3^c)D_y+im^2D_z \right ) \end{array} \right)
\end{equation}
\begin{equation}
\left( \begin{array}{c}
\mathcal D_{\bar x}  \\
\mathcal D_{\bar y}  \\
\mathcal D_{\bar z}   \end{array} \right)=\left( \begin{array}{c}
  \ca_1\left (-\zeta_1^cD_{\bar x}+ \lam_1^c - (\zeta_1^c)D_{\bar y}-im^2D_{\bar z} \right ) \\ 
  \ca_2\left (-\zeta_2^cD_{\bar x}+ (\lam_2^c - \zeta_2^c)D_{\bar y}-im^2D_{\bar z} \right ) \\
  \ca_3\left (-\zeta_3^cD_{\bar x}+ (\lam_3^c - \zeta_3^c)D_{\bar y}-im^2D_{\bar z} \right ) \end{array} \right)
\end{equation}
The solution to (\ref{eqc}) then reads
\begin{equation}
\chi_{c^+_{r}}^{\text{hol}}=e^{(x-y)\left (\zeta_1^c\bar x  - (\lam_1^c - \zeta_1^c)\bar y\right)}f_{c^+_r}\left ( (\lam_1^c - \zeta_1^c) x +\zeta_1^c y\right )
\end{equation} 
with $f_{c^+_r}$ holomorphic. Again, this solution fits into the general expression (\ref{gensolpsi}). Taking $\lam_1^c=\lam_{c_q}$ and $\zeta_{c_q}=\zeta_1^c$ we find the solution
\begin{equation}
\Psi_{c^+_q}^{\rm hol}\, =\,
 \left( \begin{array}{c}
0  \\
\frac{i\zeta_{c_q}}{m^2} \\
\frac{i (\zeta_{c_q} - \lam_{c_q})}{m^2}\\
1  \end{array} \right) 
e^{(x-y)\left (\zeta_{c_q}\bar x - (\lam_{c_q} - \zeta_{c_q})\bar y\right )}f_{c^+_r}\left ((\lam_{c_q} - \zeta_{c_q}) x + \zeta_{c_q}y\right ) E_{c_r^+}
\end{equation}


\subsubsection*{X,Y bosons}

Let us finally consider the sectors $X^\pm$, $Y^\pm$ in table \ref{t2}, arising from the breaking $SU(5) \raw SU(3) \times SU(2) \times U(1)_Y$ triggered by the presence of the hypercharge flux. Such hypercharge flux is typically chosen so that no zero modes arise from this sector, as they would correspond to $SU(5)$ exotics. However, there will be massive modes, and in particular a tower of 4d massive gauge bosons that can be identified with the $X^\pm$, $Y^\pm$ bosons of 4d $SU(5)$ GUTs.

Unlike the cases analyzed previously, the sectors $X^\pm$, $Y^\pm$ do not correspond to any matter curve, and so they contain wavefunctions of the bulk of the 4-cycle $S$. As a result the quantity $m^2$ does not appear anywhere in the flux matrix, which reads
\begin{equation}
\mathbf{M}_{{\text{X,Y}}^{\pm}}=\pm
\frac{5}{6}
 \left( \begin{array}{cccc}
0 & 0 & 0 & 0 \\
0 & -\tilde N_Y &N_Y & 0 \\
0 &N_Y & \tilde N_Y & 0 \\
0 & 0 & 0 & 0 \end{array} \right)
\end{equation}
and that as expected becomes trivial for vanishing hypercharge fluxes $N_Y$ and $\tilde N_Y$. This matrix is diagonalized by 
\begin{equation}
\mathbf{J}_{X,Y}=  \left( \begin{array}{cccc}
1 & 0 & 0 & 0 \\
0 & c  & - s & 0 \\
0 & s & c & 0 \\
0 & 0 & 0 & 1 \end{array} \right)
\end{equation}
where
\be
c\, =\, \frac{N_Y}{\sqrt{N_Y^2 + (\tilde{N}_Y - \lam)^2}}\quad \quad s\, =\, \frac{\tilde{N}_Y - \lam}{\sqrt{N_Y^2 + (\tilde{N}_Y - \lam )^2}}
\quad \quad \lam\, =\, \sqrt{N_Y^2 + \tilde{N}_Y^2}
\ee
It is easy to obtain the lowest eigenfunction of the scalar Laplacian $\Delta$. Indeed, computing the rotated Dirac operator  $\mathcal D_{\mathbf A} = \mathbf{J}^t_{X,Y} \mathbf{D_A} \mathbf{J}_{X,Y}$ gives us
\begin{equation}
\left( \begin{array}{c}
\mathcal D_x  \\
\mathcal D_y  \\
\mathcal D_z   \end{array} \right)=\left( \begin{array}{c}
c D_x + s D_y  \\
- s D_x + c D_y \\
   D_z \end{array} \right)
\ee
\be
   \left( \begin{array}{c}
\mathcal D_{\bar x}  \\
\mathcal D_{\bar y}  \\
\mathcal D_{\bar z}   \end{array} \right)=\left( \begin{array}{c}
 c D_{\bar x}+s D_{\bar y} \\ 
 - s D_{\bar x}+c D_{\bar y} \\
  D_{\bar z} \end{array} \right)
\end{equation}
Following appendix A of \cite{afim} one can convince oneself that the eigenfunctions of the Laplacian with lowest eigenvalue satisfy
\begin{equation}\label{eqn}
\begin{array}{c}
 \mathcal D_x \chi_{\text{X,Y}^+}=\mathcal D_{\bar y}\chi_{\text{X,Y}^+}=\mathcal D_{\bar z}\chi_{\text{X,Y}^+}=0\\
 \mathcal D_{\bar x}\chi_{\text{X,Y}^-}=\mathcal D_{y}\chi_{\text{X,Y}^-}=\mathcal D_{\bar z}\chi_{\text{X,Y}^-}=0 \end{array}
\end{equation}
Notice that if we define the rotated holomorphic coordinates $u$ and $v$ as 
\be
u=cx +sy \qquad\qquad v= -sx+c y
\ee
then in the holomorphic gauge the above operators read
\be
\begin{array}{ccc}
\mathcal D_x \, =\, \p_u \pm \frac{5}{6} \lam \bar{u} & & \mathcal D_{\bar x} \, =\, \p_{\bar{u}}\\
\mathcal D_y \, =\, \p_v \mp \frac{5}{6} \lam \bar{v} & & \mathcal D_{\bar y} \, =\, \p_{\bar{v}}
\end{array}
\ee
for the sectors $X,Y^\pm$ respectively. Hence, the solution to (\ref{eqn}) in the holomorphic gauge is given by
\be
\chi_{\text{X,Y}^{+}}^{\text{hol}}=e^{- \frac{5}{6}\lam |u|^2}f_{\text{X,Y}^{+}}\left (\bar u, v \right )\qquad\qquad\chi_{\text{X,Y}^{-}}^{\text{hol}}=e^{- \frac{5}{6}\lam |v|^2}f_{\text{X,Y}^{-}}\left (u, \bar v \right )
\label{waveXY}
\ee
with $f_{\text{X,Y}^{\pm}}$ depending on one holomorphic and one anti-holomorphic variable.
Finally, it is easy to see that the lightest 4d gauge boson of this sector
\begin{equation}
 \Psi_{X,Y^\pm}= \left( \begin{array}{c}
1  \\
0  \\
0  \\
0  \end{array} \right) \chi_{X,Y^\pm}  E_{X,Y^\pm}
\end{equation}
has (\ref{waveXY}) as scalar wavefunction and satisfies the equation (\ref{eigenferm}) with
\be
|m_{X,Y}|^2\, =\ \frac{5}{6} \lam = \frac{5}{6}\left (N_Y^2+\tilde N_Y^2\right )^{1/2}
\label{massXY}
\ee
As $N_Y$, $\tilde{N}_Y$ are flux densities over the 4-cycle $S$, we typically have $m_{X,Y} \sim {\rm Vol}(S)^{-1/4}$.


\subsection{Zero modes in the real gauge and normalization}
\label{zmnorm}

While the holomorphic gauge simplifies the computation of wavefunctions, it is not obvious to see from (\ref{scalarwfa}) or (\ref{finalwfa}) if these zero modes actually exist. In this sense taking a real gauge provides more information, since one can study the local convergence of the wavefunction \cite{afim} and describe a set of wavefunctions that is locally chiral \cite{palti12}. In the following we will compute the zero mode wavefunctions in the real gauge for the sectors discussed above, and compute the normalization factors for each of them.

Instead of (\ref{aholo}) let us consider the real gauge
\begin{equation}
\begin{array}{rl}
\langle A \rangle^{\text{real}} =&\frac{i}{2}\left[\bar x (Q_P-M_{xy}Q_F) - \bar y Q_S\right] dx-\frac{i}{2}\left[\bar y (Q_P+M_{xy}Q_F) + \bar x Q_S \right]dy \\
& - \frac{i}{2}\left[ \bar x (Q_P-M_{xy}Q_F) - \bar y Q_S\right]d\bar x +\frac{i}{2}\left[\bar y (Q_P+M_{xy}Q_F) + \bar x Q_S \right]d\bar y\\
=&\langle A \rangle^{\text{hol}}  + d \Omega
\end{array}
\end{equation}
where
\be
\begin{array}{rl}
\Omega & =\, \frac{i}{2}\left[ \left( |y|^2-|x|^2 \right) Q_P + \left ( x\bar y+ y\bar x \right )Q_S + M_{xy} \left( |y|^2+|x|^2 \right) Q_F \right]\\
& = \,  \frac{i}{2}\left[ \left( M_x |x|^2+ M_y|y|^2 \right) Q_F -  \tilde{N}_Y \left( |x|^2-|y|^2 \right) Q_Y + \left ( x\bar y+ y\bar x \right )Q_S \right]
\end{array}
\ee
As discussed in \cite{fi09}, one can relate the wavefunctions in the holomorphic gauge to the ones in the real gauge by the simple formula
\begin{equation}
\Psi^{\text{real}}\, =\, e^{i\Omega} \Psi^{\text{hol}}
\end{equation}
where $Q_S$, $Q_P$ and $Q_F$ act on the roots $E_\rho$  as in (\ref{srho}) and (\ref{qrho}), respectively. This only changes the scalar wavefunctions, and so in the real gauge our zero mode solution for the sectors $a^+_p$ still reads (\ref{totalwfsola}) but now with $\chi_{a_p^+}$ given by
\be
\chi_{a^+_{p}}^{\text{real}}\,=\, e^{-\frac{M_x - q_Y \tilde{N}_Y}{2}|x|^2 -\frac{M_y+q_Y\tilde{N}_Y}{2}|y|^2 - q_S^a\, {\rm Re } (x\bar{y})}   e^{ \lam_{a_p}  x(\bar{x} - \zeta_{a_p} \bar{y})} f_{a^+_{p}}\left ( y + \zeta_{a_p} x \right ) 
\ee
Notice that the exponential prefactor of the wavefunction allows to compensate the divergent behavior of the holomorphic function $f_{a_p^+}$ whenever  $\lam_{a_p} - M_{x} + q_Y\tilde{N}_Y < 0$ and $M_y +q_Y\tilde{N}_Y > 0$. One can check that the first condition is automatic, while the second needs to be imposed in order to obtain locally convergent zero modes in the sectors $a^+_p$. The same condition $M_y +q_Y\tilde{N}_Y >0$ forbids the presence of zero modes in the conjugate sector $a_p^-$,  at least at this local level.\footnote{Our ultra-local description is insensitive to global features of the GUT 4-cycle $S$, and in principle there could be zero modes in the sectors $a_p^\pm$ that locally seem divergent but are globally convergent, and the other way round, see \cite{palti12} for a discussion of this point. In the following we will assume that the three zero modes that represent the three families of ${\bf \bar{5}}$'s and ${\bf 10}$'s are local zero modes in the language of \cite{palti12}.} Similarly, for the sectors $b_q^+$ we have
\be
\chi_{b^+_q}^{\text{real}}=  e^{\frac{M_{x}+q_Y\tilde{N}_Y}{2}|x|^2 + \frac{M_y-q_Y\tilde{N}_Y}{2}|y|^2 - q_S^b\, {\rm Re } (x\bar{y})}  e^{ \lam_{b_q}  y(\bar{y} - \zeta_{b_q} \bar{x})}  f_{b_q^+}(x+ \zeta_{b_q}y) 
\ee
and so for having locally convergent zero modes we need to impose $M_x + q_Y \tilde{N}_Y<0$. Altogether we have that $M_x + q_Y \tilde{N}_Y < 0 < M_y + q_Y \tilde{N}_Y$, yields zero modes in the sectors $a_p^+$ and $b_q^+$, as assumed in the main text. For the remaining sectors we have
\begin{equation}\label{eqs}
\begin{array}{c}
\chi_{c^+_{r}}^{\text{real}} =  e^{\left (\frac{1}{2}\tilde N_Y+\zeta_c\right )|x|^2+\left (\lam_c - \zeta_c-\frac{1}{2}\tilde N_Yq_Y\right )|y|^2-\left (\frac{1}{2}q_S^c-\zeta_c+\lam_c \right )x\bar y-\left ( \frac{1}{2}q_S^c+\zeta_c\right )\bar x y}f_{c^+_{q}}\left ( x+\frac{\zeta_c}{\lam_c - \zeta_c} y\right ) \\
\chi_{\text{X,Y}^{+}}^{\text{real}}\, =\, e^{-\frac{5}{12}\lam \left ( |u|^2+ |v|^2\right )}f_{\text{X,Y}^{+}}\left (\bar u, v \right )\\
\chi_{\text{X,Y}^{-}}^{\text{real}}\, =\, e^{-\frac{5}{12}\lam \left ( |u|^2+ |v|^2\right )} f_{\text{X,Y}^{-}}\left ( u, \bar v \right )
\end{array}
\end{equation}


Following  section \ref{sec:treeYukawas}, let us take the basis of holomorphic functions 
$f_\rho^{i}(v_\rho)\! = \! \gamma_\rho^{i} m_{*}^{4-i}v_\rho^{3-i}$ for the zero modes in $\rho = a_p^+, b_q^+$. 
Here $i=1,2,3$ labels the three different families and  $\gamma_\rho^{i}$ is a normalization factor that we want to fix in 
order to have canonical kinetic terms for our matter fields. The normalization condition is given by
\bea
\label{normcond}
\langle \vec{\psi}_{\rho \, i}^{\rm real} | \vec{\psi}_{\rho \, j}^{\rm real} \rangle  & = &
  m_*^2 \int_S \tr \,( \vec{\psi}_{\rho \, i}^{\rm real} \cdot \vec{\psi}_{\rho\, j}^{\, \dag\, {\rm real}})\, {\rm d vol}_S\\ \nonumber
& = &  2 m_*^2\, ||\vec{v}_{\rho}||^2  \int_S  \chi_\rho^i (\chi_\rho^{j})^* \text{dvol}_S
\, = \, \d_{ij}
\eea 
where we have used that in our conventions $\tr (E_\rho E_\sigma^\dag) = 2 \delta_{\rho\sigma}$, and defined the vector 
\be
\vec{v}_{\rho}\, =\, 
\left(
\begin{array}{c}
\vspace*{.1cm}
 -\displaystyle{\frac{i\lam_{\bar{x}}}{m^2}} \\ \displaystyle{\frac{i\lam_{\bar{y}}}{m^2}} \\ 1
\end{array}
\right)_\rho
\label{ap:vec}
\ee
whose entries are given in table \ref{t3} or in eq.(\ref{simvecs}).

Notice that the normalization condition is automatically satisfied for $i\neq j$. Indeed, all the family dependence of our matter fields is encoded in the scalar wavefunction $\chi_\rho^i$, which in a real gauge can be written in the form
\begin{equation}
\chi_\rho^{i}\, =\, \gamma_\rho^{i}\, m_{*}^{4-i}e^{-(a|x|^2+b|y|^2+cx\bar y+dy\bar x)}(\alpha x+\beta y)^{3-i} \quad i=1,2,3
\end{equation}
for a certain set of constants $a,b,c,d,\a,\b$ that vary for each sector. One then has that the norm of this scalar wavefunction
\begin{equation}
({\chi}^i_\rho, \chi^j_\rho) \,\equiv\, m_*^2\int_S \chi_\rho^i (\chi_\rho^{j})^*  \text{dvol}_S
\label{ap:norm}
\end{equation}
vanishes identically for $i\neq j$, as the integrand is not invariant under the $U(1)$ action $(x,y) \raw e^{i\a} (x,y)$. For $i=j$ we can compute this norm by extending the integration domain from our local patch to $\mathbb C^2$
\begin{equation}
||\chi_\rho^{i}||^2 = |\g_\rho^i|^2 (m_*^2)^{5-i}\int_{\mathbb C^2}e^{-(2a|x|^2+2b|y|^2+2(c+d)\text{Re}[x\bar y])}\left ( |\alpha x+\beta y \right|^2 ) ^{3-i}\text{dvol}_{\mathbb C^2}
\label{intdomain}
\end{equation}
One then obtains\footnote{We have made use of the following formulas arising from (\ref{intdomain})
\begin{equation}
\nonumber
||\chi_\rho^{1}||^2 |\g_\rho^1|^{-2}\!= \!\frac{8\pi^2 m_*^{8}}{(4ab -(c+d)^2)^3} \left ( a^2\beta^4+b^2\alpha^4+\alpha^2\beta^2(c+d)^2+ab\alpha^2\beta^2-2a(c+d)\alpha\beta^3-2b(c+d)\alpha^3\beta)  \right )
\end{equation}
\begin{equation}\nonumber
||\chi_\rho^{2}||^2|\g_\rho^2|^{-2} =\frac{2\pi^2m_*^{6}}{(4ab -(c+d)^2)^2}\left ( a\beta^2+b\alpha^2-\alpha\beta(c+d) \right )
\end{equation}
\begin{equation}\nonumber
||\chi_\rho^{3}||^2 |\g_\rho^3|^{-2} = \frac{\pi^2 m_*^{4}}{(4ab -(c+d)^2)}
\end{equation}}
\bea
||\chi_{a_p}^i||^{-2} & = & - \frac{2\lam_{a_p}  - M_x + q_Y \tilde{N}_Y + (M_y +q_Y \tilde{N}_Y) \zeta_{a_p}^2}{m_*^2 |\g_{a_p}^i|^2\pi^2 (3-i)!} \left(\frac{M_y+q_Y \tilde{N}_Y}{m_*^2} \right)^{4-i}
\label{snormga} \\
||\chi_{b_q}^i||^{-2} & = &- \frac{2\lam_{b_q} + M_y -q_Y\tilde{N}_Y - (M_x + q_Y\tilde{N}_Y) \zeta_{b_q}^2}{m_*^2 |\g_{b_q}^i|^2 \pi^2 (3-i)!} \left(-\frac{M_x+q_Y\tilde{N}_Y}{m_*^2} \right)^{4-i} 
\label{snormgb} \\
||\chi_{c_r}||^{-2} & = & -\frac{1}{\pi^4m_*^4 |\g_{c_r}^i|^2} \! \left(\!\big(2\zeta_c+\tilde N_Yq_Y\big) \! \big(\tilde N_Yq_Y+2\zeta_c-2\lam_c\big) + (q_S+\lam_c)^2 \!\right)
\label{snormgc} \\
||\chi_{X,Y^\pm}||^{-2} & = & \left(\frac{5 \lam}{6 \pi  m_*^2 |\g_{X,Y^{\pm}}^i|}\right)^2
\label{snormgxy} 
\eea
where, as above, for the sectors $\rho = c_r$, $X,Y^\pm$ we have taken the holomorphic function $f_\rho$ to be a constant. On the other hand we have that
\bea
\label{norma}
||\vec v_a||^2 & = & m^{-4} \left (m^4 + \lam_a^2 (1+\zeta_a^2)\right ) \\
||\vec v_b||^2 & = & m^{-4} \left (m^4 + \lam_b^2 (1+\zeta_b^2) \right )\\
||\vec v_c||^2 & = & m^{-4} \left (m^4 + \zeta_c^2+(\zeta_c-\lam)^2 \right )
\eea

One finally finds the following normalization factors
\bea
|\gamma_{a_p}^i|^2 & = &-\frac{1}{2\pi^2} \left(\frac{m}{m_*}\right)^4 \frac{q_P^a(2\lam_a+q_P^a (1+ \zeta_a^2))}{m^4 + \lam_a^2 (1+\zeta_a^2)} \left( \frac{q_P^a}{m_*^2} \right )^{3-i} [{(3-i)!}]^{-1}
\label{normga} \\
|\gamma_{b_q}^i|^2 & = &- \frac{1}{2\pi^2} \left(\frac{m}{m_*}\right)^4 \frac{q_P^b(-2\lam_b+q_P^b (1+ \zeta_b^2))}{m^4 + \lam_b^2 (1+\zeta_b^2)} \left( -\frac{q_P^b}{m_*^2} \right )^{3-i} [{(3-i)!}]^{-1}
\label{normgb} \\
|\gamma_{c_r}|^2 & = & - \frac{1}{2\pi^2} \left(\frac{m}{m_*}\right)^4 \frac{ (2\zeta_c+q_P^c)(q_P^c+2\zeta_c-2\lam)+(q_S^c+\lam)^2}{m^4 + \zeta_c^2+(\zeta_c-\lam)^2}
\label{normgc} \\
|\gamma_{X,Y^\pm}|^2 & = & \frac{1}{2\pi^2} \left ( \frac{5\lam}{6 m_*^2} \right )^2 
\label{normgxy} 
\eea
where for simplicity we have set $M_{xy} = 0$.



\section{Fluxes in the SO(12) model and local chirality}
\label{ap:flux}

In section \ref{s:so12} we have made some choices in order to construct the $SO(12)$ model analyzed in the rest of the paper. More precisely, given the set of matter curves that describe a point of $SO(12)$ enhancement, we have chosen a worldvolume flux $F$  can reproduce the matter spectrum of the MSSM. In this appendix we discuss and motivate the restrictions that have been imposed on such worldvolume flux, with some emphasis on the piece that corresponds to the hypercharge flux $F_Y$ which plays a particularly important role in the computation of physical Yukawa couplings. 

Recall that  the chiral spectrum of an F-theory model depends on the integral of $F$ over each matter curve $\Sigma_\a$, while the choice (\ref{totalflux}) only describes $F$ in a local patch around the point of $SO(12)$ enhancement. Hence, in principle there could be no relation between (\ref{totalflux}) and the actual spectrum that each matter curve holds. However, there is a simple way to relate the two, namely by the concept of local chiral modes discussed in \cite{palti12}. Indeed, following \cite{palti12} let us define  local zero modes as those whose internal wavefunction peak in a neighborhood of the point of $SO(12)$ enhancement. We may then demand that the MSSM spectrum localized at the curves $a$, $b$ and $c$ arises from local zero modes. This latter assumption is in fact implicit in our discussion of zero modes, since otherwise computing the wavefunction normalization as in (\ref{intdomain}) would not be accurate. 

The concept of local zero modes is important in our discussion because by assumption these modes are affected by the local density of worldvolume flux $F$ at the point of $SO(12)$ enhancement. Hence, having the appropriate spectrum of chiral modes puts local restrictions on $F$. For instance, assuming local zero modes and following the computations of subsection \ref{zmnorm}, one arrives to the conclusion that the local flux densities in (\ref{f1vev}) must satisfy $M_x + q_Y \tilde{N}_Y< 0 < M_y + q_Y \tilde{N}_Y$ to have the proper spectrum on the curves $\Sigma_a$ and $\Sigma_b$. In the following we would like to show how also having the correct local spectrum for the curve $\Sigma_c$ implies adding the extra pieces of worldvolume flux (\ref{fluxcso12}) and (\ref{fluxhyp}).

Indeed, as pointed out in the text with only the flux (\ref{f1vev}) the curve $c$ remains locally non-chiral, while in principle we would like to have a chiral spectrum that hosts the $SU(5)$ Higgs ${\bf \bar{5}}$ and then, when including the hypercharge flux $F_Y$, a chiral spectrum for the Higgs doublets and a non-chiral one for the triplets \cite{bhv2}. To have a chiral spectrum at $\Sigma_c$ we simply need to add an extra piece to the previous worldvolume flux, and a natural candidate is the non-diagonal flux $\langle F_2 \rangle$ given in (\ref{fluxcso12}) because adding this flux does not change the number of families in the curves $a$ and $b$. This should be so because
\be
\langle F_2 \rangle|_{x=0}\,=\,\langle F_2 \rangle|_{y=0}\,=\, 0
\ee 
and so adding this flux does not change the integral of $\langle F \rangle$ over $\Sigma_a$ or $\Sigma_b$. On the other hand, the pull-back of $\langle F_2 \rangle$ on $x=y$ does not vanish, and so turning on $F_2$ should have an effect on the curve $c$. 

A simple way to check how worldvolume fluxes affect the spectrum of each of the curves is by considering the operator
\be
{\bf m}_\rho\, =\, 
\left(
\begin{array}{ccc}
F_{x\bar{x}} & F_{y\bar{x}} & - D_{\bar{x}} \Phi_{xy}^\dag \\
F_{x\bar{y}} & F_{y\bar{y}} & - D_{\bar{y}} \Phi_{xy}^\dag \\
D_x \Phi_{xy} & D_y \Phi_{xy} & 0
\end{array}
\right)
\label{Lapfer}
\ee
that arises when considering (\ref{DDform}), as seen in the previous appendix. Now, det ${\bf m}_\rho$ is the product of the three eigenvalues of the matrix ${\bf m}_\rho$, which from \cite{afim} we know enter into the algebra of creation and annihilation operators for each curve. In particular, it follows from \cite{afim} that for a curve with  a (locally) chiral spectrum we need the three of these eigenvalues to be non-vanishing, and so det ${\bf m}_\rho \neq 0$. 

In fact, one can check that det ${\bf m}_\rho$ matches the definition of local chirality index of \cite{palti12}. In the following we will analyze the value of this index for each sector of the $SO(12)$ model. From such analysis we will obtain a set of restrictions that must be imposed in the local worldvolume flux in order to have an acceptable local chiral spectrum.

\subsubsection*{sector $a^+$}

For this sector, in the presence of the fluxes (\ref{f1vev}) and (\ref{fluxcso12}) only we have that
\be
{\bf m_{a^+}} \,=\,
\left(
\begin{array}{ccc}
 M_x & -N_a & -im^2\\
 -N_a & M_y &  0 \\
 im^2 & 0 & 0
\end{array}
\right)
\label{fa}
\ee
%

Hence, before adding the hypercharge flux, we see that det ${\bf m_{a^+}}$ does not depend $N_a$, and that det ${\bf m_{a^+}} = 0$ for $M_y=0$. The first observation means that, as expected, $\langle F_2 \rangle$ does not affect the chiral spectrum of this curve, and the second means that if we switch off the flux $M_y$ then the curve $a$ locally becomes a non-chiral sector. From subsection \ref{zmnorm} we know that a positive number of chiral zero modes is obtained for $M_y >0$, which corresponds to det ${\bf m_{a^+}} < 0$. 

When adding the hypercharge flux, this flux matrix becomes
\be
{\bf m_{a^+}} \,=\, 
\left(
\begin{array}{ccc}
 M_x - q_Y \tilde{N}_Y & q_S^a & -im^2 \\
 q_S^a & M_y + q_Y \tilde{N}_Y &  0 \\
 im^2 & 0 & 0
\end{array}
\right)
\label{fah}
\ee
with $q_Y(a_1^+) = -\frac{1}{3}$ and $q_Y(a_2^+) = \frac{1}{2}$. The local chirality index then reads
\be
{\rm det\, } {\bf m_{a^+}}\, =\, - m^4 (M_y + q_Y \tilde{N}_Y)
\ee
and so it has the same sign as before as long as $M_y + q_Y \tilde{N}_Y > 0$. Note that the flux density $N_Y$ that also enters into the hypercharge flux does not modify the local chiral index. However, from the computations in appendix \ref{ap:wave} one can see that both flux densities $\tilde{N}_Y$ and $N_Y$ enter into the expression for the wavefunctions of this sector, the latter one through the quantity $q_S^a$.

\subsubsection*{sector $b^+$}

With only (\ref{f1vev}) and (\ref{fluxcso12}) we have
\be
{\bf m_{b^+}} \,=\,
\left(
\begin{array}{ccc}
 -M_x & N_b & 0\\
 N_b & -M_y &  im^2_{\Phi_y} \\
0 & -im_{\Phi_y}^2 & 0
\end{array}
\right)
\label{fb}
\ee
Again, det ${\bf m_{b^+}}$ does not depend on the flux density $N_b$, but rather on $M_x$. Recall that obtaining local zero modes in this sector corresponds to taking $M_x <0$, which again translates into det ${\bf m_{b^+}} < 0$. 

After adding the hypercharge flux we have
\be
{\bf m_{b^+}}\,=\, 
\left(
\begin{array}{ccc}
 -M_x - q_Y \tilde{N}_Y & q_S^b & 0\\
 q_S^b & -M_y + q_Y \tilde{N}_Y &  im^2 \\
0 & -im^2 & 0
\end{array}
\right)
\label{fbh}
\ee
where now $q_Y(b_1^+) = \frac{2}{3}$, $q_Y(b_2^+) = - \frac{1}{6}$ and $q_Y(b_3^+) = - 1$. The local index reads
\be
{\rm det\, } {\bf m_{b^+}}\, =\,  m^4 (M_x + q_Y \tilde{N}_Y)
\ee
and so we need $M_x + q_Y \tilde{N}_Y <0$. This condition and the one for the sector $a^+$ are automatically satisfied whenever $M_x < 0 < M_y$ and  $|M_x|, |M_y| > |\tilde{N}_Y|$.

\subsubsection*{sector $c^+$}

Finally, for this sector we have that
\be
{\bf m_{c^+}} \,=\,
\left(
\begin{array}{ccc}
 0 & N_a-N_b &  im^2 \\
 N_a-N_b & 0 &  - im^2 \\
 -im^2 & im ^2 & 0
\end{array}
\right)
\label{fc}
\ee
when only the fluxes (\ref{f1vev}) and (\ref{fluxcso12}) are present.  We see that
\be
\det {\bf m_{c^+}} \, =\, \, - m^4 2 (N_a - N_b)
\label{detfc}
\ee
and so, for $N_a \neq N_b$ we should have a chiral fermion localized in the curve $c$. As before, the condition to have a chiral spectrum of left-handed chiral multiplets arising from the sector $c^+$ (as opposed to arising from $c^-$) can be expressed as det ${\bf m_{c^+}} < 0$, which by (\ref{detfc}) is equivalent to $N_a > N_b$. 

Once introduced the hypercharge flux (\ref{fluxhyp}), we have instead
\be
{\bf m}_{c^+} \,=\, 
\left(
\begin{array}{ccc}
- q_Y \tilde{N}_Y   & q_S^c &  im^2 \\
 q_S^c & q_Y \tilde{N}_Y &  - im^2 \\
 -im^2 & im^2 & 0
\end{array}
\right)
\label{fch}
\ee
and so the local chirality index becomes
\be
\det {\bf m_{c^+}} \, =\, \,-  m^4 2 q_S^c\, =\,-  m^4 2 (N_a - N_b + q_Y N_Y)
\label{detfc2}
\ee
Since $q_Y(c_1^+) = -\frac{1}{3}$ and $q_Y(c_2^+) = \frac{1}{2}$, one may achieve a locally non-chiral sector of Higgs triplets by simply imposing $N_Y = 3 (N_a - N_b)$. This constraint and the previous condition $N_a > N_b$ imply that $q_S^c > 0$ for the sector $c_2^+$, which means that there is indeed a Higgs doublet in this sector. 


\section{Non-perturbative superpotential and D-term}
\label{ap:supo}

The key ingredient that allows to obtain Yukawa matrices of rank higher than one is the deformed 7-brane superpotential (\ref{supo}) As pointed out in \cite{mm09} and further discussed in \cite{bdkkm10,dm10,afim}, this superpotential arises once that one takes into account certain non-perturbative effects present in type IIB/F-theory compactifications. The purpose of this appendix is to recall the derivation of eq.(\ref{supo}), basically following \cite{mm09,afim}, to which we refer for further details. In addition and based on the results of \cite{dm10} we will show that the same class of non-perturbative effects leave unchanged the expression for the D-term. 

\subsection{Non-perturbative superpotential}

As reviewed in section \ref{s:reviewF}, the basic setup of type IIB/F-theory GUT compactifications assumes that the GUT gauge degrees of freedom arise from a 4-cycle $S_{\rm GUT}$, a divisor of a compact threefold base $B$ that is wrapped by a stack of 7-branes. This divisor intersects another set of divisors $S_i$ also wrapped by 7-branes, and it is at the intersection curves $\Sigma_i$ where the chiral matter multiplets arise from. Near the Yukawa point where three of these matter curves meet one can understand this configuration in terms of a single stack of 7-branes wrapped on a divisor $S$ and with a higher rank gauge gauge group like $SO(12)$, $E_6$, $E_7$ or $E_8$. This last divisor and gauge group are the ones that enter into expression for the superpotential (\ref{supo7}) and D-term (\ref{FI7}), from which we can derive the zero mode equations of motion and the Yukawa couplings. The fact that a single 4-cycle $S$ describes a set of intersecting divisors is encoded in the non-trivial profile of $\langle \Phi \rangle$. Finally, as pointed out in section \ref{s:nplocal}, $S$ and $S_{\rm GUT}$ need not be the same. In particular, in the type IIB description of the $SO(12)$ model $S$ should be associated with the divisor of an O7-plane. 

In addition to all these divisors, there will be in general other set of divisors also wrapped by branes which may not be part of the MSSM sector, but instead the source of non-perturbative effects. Typical examples are 7-branes that give rise to a gauge hidden sector of the theory that undergoes a gaugino condensate, or Euclidean BPS 3-branes with the appropriate structure of zero modes to contribute the the superpotential of the 4d effective theory. In the following we will consider that one of such objects exists and that is wrapping a divisor $S_{\rm np}$, and describe how the presence of $S_{\rm np}$ affects the superpotential (\ref{supo7}) for the gauge theory on $S$.

Perhaps the most intuitive way to explain how the presence of $S_{\rm np}$ modifies the 7-brane superpotential (\ref{supo7}) is by considering the case of a gaugino condensate on $S_{\rm np}$. There we know that a non-perturbative superpotential of the form
\be
W_{\rm np}\, =\, \mu^3 e^{-f_{\rm np}/r}
\label{4dsuponp}
\ee
is generated at the level of the 4d effective theory. Here $f_{\rm np}$ is the gauge kinetic function of the stack of 7-branes on  $S_{\rm np}$ and $r$ is the rank of their gauge group. Finally, $\mu \sim m_*$ is the UV scale at which $f_{\rm np}$ is defined. This expression and all those below are also valid for the case where $S_{\rm np}$ hosts an Euclidean 3-brane, by simply replacing $f_{\rm np}$ by the complexified action of the instanton and setting $r=1$.

In general, $f_{\rm np}$ will be a holomorphic function of the 4d chiral multiplets of the theory, which arise either from the bulk or from the 7-brane sectors of the compactification
\be
f_{{\rm np}}\, =\, T_{\rm np} + \, 
f^{\rm 1-loop}_{{\rm np}}
\ee
where the first term is the gauge kinetic function $f_{7_{\rm np}}$ computed at tree-level, given by  the complexified K\"ahler modulus  $T_{\rm np} = {\rm Vol\, }({S_{\rm np}}) + i \int_{S_{\rm np}} C_4$ that corresponds to $S_{\rm np}$. The second term arises from threshold effects, and is a holomorphic function of the bulk/closed string fields and of the 7-brane fields.\footnote{In the case of instanton corrections to the superpotential, this term is identified with a moduli-dependent instanton prefactor. See \cite{cdhm12} for a recent discussion in the context of F-theory GUTs.} More precisely, if we consider the presence of a 7-brane in the divisor $S$ we have that \cite{mm09}
\be
f^{\rm 1-loop}_{{\rm np}}\, =\, - r\, {\rm log \, } \ca \, - \frac{1}{8\pi^2} \int_S \str ({\rm log\, } h\, F \wedge F)
\label{f1loopa}
\ee
where ${\cal A}$ is a function of the bulk/closed string fields, and all the dependence of the 7-brane fields enters into the second term of this expression. There the key quantity is $h$, which is the holomorphic divisor function of the 4-cycle $S_{\rm np} = \{h = 0\}$ sourcing the non-perturbative effect. While ${\rm log\, } h$ is a scalar bulk quantity, when plugged into the expression (\ref{f1loopa}) one should follow the prescription of \cite{Myers} and consider its non-Abelian pull-back into $S$. That is
\be
{\rm log\, } h  \, =\, {\rm log\, } h|_{S} + m_{st}^{-2}\, \Phi_{xy} [\p_z {\rm log\, } h]_S  +\, m_{st}^{-4}\, \Phi_{xy}^2 [\p_z^2 {\rm log\, } h]_S  + \dots  
\label{napb}
\ee
where as in the main text $z$ is the holomorphic coordinate transverse to $S$, $\Phi_{xy}$ the 7-brane position field related to it and $m_{st}^{-2} = 2\pi \alpha'$. Plugging this expression back to (\ref{f1loopa}) we obtain
\be
- f^{\rm 1-loop}_{7_{\rm np}}\, =\, r\, {\rm log \, } \ca \, + N_{\rm D3}\, {\rm log\, } h_0 +  \frac{1}{8\pi^2} \sum_n m_{st}^{-2n} \int_S [\p_z^n {\rm log\, } (h/h_0)]_{S} \, \str\, (\Phi_{xy}^n\, F \wedge F)
\label{f1loopb}
\ee
where $h_0 = \int_S h$ the mean value of $h$ in $S$ and $N_{\rm D3} = (8\pi^2)^{-1} \int_S \tr (F \wedge F)\in \IN$ the total D3-brane charge induced on $S$ by the presence of the magnetic field $F$. Finally, inserting (\ref{f1loopb}) into (\ref{4dsuponp}) we obtain
\bea
W_{\rm np} & = & \mu^3   (\ca\, e^{-T_{\rm np}/r} h_0^{N_{\rm D3}/r} )\, {\rm exp } \left[\sum_n  \frac{m_{st}^{-2n}}{8\pi^2 r} \int_S [\p_z{\rm log\, } (h/h_0)]_{S} \, \tr\, (\Phi_{xy}^n\, F \wedge F) \right] \\
& = & \mu^3 \ca\, e^{-T_{\rm np}/r} h_0^{N_{\rm D3}/r} \left(1 + \sum_n  \frac{m_{st}^{-2n}}{8\pi^2 r} \int_S [\p_z{\rm log\, } (h/h_0)]_{S} \, \str\, (\Phi_{xy}^n\, F \wedge F) +\dots \right)\nonumber
\eea
Hence, up to a constant term we have that
\be
W_{\rm np}\, =\, m_*^4 \frac{\eps}{2} \int_S \th_n  \str\, (\Phi_{xy}^n\, F \wedge F)
\label{ap:npsupo}
\ee
where we have defined
\bea
\eps & = & \CA\, e^{-T_{\rm np}/r} h_0^{N_{\rm D3}/r} \\
\theta_n & = & \frac{g_s^{-n/2} \mu^3/ r}{(2\pi)^{2+ \frac{3n}{2}} m_*^{4+2n}}\, [\p_z^n\, {\rm log\, } (h/h_0)]_{z=0} 
\label{ap:deftheta}
\eea
where we have used the relation $m_{st}^4 = g_s (2\pi)^3 m_*^4$ derived in \cite{afim}.\footnote{In fact, a factor of $g_s$ was missed in \cite{afim}, which we incorporate here.} Clearly, these expressions match the definitions of the text for the case $r=1$.

\subsection{Non-perturbative D-term}

Interestingly, one can derive the same result (\ref{ap:npsupo}) by replacing the presence of the non-perturbative sector in $S_{\rm np}$ by a deformation of the compactification background, or in other words by `back-reacting' the non-perturbative effect. This idea was put forward in \cite{km07} and further analyzed for gaugino condensate type IIB vacua in  \cite{bdkkm10,dm10}. As discussed in \cite{afim} considering this deformed background simplifies the computation of non-perturbative 7-brane Yukawa couplings. We will not follow this approach to rederive in detail the non-perturbative superpotential (\ref{supo}), referring the reader to \cite{mm09,afim} for the appropriate discussion, but will make use of it to derive the form of the D-term in the presence of the same non-perturbative effects. 

Indeed, in the scheme of \cite{km07} one may take the deformed background obtained from back-reacting a gaugino condensate and  compute the D-term in such background. More precisely, in the language of generalized geometry, one needs to find the polyforms or pure spinors that define such background, and them derive the D-term (or F-term) equations from them. For the case of a gaugino condensate arising from a D7-brane, such pure spinor analysis was carried out in \cite{dm10}. In particular, it was found that the pure spinor of interest to compute the D-term is given by 
\be
e^{2A} \im \Psi_1 \, =\, J + \frac{i}{8} \Omega \wedge \bar\Omega + \frac{i}{8} \iota_\beta (\Omega \wedge \bar\Omega) + \dots
\label{puresp}
\ee
which is basically eq.(5.15) of \cite{dm10}. Here $\beta$ is a real bivector 
\be
\b = \b^{(2,0)} + \b^{(0,2)} = \b^{ij} \p_i \wedge \p_j + c.c.
\ee
whose (2,0)-component is holomorphic, and specifies the so-called $\beta$-deformation of the background. In a given patch of the compactification $\beta$ specifies (up to a constant) a holomorphic function $\chi_0$, since
\be
\iota_\beta\Omega = \oh \b^{ij} \Omega_{ijk} dz^k = \eps\, d\chi_0
\ee
$\chi_0$, in turn, specifies the deformations of the superpotential of the form (\ref{ap:npsupo}). Indeed, we can define 
\be
\th_n (x,y) \, =\, m^{-2n} [\p_z^n \chi_0]_{z=0} 
\ee
so that each of these terms corresponds to a superpotential deformation of the form \cite{Martucci06}
\be
\frac{\eps}{2} \int_S \theta_n \,  \str \left( \Phi_{xy}^n F \wedge F\right)
\ee
where the 4-cycle $S$ is locally defined by the equation $\{z=0\}$. Notice that $\chi_0 \propto {\rm log\, } (h/h_0)|_S$, and so the case $\th_0 \neq 0$ corresponds to 
\be
\b^{yz}|_{S} = \eps  \p_x \th_0 \quad\quad \b^{zx}|_{S} = \eps \p_y \th_0 \quad\quad \b^{xy}|_{S} = 0
\ee

Now, in order to obtain the D-term one needs to compute
\be
P[e^{2A} \im \Psi_1]_S \wedge e^F
\label{presc}
\ee
where $P[\a]_S$ is the non-Abelian pull-back of a $p$-from $\a$ into the 4-cycle $S$. This means in particular that if $\a$ has two legs transverse to $S$, we should replace them with $[\Phi_{xy}, {\Phi}_{xy}^{\dagger}]$ (see section 2 of \cite{mms10} for more details). Using this prescription, it is easy to see that the first two terms of the rhs of (\ref{puresp}) give the usual D-term (\ref{FI7}), with $\omega = P[J]_S$. The third term of (\ref{puresp}) would in principle give a correction to such D-term expression, however one can check that it is of order $\CO(\eps^2)$.\footnote{Indeed, if we apply the prescription (\ref{presc}) to this $(1,3) + (3,1)$-form we always get an expression that contains either $F^{(2,0)}$ of $D_{x,y} \Phi^\dag$, which vanish at 0$^{th}$ order in $\eps$.}

To summarize, from the results of \cite{dm10} one can conclude that the expression for the D-term (\ref{FI7}) remains unchanged after the non-perturbative effect has been taken into account. As shown in the main text this does not mean that the fluctuations satisfy exactly the same D-term equation before and after the non-perturbative effect. Indeed, while the differential equation is given by (\ref{Dterm}) in both cases, the F-term equations shift the values of $\langle\Phi\rangle$, $\langle A\rangle$, and their conjugates, and so the actual D-term equations to be satisfied by the wavefunctions change.


\section{Non-perturbative Yukawas and residues}
\label{ap:res}

As pointed out in \cite{cchv09}, and proved more generally in \cite{cchv10},
one may compute the holomorphic piece of the Yukawa couplings, that is the one that appears in the 4d superpotential, by means of an elegant residue 
formula. The purpose of this section is to explain how the analysis of
\cite{cchv09} can be generalized to derive a residue formula for the general superpotential (\ref{supo}). 
We will focus on the case where only the holomorphic functions $\theta_0$ and $\theta_2$ do not vanish, and show that the resulting residue formula gives Yukawa couplings 
in agreement with those obtained by directly computing wavefunctions overlaps, as done in the main text.

\subsection{Non-perturbative F-terms and Yukawas}
\label{ap:res1}

One important result for deriving the residue formula for Yukawa couplings is to realize that the F-term equations are invariant under complexified gauge transformations, and that the holomorphic piece of the Yukawa couplings should be invariant under them \cite{cchv09}. In the following we will generalize these concepts for the more involved superpotential (\ref{supo}), setting the stage for the residue formula discussed below. For simplicity we will focus in the case where only one term $\theta_n$ in (\ref{supo}) is non-vanishing, discussing in detail the cases $n=0$ and $n=1$, but similar results can be derived for other values of $n$.\footnote{As a byproduct of our computations, we will show that the the only fluctuations that enter into the Yukawa couplings are $(a, \vphi)$, as opposed to their complex conjugates $(a^\dag, \vphi^\dag)$. This assumption, taken all over the main text and in \cite{afim}, is obviously true for the tree level superpotential (\ref{supo7}), but not for the non-perturbative one (\ref{supo}), as the fluctuations $a^\dag$ appear explicitly in the latter.}

\subsubsection*{The case $\theta_0$}

Let us consider the superpotential
\be\label{eq:sup}
W=W_0+W_1=\int_S \tr(\Phi\, \w\, F)+\frac{\epsilon}{2}\int_S\theta_0\tr(F\, \w\, F)
\ee
that is, eq.(\ref{supo}) with only $\theta_0 \neq 0$. From here we get the following set of equations of motion
\bea
F^{(0,2)}&=&0\\
\bar{\partial}_A\Phi+\epsilon\partial\theta_0 \,\w\, F^{(1,1)}&=&0
\eea
that should be complemented with the D-term equation (\ref{Dterm}). This yields the following BPS equations for the background 
\bea\label{eq:ba1}
\langle F^{(0,2)}\rangle &=&0\\\label{eq:ba2}
\pa\lp+\epsilon\partial\theta_0\,\w\, \langle F^{(1,1)}\rangle&=&0
\eea
and, expanding to first order in the fluctuation fields $(\vphi, a)$, the zero mode equations
\bea\label{eq:fluc1}
\pa a&=&0\\\label{eq:fluc3}
\pa\varphi-i[a,\lp]+\epsilon\partial\theta_0\,\w\,(\partial_{\langle A\rangle}a+\bar{\partial}_{\langle A\rangle}a^\dag)&=&0
\eea
while the D-term gives
\be
\omega \wedge (\ph a + \pa a^\dag) - \frac{1}{2} \left( [\langle \bar{\Phi} \rangle, \varphi] + [\vphi^\dag,\langle \Phi \rangle]\right)\, =\, 0
\label{eq:fluc2}
\ee
One can check that the equations (\ref{eq:fluc1}), (\ref{eq:fluc3}) and (\ref{eq:fluc2}) are invariant under the infinitesimal gauge transformations
\bea
a&\rightarrow& a+\bar{\partial}_{\langle A \rangle}\chi\\
a^\dag &\rightarrow& a^\dag + \partial_{\langle A \rangle}\chi\\
\varphi&\rightarrow& \varphi-i\left [\lp,\chi\,\right ]\\
\varphi^\dag &\rightarrow& \varphi^\dag + i\left [\langle \bar{\Phi} \rangle,\chi\,\right ]
\eea
where $\chi=\chi^{\dagger}$ is a real gauge transformation parameter. However, if one is only interested in the F-term equations (\ref{eq:fluc1}) and (\ref{eq:fluc3}), then $\chi$ can be arbitrary. This means that the equations are invariant under the complexified gauge group, as can be directly seen from the superpotential (\ref{eq:sup}).

Following \cite{cchv09}, one finds that the general solution of (\ref{eq:fluc1}) is 
\be
a=\pa \xi
\label{gena}
\ee
where $\xi$ is a scalar in the adjoint representation of complexified algebra. Using this result as well as the equation (\ref{eq:ba2}) one can rewrite (\ref{eq:fluc3}) as follows
\be
\label{eq:pri}\pa\left (\varphi+i[\lp,\xi]-\epsilon\partial\theta_0\,\w\,(a^\dag-\ph\xi)\right )=0
\ee
from which we get the general solution
\be\label{eq:phi}
\varphi=h-i[\lp,\xi]+\epsilon\partial\theta_0\,\w\,(a^\dag-\ph\xi)
\ee
where $h$ is an adjoint $(2,0)$-form such that $\pa h=0$.

As explained in the main text, the Yukawa couplings arise when we expand (\ref{eq:sup}) to cubic order in fluctuations, plug in the solution to the equations of motion and perform the integration. We have two contributions to the Yukawa couplings, $Y=Y_0+Y_1$, coming from $W_0$ and $W_1$, respectively. The piece coming from $W_0$ is
\bea
Y_0&=&-i\int_S \tr \, (\varphi\, \w\, a\, \w\, a)\\\nonumber
&=&-i\int_S \tr \left ( (h-i[\lp,\xi] - \epsilon\, \partial\theta_0\,\w\,\ph\xi)\,\w\, a\, \w\, a\, \right )-i\epsilon\int_S\tr\left ( \partial\theta_0\,\w\, a^\dag\,\w\,a\,\w\,a \right )
\label{Y00}
\eea
where we have used (\ref{eq:phi}). The contribution coming from $W_1$ is
\bea
Y_1=\frac{\epsilon}{2}\int_S\theta_0\tr(F\,\w\,F)_{(3)}
\eea
where the subscript $(3)$ means we only take the cubic terms in fluctuations. One can see that this expression amounts to 
\be
Y_1=-i\epsilon\int_S\theta_0  \ph\left (\tr\, (a^\dag \,\w\, a \,\w\,a)\right )-i\epsilon\int_S\theta_0\pa\left (\tr\, (a \,\w\, a^\dag \,\w\, a^\dag)\right )
\ee
Since when acting on a gauge invariant object, the operators $\ph$ and $\pa$ reduce to $\partial$ and $\bar \partial$, respectively, we can drop the subscript $\langle A\rangle $ in the last expression. Using that  $\theta_0$ is holomorphic we then obtain
\be \label{eq:Y1}
Y_1 \, =\, -i\epsilon \int_S\theta_0\partial \left (\tr\, (a^\dag \,\w\, a \,\w\,a)\right ) - i\epsilon\int_S\bar\partial\left (\theta_0\tr\, (a \,\w\, a^\dag \,\w\, a^\dag)\right )
\ee
The second term in (\ref{eq:Y1}) is a boundary term so it vanishes by compactness of $S$. Adding the first term to the last term of (\ref{Y00}) we have
\be
-i\epsilon\int_S\tr\left ( \partial\theta_0\,\w\, a^\dag\,\w\,a\,\w\,a \right )-i\epsilon \int_S\theta_0\partial \left (\tr\, (a^\dag \,\w\, a \,\w\,a)\right )
\, =\, -i\epsilon\int_S\partial\left (\tr \left (\theta_0\, a^\dag\,\w\,a\,\w\,a  \right )\right )
\ee
which also vanishes upon integration. As a result, $a^\dag$ does not enter into the expression for the Yukawa coupling, which can be expressed as
\be
Y\, = \, -i\int_S \tr \left( \left[h-i[\lp,\xi] - \epsilon\, \partial\theta_0\,\w\,\ph\xi\right]\w\, \pa \xi\, \w\, \pa \xi \, \right ) 
\ee
To sum up, the expression for the Yukawa couplings for the case $\theta_0 \neq 0$ in this case is given by 
\be
Y \, =\, -i \int_S \tr \left(\varphi\, \w\, a\, \w\, a \right) 
\ee
where $a$ is giving by (\ref{gena}) and $\vphi$ by
\be
\varphi=h-i[\lp,\xi]-\epsilon\partial\theta_0\,\w \ph\xi
\ee
which amounts to (\ref{eq:phi}) with the fluctuation $a^\dag$ set to zero. In fact, for the purposes of computing Yukawa couplings to order $\CO(\eps)$ one may forget about the presence of $a^\dag$ in the superpotential, which indeed has been the working assumption of the main text.

\subsubsection*{The case $\theta_1$}

As a slightly more involved case let us consider the superpotential
\be
W=W_0 +W_1 =\int_S \tr\left ( \Phi_{xy}F \right )\,\w\,dx\,\w\,dy+\frac{\epsilon}{2}\int_S\theta_1\,\tr\left (\Phi_{xy}F\,\w\,F\right )
\ee
that corresponds to the case analyzed in \cite{afim}. As in there, we have the following set of equations of motion\footnote{In our conventions $\{A,B\}\equiv\frac{1}{2}(AB+BA)$.}
\bea
F\,\w\,dx\,\w\,dy+\epsilon\theta F \,\w\,F&=&0\\
\bar \partial_A\Phi+\epsilon\left [ \left \{ F^{(2,0)},\bar \partial_A(\theta\Phi_{xy}) \right \}+\left \{ F^{(1,1)},\partial_A(\theta\Phi_{xy}) \right \} \right ]&=&0\\
\label{eq:sobra}\left \{ F^{(0,2)},\partial_A(\theta\Phi_{xy}) \right \}+\left \{ F^{(1,1)},\bar\partial_A(\theta\Phi_{xy}) \right \}&=&0
\eea
We now expand both 7-brane fields $\Phi$ and $A$ in the parameter $\eps$, as in (\ref{expeps}), in order to write down the equations of motion for the fluctuations $\varphi_{xy}^{(1)}$, $a^{(1)}$ and $a^{\dag\, (1)}$ as 
\bea\nonumber
\pa a^{(1)}\,\w\, dx\,\w\, dy-i[\langle A \rangle^{(1)}, a^{(0)}]\,\w\,dx\,\w\,dy+\theta\left \{\langle F\rangle ,\pa a^\dag+\ph a \right \}^{(0)}&=&0\\
\pa \varphi^{(1)}-i[a^{(1)},\langle\Phi\rangle^{(0)}]-i[\langle A\rangle^{(1)},\varphi^{(0)}]+\left ( \{\ph a+\pa a^\dag\,, \ph(\theta\langle\Phi_{xy})\rangle \}\right.&&\\
\left.+\{\langle F\rangle ,\ph(\theta\Phi_{xy})-i[a^\dag,\theta\langle F\rangle] \} \right )^{(0)} &=&0
\nonumber
\eea
where we have made use of the zeroth order equations for the background. Following \cite{afim} one can rewrite the above equations as
\be
\label{eq:pre}
\bar \partial_{\mathbf A}\vec \psi^{(1)}=-\Theta(\tilde {\mathbf K}+i\tilde {\mathbf A})^{(0)}\Theta\vec \psi^{(0)}-{\mathbf S}\vec \rho^{(0)}
\ee
where $\vec \psi$ is defined as in (\ref{npzmexp}), and $\Theta$, ${\mathbf K}$ and ${\mathbf A}$ are defined by eqs.(3.19), (3.22) and (4.10) of \cite{afim}. Finally we have that ${\mathbf S}$ and $\vec \rho$ correspond to
\be
{\mathbf S}=\left( \begin{array}{ccc}
\{F^{\theta}_{y\bar z},D_{\bar y\cdot}\}-\theta\{F_{y\bar y},D_{\bar z}\cdot\} & \theta\{F_{x\bar y},D_{\bar z}\cdot\}-\{F^{\theta}_{x\bar z},D_{\bar y}\cdot\} & 0 \\
\theta\{F_{y\bar x},D_{\bar z}\cdot\}-\{F^{\theta}_{y\bar z},D_{\bar x}\cdot\} & \{F^{\theta}_{x\bar z},D_{\bar x}\cdot\}-\theta\{F_{x\bar x},D_{\bar z}\cdot\} & 0 \\
\theta\{F_{y\bar y},D_{\bar x}\cdot\}-\theta\{F_{y\bar x},D_{\bar y}\cdot\} & \theta\{F_{x\bar x},D_{\bar y}\cdot\}-\theta\{F_{x\bar y},D_{\bar x}\cdot\} & 0 \end{array} \right)\quad\vec \rho=\left ( \begin{array}{c}
a_x^\dag\\
a_y^\dag \\
0\end{array}\right )
\ee

Notice that eq.(\ref{eq:pre}) corresponds to the zero mode equation (4.8), (4.9) of \cite{afim}, except for the extra term proportional to $\vec{\rho}$. As we will see in the following, in the computation of Yukawa couplings all terms including $a^\dag$ can be ignored, just like for the $\th_0$ case. Hence, in the equation of motion above we can ignore $\rho$, and so the zero mode analysis reduces to the one in \cite{afim}. 

Indeed, let us assume that we had found a wavefunction vector $\vec \lambda$ such that
\be
\vec\lambda^{(0)}=\vec\psi^{(0)} \quad \quad \quad
\bar \partial_{\mathbf A}\vec\lambda^{(1)}=-\Theta(\tilde {\mathbf K} + i \tilde {\mathbf A})\Theta \vec \psi^{(0)} 
\ee
which corresponds to solving the zero mode equation in \cite{afim}. Since we can write ${\mathbf S}=\bar \partial_{\mathbf A}{\mathbf H}$ with
\be
{\mathbf H}=\left( \begin{array}{ccc}
\theta\{F_{y\bar x},\cdot\} & -\theta\{F_{x\bar x},\cdot\} & 0 \\
\theta\{F_{y\bar y},\cdot\} & -\theta\{F_{x\bar y},\cdot\} & 0 \\
\{F^{\theta}_{y\bar z},\cdot\} & -\{F^{\theta}_{x\bar z},\cdot\} & 0 \end{array} \right)
\ee
we have that a general solution to (\ref{eq:pre}) is given by
\be
\vec\psi=\vec\lambda-\epsilon{\mathbf H}\vec\rho
\ee
which is analogous to the solution (\ref{eq:phi}) for the case $\theta_0$, and we can basically repeat the steps of this case to show that $\vec \rho$ only appears in the trilinear couplings through total derivatives. On the one hand, we have a trilinear term involving $\vec \rho$ that arises from $W_0$ 
\be
\label{eq:Y0}Y_0 \supset i\epsilon\int_S\tr\left (\varphi\,\w\,[\varpi, a]+\{a^\dag,\ph (\theta\langle\Phi_{xy}\rangle) \}\,\w\,a \,\w\,a\right )^{(0)}
\ee
where we have defined $\varpi\equiv(H\rho)_1d\bar x+(H\rho)_2d\bar y=\{a^\dag,F^{(1,1)} \}_{xy}$. On the other hand, $W_1$ contains the following terms involving $a^\dag$
\bea
\label{eq:YY1}Y_1 &\supset& \frac{-i\epsilon}{2}\int_S\tr \Big (\theta\,\varphi_{xy}\left ( \left \{\langle F\rangle ,\left [ a^\dag ,a \right] \right \}+i\left [\ph a ,\pa a^\dag \right ]+i\pa a^\dag \,\w\,\pa a^\dag \right ) \\
\nonumber&& +\, \theta\lpc \left (  \left \{\ph a^\dag, a\,\w\,a \right \}+\left \{\ph a , \left [ a^\dag, a \right ]\right \} +\left \{\pa a^\dag ,\left [ a^\dag, a \right ] \right \}\right )  \Big )^{(0)}
\eea
Putting all these terms together one finally obtains
\be
-\frac{i\epsilon}{2}\!\int_S \!
\tr\left(\bar\partial\left([a\,,\theta\lpc a^\dag]\w\,a^\dag+i\theta\varphi_{xy}\left (a^\dag \w\,\bar\partial a^\dag
+[ a^\dag ,\partial a ]\right)  \right) 
+2\partial\left (\left\{a^\dag,\theta\lpc\right\} \w \, a \, \w  \, a  \right)\right)^{\! (0)}
\ee
which vanishes upon integration. Therefore, one can set $a^\dag=0$ from the very beginning without loss of generality, and recover the analysis performed in \cite{afim}.

\subsection{Residue formula}
\label{ap:formu}

In this section we will derive the residue formula for the holomorphic Yukawas induced by the non-perturbative
terms in the general superpotential (\ref{supo}). We will treat separately the dependence on $\theta_0$
and $\theta_2$, which are the parameters of interest for the $SO(12)$ model described in section 3.

\subsubsection*{Yukawas due to $\theta_0$}

In section \ref{ap:res1} we have seen that the Yukawa couplings including corrections are given by
\bea
Y &=& -i m_* \int_S \Tr(\varphi \wedge a \wedge a) \nonumber \\
& = & m_* f_{abc}\int_S\left[ \varphi_c \wedge a_a \wedge a_b +  \varphi_b \wedge a_c \wedge a_a +  \varphi_a \wedge a_b \wedge a_c \right] 
\label{yuka}
\eea
These couplings can be conveniently rewritten by using the F-terms found after shifting the background as explained in section \ref{ss:eom}.
In particular, eq.~(\ref{ft02}), see also eq.~(\ref{ft04}), tells us that
\be
\bar\p\varphi_\rho  =  -im^2 q_\Phi(\rho) a_\rho dx \wedge dy - \epsilon \partial \theta_0 \wedge \partial a_\rho + \CO(\eps^2)
\label{af1}
\ee
On the other hand,  eq.~(\ref{ft01}) amounts to 
\be
a_\rho  =  \bar\p \xi_{\rho}  + \CO(\eps^2)
\label{af2}
\ee
where $\xi_\rho$ is a function with properties to be discussed shortly.
Clearly (\ref{af1}) can then be integrated to obtain
\be
\varphi_\rho = h_\rho dx \wedge dy -im^2 q_\Phi(\rho) \xi_\rho dx \wedge dy - \epsilon \partial \theta_0 \wedge \partial \xi_\rho 
\label{af3}
\ee
where $h_\rho$ is a holomorphic function.

Substituting the F-terms in the coupling $Y$ we find
\be
Y
=  m_* f_{abc}\!\! \int_S \!\! \left\{ h_c a_a \wedge a_b \wedge dx \wedge dy +  \bar\p\left[(\varphi_b \wedge a_a -  \varphi_a \wedge a_b) \xi_c \right]
-\eps\partial\left[ \theta_0 \p\left(a_a \wedge a_b \, \xi_c\right)\right]
\right\}
\label{yukb}
\ee
to $\CO(\eps^2)$. Upon integration the total derivatives lead to boundary terms that vanish because, as shown in section \ref{ss:n0}, the zero modes 
for $a_\rho$ and $\varphi_\rho$ are localized on the curve $\Sigma_\rho$. Taking into account that $h_c$ is holomorphic, and using (\ref{af2}) once more, then gives
\be
Y
=  m_* f_{abc}\int_S \bar\p\left( h_c \xi_a \bar\p \xi_b\right) \wedge dx \wedge dy 
\label{yukc}
\ee
Before further integrating we will first study the functions $\xi_\rho$ in more detail.

The idea is to determine $\xi_\rho$ recursively from (\ref{af3}). We know that $\varphi_\rho$ has an expansion (\ref{chirexp}) in 
powers of $\eps$, and in fact the corrections $\varphi_\rho^{(1)}$ are explicitly given in section \ref{ss:n0}. 
It is natural to expand $\xi_\rho$ in a similar way, namely
\be
\xi_\rho  =  \xi_\rho^{(0)} + \eps \xi_\rho^{(1)} + \CO(\eps^2) 
\label{hxiexp} 
\ee 
{}From (\ref{af3}) it then follows that
\begin{subequations}
\label{xi01}
\begin{align}
\xi_\rho^{(0)} \, = \ & \frac{i}{m^2 q_\Phi(\rho)} (\varphi_\rho^{(0)} - h_\rho)
\label{xi0}\\[2mm]
\xi_\rho^{(1)} \, = \  & \frac{i}{m^2 q_\Phi(\rho)} \left[\varphi_\rho^{(1)} 
+ \p_x \theta_0 \p_y \xi_\rho^{(0)} - \p_y \theta_0 \p_x \xi_\rho^{(0)} \right]
\label{xi1}
\end{align}
\end{subequations}
The functions $\xi_\rho$ are requested to be regular at the locus $q_\Phi(\rho)=0$. Recall that \mbox{$q_\Phi(b^+)=y$} whereas 
\mbox{$q_\Phi(a^+)=-x$}. At zeroth order, to avoid poles it must be that
$h_\rho= \varphi_\rho^{(0)}\lvert_{q_\Phi=0}$. For example, $h_a= m_* f_i(y)$ and $h_b= m_* g_j(x)$.
We will later study the regularity of the corrections $\xi_\rho^{(1)}$. 

To arrive at the residue formula it is important to realize that after integrating the total derivative in (\ref{yukc})
the terms involving $\varphi_\rho$ in $\xi_\rho$ will not contribute to the couplings. 
The justification to drop these terms is again the localization of $\varphi_\rho$ on the curve $\Sigma_\rho$. It is then useful to define auxiliary functions
$\eta_\rho$ such that $\xi_\rho$ reduces to $-\eta_\rho$ away from the curve, as it is done in \cite{cchv09} where  $\eta_a$ and $\eta_b$ are 
denoted $\zeta_{12}$ and $\zeta_{23}$ respectively. In practice $\eta_\rho$ is found discarding the $\varphi_\rho$ dependence in (\ref{xi01}). Therefore,
\be
\eta_\rho  =   -\frac{i}{m^2 q_\Phi(\rho)} h_\rho  
+\frac{\eps}{m^4 q_\Phi(\rho)} \left[ \p_x \theta_0 \p_y \left(\frac{h_\rho}{q_\Phi(\rho)}\right)  
- \p_y \theta_0  \p_x\left(\frac{h_\rho}{q_\Phi(\rho)}\right)    \right] + \CO(\eps^2)
\label{etar}
\ee
Notice that $\eta_\rho$ is holomorphic. Thus, after integrating the total derivative in (\ref{yukc}) the coupling reduces to
\be
Y =  m_* f_{abc}\int_{\p S} \!\! \bar\p\left( h_c \eta_a \xi_b\right) \wedge dx \wedge dy 
\label{yukd}
\ee
Integrating the last total derivative then gives
\be
Y =  m_* f_{abc}\! \int_{\car} \left(h_c \eta_a \eta_b\right) dx \wedge dy 
\label{yuke}
\ee
where the final integration region is effectively the product of two circles, $|x|=\d_1$ and $|y|=\d_2$, each of infinitesimal radius.
Evaluating by residues gives\footnote{We also divide by $(2i)^2$ to match the normalization of ${\rm d vol}_S$ used in evaluating
overlap integrals of zero modes.}
\be
Y = m_* f_{abc} \pi^2 \,  {\rm Res}(h_c \, \eta_a \, \eta_b, x=0, y=0)
\label{resfor}
\ee
This is the residue formula. The couplings only depend on the holomorphic functions $h_\rho$ that determine $\eta_\rho$ through 
eq.~(\ref{etar}). We will shortly present explicit examples of couplings computed via residues.

\medskip

\noindent
{\underline{Regularity of $\xi_\rho^{(1)}$}}

The $\xi_\rho^{(1)}$ are easily found inserting the corrected wavefunctions in (\ref{xi01}). We will briefly discuss 
the sectors $a^+$, which is analogous to $b^+$, and $c^+$. To keep track of the generations in the $a^+$ and $b^+$ sectors 
we add superscripts $i$ and $j$. To connect with section \ref{ss:n0} we take $\theta_0 = i m^2(\th_{00} + x \th_{0x} + y \th_{0y})$. 

Lets us consider the $a^+$ sector. The correction to the wavefunction $\varphi_{a^+}^i$ is given in (\ref{chia11}).
Substituting in (\ref{xi01}), together with \mbox{$\varphi_{a^+}^{i\, (0)}= m_* e^{\lam_a x\bar u_a} f_i(v_a)$} 
and $h_{a^+}^{i}=m_* f_i(y)$, and Taylor expanding, leads to  
\be
\xi_{a^+}^{i\, (1)} \!=\! -\frac{i m_*}{m^2 x} \left[A_i(y) 
+ \frac{\zeta_a}2 \left(2\th_{0x} - \zeta_a\th_{0y}\right)f_i^{\prime\prime}(y)  \right]  + \cdots
\label{xiares}
\ee
where the ellipsis stands for terms regular at $x=0$. Hence, to avoid the pole at $x=0$ it suffices to take
\be
A_i(y) = \frac{\zeta_a}2
\left(\zeta_a\th_{0y} - 2\th_{0x}\right)f_i^{\prime\prime}(y) 
\label{ha1}
\ee
up to a holomorphic function that vanishes at $x=0$.
Since $f_i(y)=m_*^{(3-i)} \g_a^1 y^{(3-i)}$, we see that only $A_1 \not= 0$ and moreover
\be
A_1 = \g_a^1 a_0 \quad ; \quad a_0=m_*^2\zeta_a (\zeta_a\th_{0y} - 2\th_{0x})
\label{fixa01}
\ee
as reported in section \ref{ss:n0}. 
The results for the $b^+$ sector are analogous. For $\xi_{b^+}^{1\, (1)}$ to be free of poles 
$B_2=B_3=0$, whereas $B_1=\g_b^1 b_0$, with \mbox{$b_0=m_*^2\zeta_b (\zeta_b\th_{0x} - 2\th_{0y})$}.

We now address the $c^+$ sector.
The corrected wavefunction is given in (\ref{chic11}). In this case $\varphi_{c^+}^{(0)}  = m_*\g_c  e^{\zeta_c (x-y)\bar u_c}$, 
so that $h_{c^+}  = m_*\g_c$. Inserting in (\ref{xi01}) we arrive at
\be
\xi_{c^+}^{(1)} =  \frac{i}{m^2 (x-y)} m_*\g_c C  + \cdots
\label{xic1}
\ee
where $\cdots$ represents terms regular at $x=y$. Hence, to remove the pole at $x=y$ it is enough to set $C=0$,
as we announced in section \ref{ss:n0}.

\medskip
 
\noindent
{\underline{Couplings}}

We have just seen that $h_{c^+}=m_* \g_c$. Then, according to (\ref{resfor}), to compute the couplings $Y^{ij}$ to 
$\CO(\eps^2)$ we just need to extract the residue of $\eta_{a^+}^i \eta_{b^+}^j$ at $(x,y)=(0,0)$. 
Recall that the $a^+$ and $b^+$ sectors are analogous. We will present two illustrative examples. Other couplings can be found
in a similar way.

Let us first calculate $Y^{33}$.
For the third generation we have $h_{a^+}^{3}=m_* \g_a^3$, and likewise $h_{b^+}^{3}=m_* \g_b^3$. {}From (\ref{etar}) we find
\bea
\eta_{a^+}^3  & = &  \frac{i}{m^2 x} \left(m_* \g_a^3  + \frac{\eps m_* \g_a^3\th_{0y}}{x^2}\right)
+ \CO(\eps^2)
\label{azet3} \\[2mm]
\eta_{b^+}^3  & = &  -\frac{i}{m^2 y} \left(m_* \g_b^3  + \frac{\eps m_* \g_b^3 \th_{0x}}{y^2}\right)
+ \CO(\eps^2)
\label{bzet3}
\eea
Extracting ${\rm Res}(\eta_{a^+}^3 \, \eta_{b^+}^3, x=0, y=0)$ gives
\be
Y_{a^+b^+c^+}^{33} = \pi^2 \left(\frac{m_*}{m}\right)^4 \! f_{a^+b^+c^+} \, \g_c \g_a^3 \g_b^3 +  \CO(\eps^2) 
\label{y33res}
\ee
This coupling remains then uncorrected to $\CO(\eps^2)$, as we also found calculating the overlap integral of corrected wavefunctions.

We now compute $Y^{13}$. Recall that $f_1=m_*^2\g_a^1 (y+\zeta_a x)^2$, so that $h_{a^+}^{1}(y)=m_*^3 \g_a^1 y^2$. 
Inserting in (\ref{etar}) yields
\be
\eta_{a^+}^1   =   \frac{i}{m^2 x} \left[m_*^3 \g_a^1 y^2  
+ \frac{\eps m_*^3 \g_a^1}{x}\left(2\th_{0x} y + \frac{\th_{0y}}{x} y^2 \right)\right] + \CO(\eps^2)
\label{azet1} 
\ee
$\eta_{b^+}^3$ is still given by (\ref{bzet3}).
Evaluating the residue of $\eta_{a^+}^1 \eta_{b^+}^3$ at $(0,0)$ we find that
 \be
Y_{a^+b^+c^+}^{13} = \frac{\eps m_*^6}{m^4} \pi^2 f_{a^+b^+c^+} \, \g_c \g_a^1 \g_b^3 \th_{0x} + \CO(\eps^2)
\label{y13res}
\ee
in agreement with the coupling (\ref{y13}) obtained from the overlap integral of corrected zero modes.

\subsubsection*{Yukawas due to $\theta_2$}

We now start from the superpotential
\be
W=W_0 + W_1 =\int_S \tr\left ( \Phi_{xy}F \right )\,\w\,dx\,\w\,dy+\frac{\epsilon}{2}\int_S\theta_2\,\str\left (\Phi_{xy}^2 F\,\w\,F\right )
\label{supo2}
\ee
We will take $\theta_2$ constant.
The non-perturbative piece $W_1$ includes a cubic term, shown in (\ref{w1t}), depending on the
fluctuations $(\varphi, a)$. In turn this term implies the corrected Yukawa coupling (\ref{yuk1}) which can be written as 
\mbox{$Y_1 = Y_{1\langle \Phi\rangle} + Y_{1\langle F\rangle}$}. 
In the $SO(12)$ model, taking into account the group properties (\ref{dres}), $Y_{1\langle \Phi\rangle}$ is given in simplified notation by
\be
Y_{1\langle \Phi\rangle}  = \frac{i}2 \eps m_* f_{abc} \! \int_S \hat\theta_2 \left[ Da_a \wedge Da_b \, \varphi_{c xy} +  Da_c \wedge Da_a \, \varphi_{b xy} +
 Da_b \wedge Da_c \, \varphi_{a xy}  \right] 
\label{yuk1b}
\ee
where $\hat\theta_2$ is the coordinate dependent parameter
\be
\hat\theta_2=\frac{\theta_2}3 m^2 (y-2x)
\label{ht2def}
\ee
Similarly, for $\langle F \rangle = \mff \, Q_F$, the flux piece $Y_{1\langle F\rangle}$ reduces to
\be
Y_{1\langle F \rangle}  = \frac{i}6 \eps m_* f_{abc} \! \int_S \theta_2 \left( \varphi_{a xy}  \varphi_{b xy} Da_c +
 \varphi_{c xy}  \varphi_{a xy} Da_b   + \varphi_{b xy} \varphi_{c xy}  Da_a \right) \wedge \mff
\label{yuk1c}
\ee 
The total coupling can be expanded in powers of $\eps$ as 
\be
Y = Y_0 + Y_{1\langle \Phi\rangle} + Y_{1\langle F \rangle}  + \CO(\eps^2)
\label{expap}
\ee
where $Y_0$ arises from the tree-level superpotential $W_0$ and it is actually given in (\ref{yuka}).

In the $n=0$ case discussed before we have seen that to arrive at a residue formula we can use the F-terms to 
expose the total derivatives in the integrand in the full coupling $Y$. However, in principle this approach is now more difficult 
because the F-terms (\ref{FtermA02}) and (\ref{Ftermphi02}) are formidable equations to work with. A way to overcome
the difficulties is to switch to variables $\hat A_{\bar m}$ and $\hat \Phi_{xy}$ by performing the generalized SW map introduced in (\ref{ansatz}).
We will then first take a detour to show that in the new variables the F-term equations, as well as the Yukawa couplings,
are much simpler and in fact independent of fluxes.

With $\theta_2$ constant, and other $\theta_n=0$, the generalized SW map (\ref{ansatz}) is given by
\begin{subequations}
\label{ansatz2}
\begin{align}
\hat{A}_{\bar{m}} \, =\ & A_{\bar{m}} + \tilde{A}_{\bar{m}} \ =\, A_{\bar{m}} + 
\eps \theta_2 {\rm S}  \left\{\Phi_{xy} \left[A_x (\p_y A_{\bar{m}} + F_{y\bar{m}}) - A_y(\p_x A_{\bar{m}} + F_{x\bar{m}}) \right]\right\} 
\label{ansatzA2} \\
\hat{\Phi}_{xy}\, = \ & \Phi_{xy} + \tilde{\Phi}_{xy}  \, = \, \Phi_{xy} + 
\frac{\eps}{2} \theta_2 {\rm S} \left[ A_x (\p_y + D_y) (\Phi_{xy}^2) -  A_y (\p_x + D_x) (\Phi_{xy}^2) \right]
\label{ansatzPhi2}
\end{align}
\end{subequations} 
up to $\CO(\eps^2)$ corrections. We also have to expand in fluctuations $(\hat\varphi_{xy}, \hat a_{\bar m})$ defined by
\be
\hat \Phi_{xy}\, =\, \langle \hat \Phi_{xy} \rangle + \hat\varphi_{xy} \quad ; \quad \hat A_{\bar{m}}\, =\, \langle \hat A_{\bar{m}} \rangle + \hat a_{\bar{m}}
\label{flucthat}
\ee
In turn the vevs and the fluctuations have expansions in powers of $\eps$ of the form 
\be
\langle \hat \Phi_{xy} \rangle = \langle \hat \Phi_{xy} \rangle^{(0)} +   \langle \hat \Phi_{xy} \rangle^{(1)} + \CO(\eps^2)
\label{pvevhat}
\ee 
In the holomorphic gauge $\langle A_{\bar m} \rangle^{(0)}=0$ the vevs reduce to
\be
\langle \hat \Phi_{xy} \rangle = \langle \Phi_{xy} \rangle^{(0)} + \CO(\eps^2) \quad ; \quad
\langle \hat A_{\bar m} \rangle = 0 + \CO(\eps^2)  
\label{pvevhath}
\ee
as can be proved using (\ref{solAbkg}) and (\ref{solPhibkg}). For the fluctuations we find
\be
\hat \varphi_{xy}\, =\, \varphi_{xy}  + \tilde \varphi_{xy} \quad ; \quad \hat a_{\bar{m}}\, =\, a_{\bar{m}}  + \tilde a_{\bar{m}}
\label{flucthat2}
\ee
where
\begin{subequations}
\label{aphitf}
\begin{align}
\tilde \varphi_{xy} \, =\  & 2 \eps \theta_2  {\rm S} \left[ \langle A_x \rangle \p_y (\langle \Phi_{xy}\rangle \varphi_{xy}) -  
\langle A_y \rangle \p_x (\langle \Phi_{xy}\rangle \varphi_{xy}) \right] 
\label{phitf} \\
\tilde{a}_{\bar{m}} \, =\ &  
2 \eps \theta_2 {\rm S}  \left[\langle \Phi_{xy}\rangle \left(\langle A_x \rangle  \p_y a_{\bar{m}}  - \langle A_y \rangle \p_x a_{\bar{m}} \right)\right] \nonumber \\
& + \eps \theta_2 {\rm S}  \left[\varphi_{xy} \left(\langle A_x \rangle  \langle F_{y\bar m} \rangle  - \langle A_y \rangle  \langle F_{x\bar m} \rangle  \right)\right] 
\label{atf}
\end{align}
\end{subequations} 
up to $\CO(\eps^2)$ terms.

To derive the equations of motion satisfied by $\hat \varphi_{xy}$ and $ \hat a_{\bar{m}}$ we substitute (\ref{flucthat2}), together with
(\ref{aphitf}), into (\ref{FtermA02}) and (\ref{Ftermphi02}). In the end we obtain
\begin{subequations}
\label{hatfterms}
\begin{align} 
\p_{\bar{x}} \hat a_{\bar{y}} - \p_{\bar{y}} \hat a_{\bar{x}} \, = \ &  0 
\label{fth1}\\
\p_{\bar{m}} \hat \vphi_{xy}+ i [\langle \Phi_{xy} \rangle^{(0)}, \hat a_{\bar{m}}]  \,  = \ & 
2\eps \theta_2
{\rm S}\left[\langle \Phi_{xy}\rangle \left(\p_x \hat a_{\bar m} \partial_y\langle \Phi_{xy} \rangle 
- \p_y \hat a_{\bar m}   \partial_x\langle \Phi_{xy} \rangle\right) \right]^{(0)}  
\label{fth2}
\end{align}
\end{subequations}
up to $\CO(\eps^2)$ terms. 
Notice that these equations are independent of fluxes. In fact they can be derived from (\ref{FtermA02}) and (\ref{Ftermphi02}) ignoring flux terms
and replacing  $(a_{\bar m} ,\vphi_{xy})$ by  $(\hat a_{\bar m},\hat\vphi_{xy})$. For instance, using (\ref{s3b}) from (\ref{fth2}) we deduce 
\be
\bar\p_{\bar m} \hat\varphi_{\rho xy} = - i m^2 q_\Phi(\rho) \hat a_{\rho \bar m} - 
2\eps \theta_2 m^4 \left[t_{x}(\rho) \partial_y \hat a_{\rho \bar m} 
- t_{y}(\rho) \partial_x \hat a_{\rho \bar m}  \right] + \CO(\eps^2)
\label{fth3}
\ee
where the $t_m(\rho)$ are defined in (\ref{tmrho}). This equation corresponds to the F-term (\ref{fnpsr}),
which together with the equivalent of (\ref{fth1}), was applied in section \ref{ss:n2} to find the zero modes in the absence of fluxes.
Thus, the wavefunctions for $\hat\vphi_{\rho xy}$ and $\hat a_{\rho \bar m}$ can be inferred from the results in section \ref{ss:n2}. In particular,
we see that these wavefunctions are localized on the curves $\Sigma_\rho$.

We now continue as in the $n=0$ case. To begin notice that (\ref{fth3}) can be written as
\be
\bar\p \hat\varphi_{\rho} = - i m^2 q_\Phi(\rho) \hat a_{\rho } - 
2\eps \theta_2 m^4\, t(\rho) \wedge \p \hat a_{\rho} + \CO(\eps^2)
\label{fth4}
\ee
Recall also that $\p t(\rho)=0$. Since
(\ref{fth1}) means \mbox{$\hat a_{\rho} = \bar\p \hat \xi_\rho + \CO(\eps^2)$} we can integrate (\ref{fth4}) to obtain
\be
\hat\varphi_{\rho} = \hat h_\rho\,  dx\wedge dy - i m^2 q_\Phi(\rho) \hat \xi_{\rho} \, dx \wedge dy - 
2\eps \theta_2 m^4  \, t(\rho) \wedge \p \hat \xi_{\rho} + \CO(\eps^2) 
\label{hvphir}
\ee
where $\hat h_\rho$ is a holomorphic function. We can solve for $\hat \xi_\rho$ iteratively and then define auxiliary functions $\hat \eta_\rho$ such that
$\hat \xi_\rho$ reduces to $-\hat\eta_\rho$ on the curve $\Sigma_\rho$. It follows that
\be
\hat\eta_\rho   =     -\frac{i}{m^2 q_\Phi(\rho)} \hat h_\rho
 +  \frac{2\eps\theta_2}{q_\Phi(\rho)} \left[ t_x(\rho) \p_y \left(\frac{\hat h_\rho}{q_\Phi(\rho)}\right)  
- t_y(\rho) \p_x\left(\frac{\hat h_\rho}{q_\Phi(\rho)}\right)    \right] + \CO(\eps^2)
\label{hetar}
\ee
We see that the $\hat \eta_\rho$ are also holomorphic.

Let us now discuss the Yukawa couplings. To begin we will express them in terms of the hat variables by substituting (\ref{flucthat2}) in
the full expansion (\ref{expap}). The result hinges crucially on the corrections (\ref{aphitf}) in which we can insert the 
vevs to order $\eps^0$. In the $SO(12)$ model, with $\langle A_m \rangle = \mfa_m \, Q_F$,  we find
\be
\tilde \varphi_{\rho xy}  =  -2 \eps m^2 \theta_2  \left[ \mfa_x \p_y (r_\rho \varphi_{\rho xy}) -  
\mfa_y \p_x (r_\rho \varphi_{\rho xy}) \right] 
\label{phitfb} 
\ee
Here $r_\rho=x s_{xx}(\rho) + (x+y) s_{xy}(\rho) + y s_{yy}(\rho)$, where the $s_{mn}(\rho)$ are provided in table \ref{t1}. Similarly,
\be
\tilde{a}_{\rho \bar{m}}  =   
- 2 \eps \theta_2 m^2 r_\rho  \left[\mfa_x  \p_y a_{\rho \bar{m}}  - \mfa_y \p_x a_{\rho \bar{m}} \right] 
 - \eps \theta_2 s_\rho  \left[\mfa_x \p_{\bar m} \mfa_y  - \mfa_y \p_{\bar m} \mfa_x\right]  \varphi_{\rho xy} 
\label{atfb}
\ee
where  $s_\rho=s_{xx}(\rho) + 2s_{xy}(\rho) + s_{yy}(\rho)$. Since we are working in the holomorphic gauge we have written $\mff_{n\bar m} = - \p_{\bar m} \mfa_n$
for the flux components.

After long but direct calculations we find that the full Yukawa couplings take the form
\be
Y = \hat Y_0 + \hat Y_{1\langle \Phi\rangle} + \CO(\eps^2)
\label{exphat}
\ee
where $\hat Y_0$ and $\hat Y_{1\langle \Phi\rangle}$ are obtained by replacing $(a_\rho,\vphi_\rho)$ with $(\hat a_\rho, \hat\vphi_\rho)$ in the
formulas (\ref{yuka}) for $Y_0$, and (\ref{yuk1b}) for $Y_{1\langle \Phi\rangle}$. Moreover, in $\hat Y_{1\langle \Phi\rangle}$ all flux dependence
disappears because only normal derivatives $\p \hat a_{\rho}$ do enter in the final expression. In deriving (\ref{exphat}) we have used the F-terms
(\ref{hatfterms}). We have also dropped boundary terms that vanish due to localization of the wavefunctions.  

The next step is to use repeatedly the equations satisfied by $\hat a_\rho$ and $\hat \vphi_\rho$ to show that (\ref{exphat}) implies
\be
Y = m_* f_{abc} \! \int_S \hat h_c\bigg[\hat a_a \wedge \hat a_b \wedge dx \wedge dy + \frac{i\eps\hat\theta_2}{\sm2} \p \hat a_a \wedge \p \hat a_b \bigg]  
\label{yfhat}
\ee
In the proof we have again discarded boundary terms. The result also relies on the properties
\be
\p (q_\Phi(b^+) \hat \theta_2) = 4 m^2 \theta_2 [t(a^+) - t(c^+)] \quad ; \quad \p (q_\Phi(a^+) \hat \theta_2) = 4 m^2 \theta_2 [t(c^+) - t(b^+)]
\label{propdq}
\ee
which are valid in the $SO(12)$ model. At this point we can use $\hat a_\rho = \bar \p \hat \xi_\rho + \CO(\eps^2)$ to continue integrating.
Furthermore, at the boundary we can substitute $\hat \xi_\rho$ by $\hat \eta_\rho$ thanks to localization. In this way we find 
\be
Y =  m_* f_{abc}\! \int_{\car} \hat h_c\bigg[ \hat \eta_a \hat \eta_b - \frac{i\eps\hat\theta_2}{\sm2} 
\left( \p_x \hat\eta_a \p_y\hat\eta_b -  \p_y \hat\eta_a \p_x\hat\eta_b \right)
 \bigg] dx \wedge dy 
\label{yuk2e}
\ee
Evaluating by residues as in the $n=0$ case then gives
\be
Y =  m_* f_{abc} \pi^2 \,  {\rm Res}\bigg( \hat h_c\bigg[ \hat \eta_a \hat \eta_b - \frac{i\eps\hat\theta_2}{\sm2} 
\left( \p_x \hat\eta_a \p_y\hat\eta_b -  \p_y \hat\eta_a \p_x\hat\eta_b \right) \bigg] , x=0, y=0\bigg)
\label{resfor2}
\ee
This is our final residue formula.

Couplings can now be easily calculated. To this purpose we first determine $\hat \eta_\rho$ from (\ref{hetar}). The holomorphic
functions $\hat h_\rho$ must be such that $\hat \xi_\rho^{(0)}$ is regular at $q_\Phi(\rho)=0$. This implies 
\be
\hat h_\rho = \hat \vphi_{\rho xy}^{(0)} \lvert_{q_\Phi(\rho)=0} \, =   \vphi_{\rho xy} ^{(0)} \lvert_{q_\Phi(\rho)=0} 
\label{hh0}
\ee
Therefore, $\hat h_{a^+}(y) = m_* f_i(y)$, $\hat h_{b^+}(x) = m_* g_j(x)$, and $\hat h_{c^+} =m_* \g_c$.
{}From the results in section \ref{ss:n2} we can further show that the resulting $\hat \xi_\rho^{(1)}$ are indeed
regular. 
 
Since $\hat h_{c^+}=m_* \g_c$, to compute the couplings $Y^{ij}$ to 
$\CO(\eps^2)$ it suffices to determine $\hat \eta_{a^+}^i$ and $\hat \eta_{b^+}^j$. As a first example let us consider $Y^{33}$. 
Substituting $\hat h_{a^+}^{3}(y) = m_* \g_a^3$, and $\hat h_{b^+}^{3}(x) = m_*\g_b^3$ in (\ref{hetar}) yields
\be
\hat \eta_{a^+}^3   =   \frac{i m_*\g_a^3}{m^2 x} - \frac{\eps m_* \theta_2\g_a^3}{2x^3} (x-y)
\quad ; \quad
\hat \eta_{b^+}^3   =   -\frac{i m_*\g_b^3}{m^2 y} 
\label{hh3}
\ee
to $\CO(\eps^2)$. Plugging in (\ref{resfor2}) readily reproduces the uncorrected coupling in (\ref{y33res}).
For a more interesting example we calculate $Y^{32}$. With $\hat h_{b^+}^{2} = m_*^2 \g_b^2 x$ we find
\be
\hat \eta_{b^+}^2   =   m_*^2 \g_b^2\bigg(\frac{-ix}{m^2 y} - \frac{\eps \theta_2}{6 y} \bigg)
\label{hh2}
\ee
{}From (\ref{resfor2}) we now obtain 
\be
Y_{a^+b^+c^+}^{32} = \frac{i\eps m_*^5 \pi^2}{6m^2}  f_{a^+b^+c^+} \, \g_c \g_a^3 \g_b^2 \, \th_2 + \CO(\eps^2)
\label{y32res}
\ee
which matches (\ref{y32}) taking $\theta_2=i\theta_{20}$. It is easy to check that the coupling $Y^{23}$ derived from the residue formula
agrees with (\ref{y23}), and that all other couplings vanish to $\CO(\eps^2)$.


\begin{thebibliography}{10}

\bibitem{thebook}
For an overview of string phenomenology see   L.E. Ib\'a\~nez and A.M. Uranga, 
  {\it String Theory and Particle Physics. An Introduction to String Phenomenology},
  Cambridge University Press (2012).

  
\bibitem{aiqu}
  G.~Aldaz\'abal, L.~E.~Ib\'a\~nez, F.~Quevedo and A.~M.~Uranga,
  {\em ``D-branes at singularities: A Bottom up approach to the string embedding of
  the standard model,''}
  JHEP {\bf 0008}, 002 (2000)
  [arXiv:hep-th/0005067].


\bibitem{dw1}
  R.~Donagi, M.~Wijnholt,
  {\em ``Model Building with F-Theory,''}
  [arXiv:0802.2969 [hep-th]].

\bibitem{bhv1}
  C.~Beasley, J.~J.~Heckman and C.~Vafa,
 {\em ``GUTs and Exceptional Branes in F-theory - I,''}
  JHEP {\bf 0901} (2009) 058
  [arXiv:0802.3391 [hep-th]].

\bibitem{bhv2}
  C.~Beasley, J.~J.~Heckman and C.~Vafa,
 {\em ``GUTs and Exceptional Branes in F-theory - II: Experimental Predictions,''}
  JHEP {\bf 0901} (2009) 059
  [arXiv:0806.0102 [hep-th]].
  
\bibitem{dw2}
  R.~Donagi and M.~Wijnholt,
  {\em ``Breaking GUT Groups in F-Theory,''}
 [ arXiv:0808.2223 [hep-th]].
 
 \bibitem{ftheoryreviews}
   J.~J.~Heckman,
  {\em ``Particle Physics Implications of F-theory,''}
  arXiv:1001.0577 [hep-th]; 
      T.~Weigand,
  {\em ``Lectures on F-theory compactifications and model building,''}
  Class.\ Quant.\ Grav.\  {\bf 27}, 214004 (2010)
  [arXiv:1009.3497 [hep-th]];
  L.~E.~Ib\'a\~nez,
  {\em ``From Strings to the LHC: Les Houches Lectures on String Phenomenology,''}
  arXiv:1204.5296 [hep-th];
  G.~K.~Leontaris,
  {\em ``Aspects of F-Theory GUTs,''}
  PoS CORFU {\bf 2011} (2011) 095
  [arXiv:1203.6277 [hep-th]].
  
  
\bibitem{hktw}
 H.~Hayashi, T.~Kawano, R.~Tatar and T.~Watari,
  {\em ``Codimension-3 Singularities and Yukawa Couplings in F-theory,''}
  Nucl.\ Phys.\  B {\bf 823}, 47 (2009)
  [arXiv:0901.4941 [hep-th]].
  
\bibitem{hktw2}
  H.~Hayashi, T.~Kawano, Y.~Tsuchiya and T.~Watari,
  {\em ``Flavor Structure in F-theory Compactifications,''}
  JHEP {\bf 1008}, 036 (2010)
  [arXiv:0910.2762 [hep-th]].

 
  \bibitem{hv08}
 J.~J.~Heckman and C.~Vafa,
  {\em ``Flavor Hierarchy From F-theory,''}
  Nucl.\ Phys.\  B {\bf 837} (2010) 137
  [arXiv:0811.2417 [hep-th]].
  
  
\bibitem{fi1}
  A.~Font and L.~E.~Ib\'a\~nez,
  {\em ``Yukawa Structure from U(1) Fluxes in F-theory Grand Unification,''}
  JHEP {\bf 0902}, 016 (2009)
  [arXiv:0811.2157 [hep-th]].
  
  
  
\bibitem{DudasPalti}
  E.~Dudas and E.~Palti,
  {\em``Froggatt-Nielsen models from E(8) in F-theory GUTs,''}
  JHEP {\bf 1001}, 127 (2010)
  [arXiv:0912.0853 [hep-th]].
  
\bibitem{Ross}
G.~K.~Leontaris and G.~G.~Ross,
  {\em ``Yukawa couplings and fermion mass structure in F-theory GUTs,''}
  JHEP {\bf 1102}, 108 (2011)
  [arXiv:1009.6000 [hep-th]].

\bibitem{Krippendorf}
  S.~Krippendorf, M.~J.~Dolan, A.~Maharana and F.~Quevedo,
  {\em ``D-branes at Toric Singularities: Model Building, Yukawa Couplings and
  Flavour Physics,''}
  JHEP {\bf 1006}, 092 (2010)
  [arXiv:1002.1790 [hep-th]].

  \bibitem{yukint}
  D.~Cremades, L.~E.~Ib\'a\~nez and F.~Marchesano,
  {\em ``Yukawa couplings in intersecting D-brane models,''}
  JHEP {\bf 0307}, 038 (2003)
  [arXiv:hep-th/0302105].
  
  \bibitem{magnus}
  D.~Cremades, L.~E.~Ib\'a\~nez and F.~Marchesano,
  {\em ``Computing Yukawa couplings from magnetized extra dimensions,''}
  JHEP {\bf 0405}, 079 (2004)
  [arXiv:hep-th/0404229].
 
   \bibitem{ms04}  
  F.~Marchesano and G.~Shiu,
  {\em ``MSSM vacua from flux compactifications,''}
  Phys.\ Rev.\  D {\bf 71}, 011701 (2005)
  [arXiv:hep-th/0408059].
   {\em ``Building MSSM flux vacua,''}
  JHEP {\bf 0411}, 041 (2004)
  [arXiv:hep-th/0409132].
 
 
\bibitem{cchv09}
  S.~Cecotti, M.~C.~N.~Cheng, J.~J.~Heckman and C.~Vafa,
  {\em ``Yukawa Couplings in F-theory and Non-Commutative Geometry,''}
  [arXiv:0910.0477 [hep-th]].


\bibitem{cp09}
  J.~P.~Conlon and E.~Palti,
  {\em ``Aspects of Flavour and Supersymmetry in F-theory GUTs,''}
  JHEP {\bf 1001}, 029 (2010)
  [arXiv:0910.2413 [hep-th]].

   \bibitem{fi09}
  A.~Font and L.~E.~Ib\'a\~nez,
  {\em ``Matter wave functions and Yukawa couplings in F-theory Grand Unification,''}
  JHEP {\bf 0909}, 036 (2009)
  [arXiv:0907.4895 [hep-th]].

    
\bibitem{mm09}
  F.~Marchesano and L.~Martucci,
  {\em ``Non-perturbative effects on seven-brane Yukawa couplings,''}
  Phys.\ Rev.\ Lett.\  {\bf 104}, 231601 (2010)
  [arXiv:0910.5496 [hep-th]].


\bibitem{afim}
  L.~Aparicio, A.~Font, L.~E.~Ib\'a\~nez, F.~Marchesano,
  {\em ``Flux and Instanton Effects in Local F-theory Models and Hierarchical Fermion Masses,''}
  JHEP {\bf 1108} (2011) 152
  [arXiv:1104.2609 [hep-th]].

  
\bibitem{collinucci09}
  A.~Collinucci,
  {\em ``New F-theory lifts. II. Permutation orientifolds and enhanced singularities,''}
  JHEP {\bf 1004}, 076 (2010)
  [arXiv:0906.0003 [hep-th]].


\bibitem{bgjw09}
 R.~Blumenhagen, T.~W.~Grimm, B.~Jurke and T.~Weigand,
  {\em ``F-theory uplifts and GUTs,''}
  JHEP {\bf 0909}, 053 (2009)
  [arXiv:0906.0013 [hep-th]].
  {\em ``Global F-theory GUTs,''}
  Nucl.\ Phys.\  B {\bf 829}, 325 (2010)
  [arXiv:0908.1784 [hep-th]].

\bibitem{mss2}
  J.~Marsano, N.~Saulina and S.~Schafer-Nameki,
  {\em ``Monodromies, Fluxes, and Compact Three-Generation F-theory GUTs,''}
  JHEP {\bf 0908}, 046 (2009)
  [arXiv:0906.4672 [hep-th]].
  
 
\bibitem{Cordova}
  C.~Cordova,
  {\em ``Decoupling Gravity in F-Theory,''}
  Adv.\ Theor.\ Math.\ Phys.\  {\bf 15}, 689 (2011)
  [arXiv:0910.2955 [hep-th]].
  
\bibitem{mss}
  J.~Marsano, N.~Saulina and S.~Schafer-Nameki,
  {\em ``Compact F-theory GUTs with U(1) (PQ),''}
  JHEP {\bf 1004}, 095 (2010)
  [arXiv:0912.0272 [hep-th]].


\bibitem{gkw09}
  T.~W.~Grimm, S.~Krause and T.~Weigand,
  {\em ``F-Theory GUT Vacua on Compact Calabi-Yau Fourfolds,''}
  JHEP {\bf 1007}, 037 (2010)
  [arXiv:0912.3524 [hep-th]].
  
  
\bibitem{grimm}
  T.~W.~Grimm,
  {\em ``The N=1 effective action of F-theory compactifications,''}
  Nucl.\ Phys.\  B {\bf 845}, 48 (2011)
  [arXiv:1008.4133 [hep-th]].
  
 
  
\bibitem{dp10}
  E.~Dudas and E.~Palti,
  {\em ``On hypercharge flux and exotics in F-theory GUTs,''}
  JHEP {\bf 1009}, 013 (2010)
  [arXiv:1007.1297 [hep-ph]].

  
  \bibitem{japan}  
  H.~Abe, T.~Kobayashi and H.~Ohki,
  {\em ``Magnetized orbifold models,''}
  JHEP {\bf 0809}, 043 (2008)
  [arXiv:0806.4748 [hep-th]]; \\
  H.~Abe, K.~S.~Choi, T.~Kobayashi and H.~Ohki,
  {\em ``Three generation magnetized orbifold models,''}
  Nucl.\ Phys.\  B {\bf 814}, 265 (2009)
  [arXiv:0812.3534 [hep-th]].


\bibitem{ConlonWF}
  J.~P.~Conlon, A.~Maharana and F.~Quevedo,
  {\em ``Wave Functions and Yukawa Couplings in Local String Compactifications,''}
  JHEP {\bf 0809}, 104 (2008)
  [arXiv:0807.0789 [hep-th]].


\bibitem{DiVecchia}
  P.~Di Vecchia, A.~Liccardo, R.~Marotta, F.~Pezzella,
  {\em ``K\"ahler Metrics and Yukawa Couplings in Magnetized Brane Models,''}
  JHEP {\bf 0903 } (2009)  029.
  [arXiv:0810.5509 [hep-th]].


\bibitem{cm09}
 P.~G.~C\'amara and F.~Marchesano,
  {\em ``Open string wavefunctions in flux compactifications,''}
  JHEP {\bf 0910}, 017 (2009)
  [arXiv:0906.3033 [hep-th]].
   {\em ``Physics from open string wavefunctions,''}
  PoS E {\bf PS-HEP2009}, 390 (2009).
  
  
 \bibitem{georgi}
 H. Georgi, {\em Lie algebras in Particle Physics: from isospin to unified theories},
 Westview Press, 1999.
  
  
   \bibitem{palti12}
    E.~Palti,
  {\em ``Wavefunctions and the Point of $E_8$ in F-theory,''}
   JHEP {\bf 1207}, 065 (2012)
  [arXiv:1203.4490 [hep-th]].
  
  \bibitem{nosusy}
  L.~E.~Ib\'a\~nez, F.~Marchesano, D.~Regalado and I.~Valenzuela,
  {\em ``The Intermediate Scale MSSM, the Higgs Mass and F-theory Unification,''}
  JHEP {\bf 1207}, 195 (2012)
  [arXiv:1206.2655 [hep-ph]].
  
  \bibitem{bdkmmm06}
  D.~Baumann, A.~Dymarsky, I.~R.~Klebanov, J.~M.~Maldacena, L.~P.~McAllister and A.~Murugan,
  {\em ``On D3-brane Potentials in Compactifications with Fluxes and Wrapped D-branes,''}
  JHEP {\bf 0611}, 031 (2006)
  [hep-th/0607050].
  
  \bibitem{ag06}
  S.~A.~Abel and M.~D.~Goodsell,
  {\em ``Realistic Yukawa couplings through instantons in intersecting brane worlds,''}
  JHEP {\bf 0710}, 034 (2007)
  [arXiv:hep-th/0612110].
  
\bibitem{bcm11}
  M.~Bianchi, A.~Collinucci and L.~Martucci,
  {\em ``Magnetized E3-brane instantons in F-theory,''}
  JHEP {\bf 1112}, 045 (2011)
  [arXiv:1107.3732 [hep-th]].
  
\bibitem{cgh11}
  M.~Cveti\v c, I.~Garc\'ia-Etxebarria and J.~Halverson,
  {\em ``Three Looks at Instantons in F-theory -- New Insights from Anomaly Inflow, String Junctions and Heterotic Duality,''}
  JHEP {\bf 1111}, 101 (2011)
  [arXiv:1107.2388 [hep-th]].
  
  \bibitem{dm10}
  A.~Dymarsky and L.~Martucci,
  {\em ``D-brane non-perturbative effects and geometric deformations,''}
  arXiv:1012.4018 [hep-th].

\bibitem{sw99}
  N.~Seiberg and E.~Witten,
  {\em ``String theory and noncommutative geometry,''}
  JHEP {\bf 9909}, 032 (1999)
  [arXiv:hep-th/9908142].

\bibitem{Ross:2007az}
  G.~Ross and M.~Serna,
  {\em ``Unification and fermion mass structure,''}
  Phys.\ Lett.\ B {\bf 664} (2008) 97
  [arXiv:0704.1248 [hep-ph]].
  
\bibitem{Elor:2012ig}
  G.~Elor, L.~J.~Hall, D.~Pinner and J.~T.~Ruderman,
  {\em ``Yukawa Unification and the Superpartner Mass Scale,''}
  JHEP {\bf 1210} (2012) 111
  [arXiv:1206.5301 [hep-ph]].
  
\bibitem{GJ}
  H.~Georgi, C.~Jarlskog,
 {\em ``A New Lepton - Quark Mass Relation in a Unified Theory,''}
 Phys.\ Lett.\  {\bf B86 } (1979)  297-300.
  
  \bibitem{bora}
  K.~Bora,
  {\em ``Updated values of running quark and lepton masses at GUT scale in SM, 2HDM and MSSM,''}
  arXiv:1206.5909 [hep-ph].

 \bibitem{cdp11}
  P.~G.~C\'amara, E.~Dudas and E.~Palti,
  {\em ``Massive wavefunctions, proton decay and FCNCs in local F-theory GUTs,''}
  JHEP {\bf 1112}, 112 (2011)
  [arXiv:1110.2206 [hep-th]].
  
   
  \bibitem{bdkkm10}
  D.~Baumann, A.~Dymarsky, S.~Kachru, I.~R.~Klebanov and L.~McAllister,
  {\em ``D3-brane Potentials from Fluxes in AdS/CFT,''}
  JHEP {\bf 1006}, 072 (2010)
  [arXiv:1001.5028 [hep-th]].

\bibitem{cdhm12}
  M.~Cvetic, R.~Donagi, J.~Halverson and J.~Marsano,
  JHEP {\bf 1211}, 004 (2012)
  [arXiv:1209.4906 [hep-th]].

\bibitem{Myers}
  R.~C.~Myers,
  {\em ``Dielectric branes,''}
  JHEP {\bf 9912}, 022 (1999)
  [arXiv:hep-th/9910053].



 \bibitem{km07}
  P.~Koerber and L.~Martucci,
  {\em ``From ten to four and back again: how to generalize the geometry,''}
  JHEP {\bf 0708}, 059 (2007)
  [arXiv:0707.1038 [hep-th]].


  \bibitem{Martucci06}
  L.~Martucci,
  {\em``D-branes on general N=1 backgrounds: Superpotentials and D-terms,''}
  JHEP {\bf 0606}, 033 (2006)
  [hep-th/0602129].

 \bibitem{mms10}
  F.~Marchesano, P.~McGuirk and G.~Shiu,
  {\em ``Chiral matter wavefunctions in warped compactifications,''}
  JHEP {\bf 1105}, 090 (2011)
  [arXiv:1012.2759 [hep-th]].
    

\bibitem{cchv10}
  S.~Cecotti, C.~Cordova, J.~J.~Heckman and C.~Vafa,
 {\em ``T-Branes and monodromy,''}
  JHEP {\bf 1107}, 030 (2011)
  [arXiv:1010.5780 [hep-th]].

  
\end{thebibliography}
\end{document}